%% file: main.tex
\documentclass[conference]{IEEEtran}
\IEEEoverridecommandlockouts
\usepackage[T1]{fontenc}

\usepackage{cite}
\usepackage{amsmath,amssymb,amsfonts}
\usepackage{graphicx}
\usepackage{textcomp}
\usepackage{xcolor}
\def\BibTeX{{\rm B\kern-.05em{\sc i\kern-.025em b}\kern-.08em
    T\kern-.1667em\lower.7ex\hbox{E}\kern-.125emX}}

 \usepackage{graphicx}
\usepackage[normalem]{ulem}

\input{macros}

\begin{document}

\title{Allegories of Symbolic Manipulations
 }

\author{\IEEEauthorblockN{Francesco Gavazzo}
\IEEEauthorblockA{
%\textit{dept. name of organization (of Aff.)} \\
\textit{University of Pisa}\\
francesco.gavazzo@unipi.it
%\href{mailto:francesco.gavazzo@unipi.it}{francesco.gavazzo@unipi.it}
}
}

\maketitle

\renewcommand{\a}{R}
\renewcommand{\b}{b}
\newcommand{\s}{s}
\renewcommand{\G}{\Gamma}
\newcommand{\dd}{\meno}
\newcommand{\dddd}{^{\scriptstyle \circ \circ}}

\newcommand{\ddstar}{^{{\scriptstyle \circ}*}}
\newcommand{\stardd}{^{*{\scriptstyle \circ}}}

\newcommand{\meno}{^{\scriptstyle \circ}}

\newcommand{\algone}{\alpha}

\renewcommand{\leqq}{\leq}

\maketitle

% \epigraph{\emph{On May 2, the first author became a father. 
% The theory presented in this work is dedicated to his son, Giulio Febo.
% }}{}

\begin{abstract}
Moving from the mathematical theory of (abstract) syntax, we develop a
general \emph{relational} theory of symbolic manipulation parametric with respect to, 
and accounting for, general notions of syntax. 
We model syntax relying on categorical notions, 
such as free algebras and monads, and show that a general theory of 
symbolic manipulation in the style of rewriting systems can be obtained 
by extending such notions to an allegorical setting. 
This way, we obtain an augmented calculus of relations accounting for syntax-based rewriting.
We witness the effectiveness of the relational approach  
by generalising and unifying milestones results 
in rewriting, such as the parallel moves and the Tait-Martin-L\"of techniques.
\end{abstract}
\begin{IEEEkeywords}
Rewriting, Relational Reasoning, Confluence
\end{IEEEkeywords}

\section{Introduction}
\label{section:introduction}
The study of symbolic expressions and their manipulation has always been 
one of the prime goals of mathematics and, even more, of theoretical computer science. 
Indeed, it is precisely its ``symbol pushing'' that makes \emph{symbolic reasoning}, 
and the associated notion of \emph{symbolic computation},
so effective. Charles Wells\footnote{See \url{https://abstractmath.org/MM/} 
(The Symbolic Language of Math).} 
greatly summarises the deep and 
fundamental relationship 
between symbolic expressions and their manipulation by writing
\emph{symbolic terms are encapsulated computations}.

The study of symbolic expressions as used by mathematicians, logicians, and 
computer scientists has been of little interest to mathematicians and logicians
for a long time. Remarkable achievements, instead, have been achieved by 
computer scientists starting from the seventies~\cite{Goguen/1977,pitts-nominal-sets,pitts-nominal-1,pitts-nominal-2,pitts-nominal-3,pitts-nominal-4,fiore-plotkin-turi-99,fiore-polynomial-functors-simple-type-theories,hamana-free-sigma-monoids,hamana-polymorphic-syntax,Hofmann-higher-order-abstract-syntax,pfenning-higher-order-abstract-syntax,Miller}, mostly in the field of 
programming language theory, where algebraic theories of (abstract) 
syntax have been developed in terms of initial algebras and 
free monads~\cite{Goguen/1977}.
% and important syntactic notions -- such as variable binding, (capture-avoiding) 
% substitution, and private name  -- 
% have been formalised. 
Altogether, these results gave raise to a new research field 
which is generically referred to  
as \emph{mathematics of syntax}, where syntax is tacitly understood as being abstract.

Symbolic syntax, however, is largely sterile without 
a (symbolic) \emph{dynamics} or \emph{operational semantics}.\footnote{
Operational semantics usually covers a large class of semantic behaviours, symbolic manipulation being just one of them.} 
%The theory of such a symbolic, discrete transformation 
Traditionally, 
\emph{rewriting theory}~\cite{newman,terese,term-rewriting-and-all-that,Huet80} 
is the 
discipline studying such symbolic, discrete transformations between expressions. 
However, even if considerably older than the aforementioned mathematics of 
syntax, rewriting has arguably not yet reached the same level of generality. 
Symbolic expressions, in fact, come in different flavours: 
they can be first-order, as in algebra, or
higher-order, as in calculus;  
typed, as in programming language theory, 
or untyped, as in logic; one-dimensional, as in traditional mathematics, 
or two-dimensional, as in category theory, etc.
This simple observation is at the very heart of the mathematical 
theory of syntax previously mentioned: there, different forms of syntax 
are obtained by different data structures which are 
uniformly understood in terms of algebras~\cite{Goguen/1977}. 

\subsubsection{Rewriting} 
Does anything similar happen to rewriting? Not really! Indeed, 
each kind of expressions previously mentioned leads to the 
development of a specific, syntax-based rewriting theory. Term rewriting~\cite{terese,term-rewriting-and-all-that,Huet80}, for instance,  
studies the symbolic manipulation of first-order expressions, 
whereas higher-order~\cite{terese,klop} and nominal~\cite{nominal-rewriting} rewriting focus on 
expressions with variable binding. 
This proliferation of \emph{ad hoc} rewriting 
formalism has prevented rewriting from qualifying as a general theory 
of symbolic manipulation. 

Actually, a general theory of symbolic manipulation, the so-called 
theory of \emph{Abstract Reduction Systems} (ARSs) \cite{newman,Huet80},
has been proposed almost one century ago (although its current 
formulation is due to Huet~\cite{Huet80}). Such a theory 
moves from the conceptual assumption 
that ``symbol pushing'' is a \emph{relational} notion and studies 
abstract properties of such a relation, such as confluence and termination. 
ARSs achieve their generality by simply ignoring the syntactic structure 
of expressions and they thus provide a limited (albeit not negligible) contribution 
to a general theory of symbolic manipulation. Unfortunately, 
being essentially syntax-free, ARSs can only account for 
symbolic manipulation as discrete transformations, this way giving no 
information 
on how such transformations interact with the syntactic structure of expressions.
Quoting Ghani and L\"uth~\cite{ghani-luth-coinserters}:
\begin{quote}
[$\hdots$] ARSs lack sufficient structure to adequately model key
concepts such as substitution, context and the layer structure whereby terms from
one system are layered over terms from another in modularity problems. Hence
ARSs are mainly used as an organisational tool with the difficult results proved
directly at the syntactic level.
\end{quote}

The proliferation of \emph{ad hoc} syntax-based 
theories of rewriting  
is precisely caused by the aforementioned deficiencies of ARSs. 
To overcome all of that and move towards a \emph{general} 
theory of symbolic reduction parametric with respect to, 
but at the same time accounting for, the syntactic structure 
 of expressions, several authors 
have proposed to rebuild rewriting on a \emph{categorical}, 
rather than \emph{relational}, basis, notable examples 
being theories based on polygraphs~\cite{burroni-polygraph}, 
Lawvere theories~\cite{power-rewriting}, 
$2$-categories (and variations thereof)~\cite{reichel1991,rydeheard1987foundations,stell1994modelling,seely1987modelling}, and 
(enriched) monads~\cite{luth1997monads,luth1998categorical,ghani-luth-coinserters}.

Although employing different categorical frameworks, all these theories 
share two common traits: first, they loose 
the relational understanding of symbolic manipulation 
replacing it with categorical constructions, such as 
suitable natural transformations, that are conceptually farther
from the everyday practice and understanding of symbolic manipulation; 
second, they are \emph{denotational}, rather than \emph{operational}, meaning that 
symbolic manipulation is defined not on syntactic expressions, but on denotations 
thereof (e.g. arrows in suitable categories).
Generality is thus achieved only denotationally, and at 
the expense of intuition.
% \footnote{Furthermore, a main advantage given by 
% a relational approach to rewriting is in terms of proof formalisation. In fact, 
% relations come with a rich (algebraic) calculus which is well-suited for 
% formalisation~\cite{relational-mathematics,pous}. A remarkable example of that is 
% given by the proof of Newman's Lemma: non-relational proofs require the formalisation of 
% first-order reasoning (which in turn requires dealing with variable binders) and 
% well-founded induction; relational proofs, instead, can be given as short equational 
% calculations in a pointfree and algebraic fashion~\cite{backshouse-calculational-approach-to-mathematical-induction}, 
% with consequent advantages in formalisation \cite{pous}.}
The question is now obvious: can the same level of generality be achieved 
\emph{operationally}, maintaining the intuitive, relational understanding 
of symbolic manipulation? 

\subsection{Contribution} 
% several research lines have proposed 
% to rebuild rewriting on a \emph{non-relational} basis 
% using the language of category theory. 
% % To overcome the deficiencies of ARSs and moving towards a general theory 
% % of symbolic manipulations accounting for the (many) syntactic structure(s) 
% % of expressions several approaches have been proposed in the last decades. 
% Notable examples of categorical theories of rewriting include 
% those based on polygraphs~\cite{}, Lawvere theories~\cite{}, and monads~\cite{}. 
% Even if remarkably successful, all these theories, as already stated, 
% loose the relational understanding of symbolic manipulation
% (which is replaced by, e.g., suitable natural transformations), this way achieving 
% generality at the expense of intuition.\footnote{Even more, a main advantage given by 
% a relational approach to rewriting is in terms of proof formalisation. In fact, 
% relations come with a rich (algebraic) calculus which is well-suited for 
% formalisation~\cite{relational-mathematics,pous}. A remarkable example of that is 
% given by the proof of Newman's Lemma: non-relational proofs require the formalisation of 
% first-order reasoning (which in turn requires dealing with variable binders) and 
% well-founded induction; relational proofs, instead, can be given as short equational 
% calculations in a pointfree and algebraic fashion~\cite{backshouse-calculational-approach-to-mathematical-induction}, 
% with consequent advantages in formalisation \cite{pous}.}
%\subsubsection{Contribution}
In this paper, we answer the above question in the affirmative 
by showing that the relational approach 
 at the heart of \ars{s} 
can account for a large class of syntactic expressions 
in a highly general, modular, and remarkably natural way that complements 
the previously mentioned mathematics of (abstract) syntax. 
This allows us to establish a new bridge between what one may ambitiously 
refer to as the mathematics of (formal) expressions and the mathematics of their 
(symbolic) manipulation. Whereas the former is algebraic and builds upon 
category theory, the latter is relational and builds upon \emph{allegory theory}~\cite{scedrov-freyd}. Remarkably, what is needed to define 
the relational theory of rewriting is exactly the allegorical 
counterpart of the categorical 
notions used to define syntax, namely initial algebras, free monads, etc.
This outlines a general framework where categorical 
notions are used to define syntactic expressions, and their allegorical extensions 
provide symbolic manipulations for such expressions. 
We achieve these results proceeding 
as follows. 
% following ways (the first two points 
% are essentially the core of the mathematical theory 
% of abstract syntax viewed, perhaps, from a different conceptual 
% perspective): 
% To achieve such a goal, we actively build upon the 
% mathematical theory of abstract syntax
% \cite{fiore-plotkin-turi-99,pitts-nominal-sets,more-on-pitts,Goguen/1977}, which 
% we present from a (perhaps) different conceptual angle.\footnote{Largely inspired by 
% considerations by Fiore.} Accordingly, we proceed as follows: 
\begin{varenumerate}
    \item We use suitable categories (viz. Grothendieck toposes~\cite{sketches-of-an-elephant}) 
        $\topos$ to model  
        the different kinds of expressions we are interested in (e.g. sets 
        for first-order expressions~\cite{Goguen/1977} and 
        presheaves for expressions with binders~\cite{fiore-plotkin-turi-99,pitts-nominal-sets}).
    \item Syntax specification is then given by suitable signature functors 
        on $\topos$, 
        whereas syntax itself is defined by free monads
        over such functors.
    \item We then apply the relational methodology 
        observing that if a topos $\topos$ models a universe of expressions, then 
        its induced \emph{allegory}~\cite{scedrov-freyd,pitts-quantaloids} 
        $\rell{\topos}$ models
        symbolic transformations between such expressions. 
    \item By extending functors, monads, and (initial) algebras  
        from $\topos$ to $\rell{\topos}$ -- something that, under suitable conditions, 
        can be always done~\cite{Barr/LMM/1970,carboni-kelly-wood,eilenberg-wright,algebra-of-programming} 
        -- we define 
        new syntax-based operators on relations that allow us to 
        define interesting notions of rewriting 
        in a purely relational way. 
    \item By exploiting the algebraic laws of the aforementioned operators, 
        we model and prove nontrivial rewriting properties.
        For instance, we prove confluence of orthogonal systems 
        by giving relational generalisation of the so-called 
        \emph{parallel moves}~\cite{klop} and 
        \emph{Tait and Martin-L\"of}~\cite{Barber96,aczel-general-church-rosser,Takahashi1995} techniques.
    % \item Since whenever $\topos$ is a (Grothendieck) topos, $\rell{\topos}$ 
    %     is not only an allegory, but a (locally complete) power 
    %     allegory~\cite{scedrov-freyd,pitts-quantaloids}, 
    %     $\rell{\topos}$ come with a rich calculus of relations 
    %     that allows us to prove many theorems in rewriting in a purely 
    %     abstract and relational setting. Among those, 
    %     we shall give abstract proof techniques for confluence 
    %     generalising, among other, those based on parallel moves~\cite{klop} 
    %     and critical pairs~\cite{Huet80}, as well as 
    %     the so-called Tait-Martin-L\"of technique~\cite{Barber96,aczel-general-church-rosser,Takahashi1995}.
\end{varenumerate} 

\subsubsection{The Augmented Calculus of Relations} 
By extending the categorical constructions defining syntax to allegories, 
we obtain a rich and novel relational vocabulary that allows us to define many rewriting notions 
in an allegorical setting. Even if such a vocabulary is obtained throughout a relational 
analysis of (the mathematics of) syntax, it turns out that all that matters for our purposes 
is the collection of new operators given by such an analysis, 
together with the (algebraic) laws governing their 
(operational) behaviour. At this point, the abstraction step is natural: we can forget about the 
syntax structure (and its relational counterpart) and work in a completely axiomatic fashion 
within a calculus of relations \cite{tarski-1941,relational-mathematics,DBLP:books/sp/97/Maddux97} 
\emph{augmented} with the aforementioned operators and their laws. 
Much in the same way as the ordinary calculus of relations provide an elegant framework for the 
study of abstract reduction systems, the aforementioned augmented calculus of relations provide a similar 
framework for the study of (several kinds of) syntax-based reduction systems in a rather 
syntax-independent way. 

\subsubsection{A Bridge Between Rewriting and Program Equivalence}
Perhaps surprisingly, this augmented calculus of relations is not entirely new: it 
can be seen as a (nontrivial) generalisation of the calculus of $\lambda$-term 
relations developed by 
Lassen~\cite{Lassen/PhDThesis,Lassen/RelationalReasoning} in the context 
of higher-order program equivalence. This way, we obtain 
a novel connection between rewriting and program equivalence. 
A first (and surprising) consequence of that is the observation that 
the construction of the so-called multi-step
reduction~\cite{Barber96,aczel-general-church-rosser,Takahashi1995} as used in the Tait and Martin-L\"of 
confluence technique coincides with Howe's construction~\cite{Howe/IC/1996,Pitts/ATBC/2011} 
of the pre-congruence candidate, the key notion in the (operational) proof of congruence 
of applicative bisimilarity~\cite{Abramsky/RTFP/1990}.
This connection not only allows us to import results and techniques from the field of 
program equivalence to rewriting (as we shall do in this work), 
but also sheds new light on operational notions 
(for instance, we will see that the aforementioned Howe's construction is obtained 
via initial relational algebras).
% , hence suggesting that formalisms such as 
% the augmented calculus of relations can be used as a relational foundation for 
% operational reasoning (see the discussion on \francesco{\emph{relational operationalism}} in 
% \autoref{section:conclusion}).

\subsection{Related Work}
Looking at the literature on rewriting, it is quite natural to classify theories 
and results according to three distinct schools of thought~\cite{gadducci-1} 
reflecting mainstream divisions in 
program semantics:
\emph{operational}~\cite{terese,term-rewriting-and-all-that}, 
\emph{logical}~\cite{MESEGUER2012721}, and \emph{denotational}~\cite{burroni-polygraph,power-rewriting,reichel1991,rydeheard1987foundations,stell1994modelling,seely1987modelling,luth1997monads,luth1998categorical,ghani-luth-coinserters}. 
There is, however, a fourth, albeit less known, school of thought, namely the \emph{relational} one. 
% The relational approach is close in spirit to operational rewriting, in the sense that 
% it does study expressions looking at their denotations, although its mathematical and 
% methodological development is similar to the one of denotational rewriting. 

The first observation of the relevance of relational reasoning in rewriting is due to 
B\"aumer~\cite{baumer} and since then several relational calculi (and alike) have been employed 
to study \emph{abstract} reduction systems. Among the many results achieved in this line of research, 
we mention relational proofs of Church-Rosser theorem~\cite{struth-1}, of Newman's Lemma~\cite{backshouse-calculational-approach-to-mathematical-induction,Struth-abstract-abstract-reduction}, 
and modularity theorems for termination~\cite{backshouse-calculational-approach-to-mathematical-induction,bachmair1986commutation}. 
Relational rewriting has been also extended to non-traditional notions of relations, such 
as monadic relations~\cite{Gavazzo-Faggian-2021} and fuzzy and quantitative 
relations~\cite{fuzzy-rewriting-1,fuzzy-rewriting-2,fuzzy-rewriting-3,10.1145/3571256}.
To the best of the author's knowledge, the literature offers no relational analysis of
syntax-based systems. 
Finally, we mention the axiomatic approach to rewriting  
\cite{axiomatic-rewriting-1,axiomatic-rewriting-2,axiomatic-rewriting-3,axiomatic-rewriting-4,axiomatic-rewriting-5,axiomatic-rewriting-6}
which, even if non-relational, is close in spirit to the present work.

\section{Prologue: (Term) Rewriting Without Syntax}
\label{section:informal-intro}
Before moving to the general theory of rewriting, 
we gently introduce the reader to some of the main ideas behind this work 
by studying a simple, yet instructive example: \emph{term rewriting}. 
In their broadest sense, term rewriting systems~\cite{terese,Huet80,term-rewriting-and-all-that} 
(\trs{s}) specify how \emph{first-order terms} 
can be syntactically manipulated. 
Given a signature $\signatureop$, i.e. a set containing operation symbols $\op$ and 
their arity, and a set $X$ of variables, 
recall that the set
$\terms{\signatureop}{X}$ of ($\signatureop$-)terms is inductively defined thus:
\[
\vspace{-0.1cm}
\infer{\val{x} \in \terms{\signatureop}{X}}{x \in X}
\qquad 
\infer{\op(\termone_1, \hhh, \termone_n) \in \terms{\signatureop}{X}}
{\termone_1 \in \terms{\signatureop}{X} \cc \termone_n \in \terms{\signatureop}{X}
&
(\op, n) \in \signatureop}
\]
A \trs{} is then given by a binary relation $\stepto$ on  $\terms{\signatureop}{\variables}$. We refer to relations 
such as $\stepto$ as
\emph{reduction relations}.

Viewed that way, there is no conceptual difference between \trs{s} and \ars{s}, 
the latter being sets endowed with a binary endorelation. 
It takes only a few seconds, 
however, to realise that the relation $\stepto$ alone is of little interest, as 
it says nothing on how to instantiate 
$\stepto$ on arbitrarily complex expressions, 
nor on how to propagate it along their structures. 
For instance, consider the following relation 
modelling natural number addition:
% $\code{f}(\varone) \to \code{g}(x,x)$ if we consider the 
% signature of natural numbers consists of a constant  
% $\code{z}$ (for zero), a unary operation $\code{succ}$ (for successor), 
% and a binary operation $\code{add}$ (for addition), and 
% the \trs{} given by: 
$\code{add}(\code{0}, \val{y}) \stepto \val{y}$,  
$\code{add}(\code{succ}(\val{x}), \val{y}) \stepto \code{succ}(\code{add}(\val{x},\val{y}))$.
% together with the rewriting relation:
% \begin{align*}
% \code{add}(\code{z}, y) &\stepto y
% & 
% \code{add}(\code{succ}(x), y) &\stepto \code{succ}(\code{add}(x,y))
% \end{align*}
%As it is, the relation $\stepto$ is not very useful since, e.g., 
Relying on $\stepto$ only, there is no way 
to reduce, e.g., the term $\code{succ}(\code{add}(\code{0}, \code{0}))$ to 
$\code{succ}(\code{0})$, as the former is 
not a redex,\footnote{Recall that a redex is a term that can be $\stepto$-reduced.} 
although it contains a substitution instance of one. 

Consequently, to obtain useful notions of reduction on $\signatureop$-terms we 
need \emph{(i)} to consider substitution instances of $\mapsto$
and \emph{(ii)} to specify how such instances can be propagated along term syntax. 
The first point is handled by working with the
substitution closure $\rightarrowtriangle$ of $\stepto$, 
whereby we consider (substitution) \emph{instances} of $\stepto$.\footnote{
Formally, we have $\termone[\vect{\valone}/\vect{\varone}] \rightarrowtriangle \termtwo[\vect{\valone}/\vect{\varone}]$ whenever $\termone \stepto \termtwo$.
}
For the second point, there are many possible natural extensions of 
$\rightarrowtriangle$, depending on the notion of reduction 
one has in mind. For example:
%Here are some possibilities. 
\begin{varitemize}
\item \emph{Sequential reduction} 
is the relation $\to$ that
$\rightarrowtriangle$-reduces exactly one redex at a time.
\item \emph{Parallel reduction} is the relation 
$\Rightarrow$ that $\rightarrowtriangle$-reduces an arbitrary number of non-nested redexes 
in parallel.
\item \emph{Full reduction} is the relation $\Rrightarrow$ 
that $\rightarrowtriangle$-reduces an arbitrary number of possibly nested redexes 
in parallel.
\end{varitemize}

A formal definition of all these notions requires to first introduce 
several specific syntactic notions on terms, 
such as positions, occurrences, contexts, etc. Sequential reduction, for example,
is defined by stipulating that 
$C[\termone]_p \to C[\termtwo]_p$ holds if and only if
$\termone \rightarrowtriangle \termtwo$ does, where 
$p$ is a position and $C[-]_p$ is a context with one hole 
at position $p$.
% whereas weak parallel reduction
% considers finitely many disjoint positions.
As a main consequence of that, \trs{s} 
become intrinsically \emph{term-dependent}, this 
way loosing the relational understanding of reduction given 
by \ars{s}.
The shift from reduction relations \emph{per se} to the 
\emph{syntactic structure} of the objects reduced 
massively impacts the way one reasons about \trs{s}, with 
relational reasoning 
leaving the place to \emph{syntactic} arguments on terms. 
Such  arguments have several well-known drawbacks: 
they are error-prone, difficult to formalise, and, most importantly, 
they lack modularity. Small changes in the syntactic structure of terms 
require to (re)develop 
the underlying rewriting theory from scratch.

\subsubsection{Term Rewriting, \sout{Syntactically} Relationally}  
The just described scenario shows a conceptual gap between abstract and term (and, more generally, syntax-based) 
rewriting systems.
This gap, however, is by no means substantial and it is possible to 
develop a fully relational theory of \trs{s}, as we shall show in this paper. 

First, we observe that 
parallel and full reduction
can obtained by means of suitable relational constructions applied on 
$\rightarrowtriangle$.\footnote{
The same can be said about sequential reduction too, although we leave its formal analysis 
for future work (see \autoref{section:conclusion}).} 
For instance, the following relational construction, known as 
\emph{compatible refinement} \cite{Lassen/PhDThesis,Gordon-1995},
defines relations $\compref{\relone}$ between terms with 
the same outermost syntactic constructs and argumentiwse $\relone$-related 
terms.
% on relations on $\signatureop$-terms 
% precisely defines 
% parallel reduction as $\widehat{{\rightarrowtriangle} \cup \idrel}$, 
% where $\idrel$ is the identity relation (the presence of $\idrel$ ensures 
% that one can choose how many redexes to reduce):
% \begin{varenumerate}
%     \item $\val{x} \mathrel{\compref{\relone}} \val{x}$, for any $x \in X$;
%     \item If $\termone_i \mathrel{\relone} \termtwo_i$, for any $i \leq n$, 
%     then $\op(\termone_1, \hhh, \termone_n) \mathrel{\compref{\relone}} \op(\termtwo_1, \hhh, \termtwo_n)$
% \end{varenumerate}
\vspace{-0.2cm}
\[
\vspace{-0.2cm}
\infer{\val{x} \mathrel{\compref{\relone}} \val{x}}{x \in X}
\qquad 
\infer{\op(\termone_1, \hhh, \termone_n) \mathrel{\compref{\relone}} \op(\termtwo_1, \hhh, \termtwo_n)}
{\termone_1 \mathrel{\compref{\relone}} \termtwo_1 & \cc & \termone_n \mathrel{\compref{\relone}}  \termtwo_n}
\]
Using compatible refinement, we obtain a relational (inductive) definition of $\Rightarrow$ 
as $\lfp{x}{{\rightarrowtriangle} \vee \compref{x}}$. 
This change of perspective is not just a way to give more compact definitions. 
Having separated the definition of $\Rightarrow$ from the (simplest) one of $\compref{-}$, 
it is natural to focus on the algebraic properties of 
the latter, rather than on the syntax of terms. 
For instance, it is easy to prove that $\compref{-}$ is functorial 
(it preserves relation composition and the identity relation) 
and ($\omega$-)continuous: this way, not only the aforementioned relational definition of 
$\Rightarrow$ is well-given, but it also allows us to reason about parallel reduction 
using algebraic calculations and fixed point induction~\cite{Backhouse-fixed-point-and-galois-connection}.

Moreover, we can use algebraic and relational reasoning to prove rewriting properties of 
$\Rightarrow$. As an example, in \autoref{section:parallel-reduction} 
we shall prove the diamond property of $\Rightarrow$
for orthogonal systems~\cite{Huet80} by showing
% For instance, we shall prove that the identity relation $\idrel$ is the least 
% compatible relation (i.e. $\compref{\idrel} \leq \idrel$). Consequently, since 
% $\Rightarrow$ is compatible, it must contain the identity, and thus it is reflexive. 
% This ensures that indeed $\Rightarrow$ reduces redexes \emph{at will}. 
% Furthermore, we can prove nontrivial confluence results about $\Rightarrow$ 
% in the same spirit. In \autoref{section:parallel-reduction}, we shall prove 
the following Kleisli-like extension lemma (here properly instantiated to 
$\Rightarrow$) akin to a semi-confluence proof technique: 
${\leftarrowtriangle}; {\Rightarrow} \subseteq 
{\Rightarrow}; {\Leftarrow}$ implies 
${\Leftarrow}; {\Rightarrow} \subseteq 
{\Rightarrow}; {\Leftarrow}$. 
% Such a result can be read as semi-confluence proof technique, whereby 
% whenever we can $\Rightarrow$-close a (semi-local) peak of the form 
% $\termtwo_1 \leftarrowtriangle \termone \Rightarrow \termtwo_2$, 
% we can also close (non-local) peak of the form 
% $\termtwo_1 \Leftarrow \termone \Rightarrow \termtwo_2$.
We will prove such a result relying on algebraic laws only, and then show that the inclusion in the aforementioned extension lemma
holds for a large class of (relationally-defined) 
reduction systems (i.e. \emph{orthogonal systems}) using the same methodology.
All of that is done relationally by decomposing $\rightarrowtriangle$ 
throughout a further relational construction, viz. \emph{relation substitution} \cite{Lassen/PhDThesis}, 
and its algebraic laws. The reader familiar with \trs{s} can see such a result as a generalisation 
of the well-known \emph{parallel moves technique}~\cite{klop}.

\subsubsection{Hello Syntax, My Old Friend}
The discussion made so far hints that a relational analysis of \trs{s} is possible by 
introducing suitable relational operators describing, at an algebraic level, 
how syntax act on reductions. This makes the operational analysis of \trs{s} indeed closer 
to the corresponding one of \ars{s}: for the former one simply needs a more powerful calculus of relations 
than the one needed for \ars{s} (we shall come back to this point later). 
All of that significantly improves reasoning, but does not overcome syntax-dependency: 
the definition of the new relational 
constructions, such as $\compref{-}$, still relies on the syntax at hand. 

To make the relational framework truly general, 
we notice that the actual 
syntax of term is not really needed for our purposes. In fact, 
it is well-known that notions of syntax can be modelled 
as free algebras (and their associated monads) over suitable signature functors~\cite{Goguen/1977}. In 
the case of first-order terms, any signature $\signatureop$ 
induces a (signature) functor $\signature$ on the category $\set$ of sets and functors. 
Such a functor acts as syntax specification and induces a functor $\freemonad$ that, 
given a set $X$, returns the set of $\signatureop$-terms over $X$. 
The functor $\freemonad$ is the carrier of a monad $(\freemonad, \unit, \multiplication)$, 
which is the free monad over $\signature$, in the sense that
the structure $X \xrightarrow{\eta} \freemonad X \xleftarrow{\sigma} \signature(\freemonad X)$ 
gives the free $\signature$-algebra over $X$. The map $\unit$ 
acts as the variable constructor mapping $x$ to $\val{x}$, whereas the map 
$\sigma$ describe the inductive step in the definition of $\signatureop$-terms, whereby 
terms are closed under operation symbols; the multiplication $\multiplication$, finally, 
flattens a `term of terms' into a term, and it is \emph{de facto} recursively defined 
relying on $\unit$ and $\sigma$.
% has a free monad 
% $(\freemonad, \unit, \multiplication)$ mapping each set $X$ to the set of $\signatureop$-terms 
% over it, with the unit $\unit$ mapping each variable $x$ to $\val{x}$, and the multiplication 
% $\multiplication$ flattening a `term of terms' into a term. 

The key observation now is to notice that these data acting on sets (of terms) 
are precisely what is needed to define (the relational operators behind) reduction relations, provided that we 
can extend their action from functions to \emph{relations}, hence to the category
$\rel$ of sets and relations. 
For instance, given a relation $\relone$ on $\freemonad X$, we 
recover $\compref{\relone}$ as $\eta\dd;\eta \vee \sigma\dd;\barr{\signature}\relone; \sigma$
where, for a relation\footnote{
We use the notation $\relone: A \torel B$ in place of $\relone \subseteq A \times B$.} 
$\relone: A \torel B$, we denote by $\relone\dd: B \torel A$ the converse of 
$\relone$ and by $\barr{\signature}$ the \emph{relational 
extension} of $\signature$, namely a functor-like mapping $\relone$ to 
$\barr{\signature}\relone: \signature A \torel \signature B$
(of course, we have to check that using $\barr{\signature}$ we indeed 
recover $\compref{\relone}$, but we will see that this is the case; even more, 
this is \emph{canonically} the case).

The question now is: how can we define $\barr{\signature}$? 
Luckily, the answer has been given long ago, and we can now rely on a 
mature theory of relational extensions 
of 
set-constructions~\cite{Hoffman/Cottage-industry/2015,backhouse-relational-theory-of-datatypes,Kurz/Tutorial-relation-lifting/2016}. In particular,
by a celebrated result by Barr~\cite{Barr/LMM/1970}, 
$\signature$ extends to a converse-preserving monotone functor 
$\barr{\signature}$ on $\rel$. Moreover, $\barr{\signature}$ -- which is usually generically 
referred to as the \emph{Barr extension} of $\signature$ -- is \emph{unique} 
and it furthermore induces a converse-preserving monotone functor $\barr{\freemonad}$ which, 
as suggested by the notation, is the Barr extension of $\freemonad$. 

Converse-preserving monotone functors are known
as \emph{relator}~\cite{kawahara1973notes,backhouse-polynomial-relators,carboni-kelly-wood}
and there is a rich theory both on relators and on how to extend functors to relators. 
We view relators as describing how (reduction) relations 
are propagated along the syntactic structure given by the underlying functor.
For instance, $\barr{\signature}$ indeed defines $\compref{-}$ as sketched above. 
Moreover, looking at $\freemonad X$ monadically, we can use $\barr{\freemonad}$ 
to obtain a further decomposition of $\Rightarrow$ as 
% Using this result, we can decompose the definition of a rewriting relation 
% parametrically on the syntax monad $\freemonad$. 
% In fact, we precisely recover $\widehat{\relone}$ 
% as\footnote{For a relation $\relone$, 
% we write $\relone\dd$ for its converse.} 
$\multiplication\dd; \barr{\freemonad}({\rightarrowtriangle}); \multiplication$. 
From a syntactic perspective, this definition corresponds to the `context-based' 
definition of parallel reduction, where using relators we can 
talk about contexts in a syntax-free way.  
Finally, we can show that the two definitions of parallel reduction hereby sketched 
indeed coincide.

% (see diagram below) 
% and prove, e.g., its confluence properties relying on 
% the algebraic laws of $\widehat{\freemonad}$ only.  
% \[ 
% \xymatrix{
% %\laxcommuterel
%  \freemonad \freemonad\variables \ar[r]^{\widehat{\freemonad} \relone} 
%  &  \freemonad \freemonad\variables \ar[d]^{\mi}
% \\
%  \freemonad \variables \ar[u]^{\mi\dd} \ar@{.>}[r] &  \freemonad\variables
% }
% \]
% \[ 
% \xymatrix{
% %\laxcommuterel
% \freemonad \variables \ar[r]^{\mi\dd}  &
% \freemonad \freemonad\variables \ar[r]^{\widehat{\freemonad} \relone} 
%  &  \freemonad \freemonad\variables \ar[r]^{\mi}
% &
% \freemonad\variables
% }
% \]

% has a free monad 
% $(\freemonad, \unit, \multiplication)$ mapping each set $X$ to the set of $\signatureop$-terms 
% over it, with the unit $\unit$ mapping each variable $x$ to $\val{x}$, and the multiplication 
% $\multiplication$ flattening a `term of terms' into a term. 

\subsubsection{From Algebra to Program Equivalence}
This (informal) analysis shows that parallel reduction
can be fully understood relationally; moreover, the role played by relators 
(and by Barr extensions, in particular) hints that 
parallel reduction is \emph{the} 
canonical notion of reduction induced by the syntax. Is that really the case? 
And what about other notions of reduction?

The answer to the first question is in the affirmative, at least as long as 
we think about $\freemonad$ as a \emph{monad}. 
At the same time, however, parallel reduction is not the only canonical notion of 
reduction induced by the syntax. In fact, $\freemonad X$ being free, 
the theory of initial algebra tells us that
we can equivalently described it as the \emph{initial algebra} of the ($\set$) functor 
$X + \signature(-)$, so that we can think about syntax also as an initial algebra. 

% Initial algebras provide an algebraic and high-level account to both 
% definition by recursion and proofs by (structural) induction. 
% This result, generically referred to as \emph{initiality}, applies 
% to functions, in the sense that, in our example, initial algebras 
% are given (essentially) in $\set$ and the algebra structure is given 
% by a function. In our concrete case, 

% More specifically, we have an algebra function 
% $\xi: X + \signature(\freemonad X) \to \freemonad X$ which, by the 
% so-called Lambek Lemma~\cite{lambek-1968}, 
% is the component of an isomorphism $X + \signature(\freemonad X)  \cong 
% \freemonad X$. 
% Initiality gives tells that whenever we have a set $A$ together with 
% an algebra map $\alpha: X + \signature A \to A$, then there is a unique extension 
% $\cata{\alpha}: \freemonad X \to A$ that respects the algebra structure. 

How is that relevant for rewriting?  
The so-called Eilenberg-Wright Lemma~\cite{eilenberg-wright} states 
that an initial algebra in $\set$ is such also in $\rel$. 
In particular, whenever we have a relation 
$\relone: X + \signature A \torel A$, then initiality 
gives a unique relation 
$\cata{\relone}: \freemonad X \torel A$ such that\footnote{By
 Lambek Lemma~\cite{lambek-1968}, we have $[\eta,\sigma]: X + \signature(\freemonad X) \cong \freemonad X$.}
$[\eta, \sigma]; \cata{\relone} = \barr{\signature}\cata{\relone}; \relone$. 
%The relation $\cata{\relone}$ is precisely the relation defined by structural recursion 
%out of $\relone$.
Thinking about $\relone$ as a reduction,  
then $\cata{\relone}$ recursively 
$\relone$-reduces along the syntactic structure of terms. 

Moving from this intuition, we discover that $\Rrightarrow$ is precisely 
$\cata{[\eta, \sigma]; {\rightarrowtriangle}}$. 
But this is not the end of the story. In fact, even if this initial algebra-based definition of 
$\Rrightarrow$ goes `beyond' the relational operators used to define 
(the least fixed point characterisation of) parallel reduction, we 
can give an inductive characterisation of $\Rrightarrow$ using that 
vocabulary, viz. as $\lfp{x}{\compref{x};\rightarrowtriangle}$.
This should ring a bell to the reader familiar with program equivalence: in fact, 
relations of 
the form $\lfp{x}{\compref{x};\relone}$ are not new, as they are precisely the 
so-called pre-congruence candidates (on $\relone$) used in the well-known Howe's technique~\cite{Howe/IC/1996,Pitts/ATBC/2011} to prove congruence of applicative 
bisimilarity~\cite{Abramsky/RTFP/1990}. 
% All of that not only sheds new light both on deep reduction and on Howe's method, but 
% it also allows us to rely on powerful relational results to prove rewriting properties of 
% $\Rrightarrow$. 
As for parallel reduction, this relational machinery is powerful enough also 
to prove interesting rewriting properties of $\Rrightarrow$. We will witness that 
by giving a fully relational  
generalisation of the so-called Tait-Martin-L\"of 
technique~\cite{Barber96,aczel-general-church-rosser,Takahashi1995}. 

\subsubsection{Beyond \trs{s}}
The discussion conducted so far hints that the theory of \trs{s} can be given 
within a relational framework. This observation builds upon three crucial points: 
\emph{(i)} the syntax of \emph{first-order} expressions 
can be modelled \emph{categorically} on $\set$; \emph{(ii)} 
the latter category has a rich category of relations, viz. $\rel$, 
that models symbolic manipulations of first-order expressions;
\emph{(iii)} the categorical notions modelling syntax can be extended to $\rel$, 
and such notions are precisely what is needed to define reduction relations
and to prove theorems about them (\emph{(iv)} as a bonus point that we will discuss later, we notice 
that such extensions allow us to extend the rich calculus of relations given by $\rel$ 
with suitable operators, this way giving a kind of extended calculus of relations 
within which rewriting theories can be expressed). 
% building upon the following observations:
% \begin{varenumerate}
%     \item \trs{s} act on \emph{first-order} expressions, and the latter 
%         can be algebraically described in terms of \emph{free monads} 
%         for suitable signature functors (viz. polynomial functors) on 
%         $\set$.
%     \item The category $\set$ has an associated category of relations, which is nothing 
%         but $\rel$, that can be used to define reductions between terms.
%     \item Signature functors and their free monads extend to $\rel$, and 
%         their (relational) extensions are used to model notions of reductions 
%         and prove theorems about them.
% \end{varenumerate}

The realm of symbolic expressions, however, is far richer than first-order terms: 
there are expressions with names and binders~\cite{fiore-plotkin-turi-99,pitts-nominal-sets}, 
sorted and typed expressions, diagrams and two-dimensional 
expressions~\cite{Selinger2011}, etc.
% \begin{enumerate}
%     \item $x + (y + z)$
%     \item $\int (\int f(x,y)\ dx) dy$
%     \item $f: \code{int} \to \code{int}, y: \code{int} \imp \code{let}\ x = y\ \code{in}\ (f\, x): \code{int}$
%     \item $\includegraphics[height=0.8cm]{string-diagram.png}$.
% \end{enumerate}
% The first one is a standard first-order term; 
% the second and the third ones are second-order expressions, as they involves names and binders. 
% The third expression, in addition, is a typed expression in context 
% \cite{fiore-plotkin-turi-99}, in the sense that we usually consider 
% $\code{let}\ x = y\ \code{in}\ (f\, x)$ as the actual expression regarding 
% $f: \code{int} \to \code{int}, y: \code{int}$ as contextual information.
% Finally, the fourth expression is a two-dimensional 
% diagram \cite{Selinger2011}
All these expressions come with suitable notions of symbolic manipulation 
between them (e.g. higher-order rewriting~\cite{klop,terese}, nominal rewriting~\cite{nominal-rewriting}, 
diagrammatic rewriting~\cite{Bonchi-algebraic-semantics-string-diagrams,Bonchi_2022_I,Bonchi_2022_II,Bonchi_2022_III}, etc.),
and to qualify as a general theory of symbolic manipulation, 
the relational theory we are going to develop has to account for all these examples. 

To achieve this goal, we notice that the aforementioned key points are not 
at all specific to $\set$, $\rel$, and first-order terms. 
All the expressions mentioned so far can be understood in terms of initial algebras 
and free monads, provided that one moves to categories other than $\set$. 
For instance, expressions with binders are modelled on variable (i.e. presheaves)~\cite{fiore-plotkin-turi-99} 
and
nominal sets~\cite{pitts-nominal-sets,pitts-nominal-1}, whereas diagrammatic expressions 
rely on categories of spans~\cite{Bonchi-algebraic-semantics-string-diagrams}. 

This allows us to recover point \emph{(i)} above. 
If we recover points \emph{(ii)} and \emph{(iii)} too, 
then we can give relational theories of symbolic manipulation for 
all the classes of symbolic expressions at issue. 
This is indeed the case, as each category $\topos$ mentioned so far 
induces a rich category of relations $\rell{\topos}$ over it together 
with extensions of syntax functors and monads on $\topos$ to 
$\rell{\topos}$. 
The last step we need to take to achieve a truly general theory is to 
crystallise the above procedure by means of a suitable axiomatics 
that captures the essential structure $\topos$ and $\rell{\topos}$ need to have. 
We will achieve that goal by taking a non-minimal yet effective axiomatisation 
whereby $\topos$ is a Grothendieck topos~\cite{sketches-of-an-elephant}. 
In fact, any Grothendieck topos $\topos$ induces a category of relations 
$\rell{\topos}$ that has the structure of a \emph{locally complete power allegory}~\cite{scedrov-freyd,pitts-quantaloids}. The latter allegories 
provide a powerful and highly expressive calculus of relations 
that allows us to develop a general theory of rewriting in a remarkably 
clear and elegant way. Although slogans should be avoided, 
the allegorical theory of rewriting we develop in this paper and 
its deep connection with the mathematical theory of syntax, seem to 
suggest that syntax is  categorical, and syntax manipulation is 
 allegorical.

\subsubsection{An Axiomatic Approach} 
Let us summarise what we have achieved so far. 
Looking at syntax as a categorical construction  
and considering its relational extension, we have recovered notions of reduction
in fully relational ways. This process can be organised into two complementary 
approaches, both of which define notions of reduction and prove properties about them. 
The first approach proceeds in 
an \emph{algebraic} fashion by enriching traditional relational calculi 
with suitable operators on relations (compatible refinement, 
relation substitution, etc.) and relying on their algebraic laws to 
prove rewriting properties.
The second approach, instead, is \emph{structural} and builds upon the relationally 
extended categorical properties of syntax 
to give definitions of reduction, relying on their universality to 
prove rewriting properties. 

A natural further abstraction step is to make the first approach completely \emph{axiomatic}. 
That is, rather than building upon signature functors, relators, etc. to build 
relational operators, we simply add them (as well as their algebraic laws) 
to the traditional calculus of relations (or variations thereof)~\cite{tarski-1941,DBLP:books/sp/97/Maddux97,relational-mathematics} 
in an axiomatic fashion. We can then develop theory of rewriting systems within such an \emph{augmented 
calculus of relations}, this way giving a truly relational foundation to rewriting. 
The structural approach previously mentioned can then be seen as a way to build models of such a calculus. 
This axiomatic approach have several advantages: for instance, it allows us to establish 
novel and deep connections between rewriting and program equivalence, 
and opens the door to enhance proof formalisation of rewriting 
theories.\footnote{Relational calculi turned out to be well-suited  
for proof 
formalisation~\cite{relational-mathematics,pous}, with remarkable example of that 
in rewriting being
given by the proof of Newman's 
Lemma~\cite{backshouse-calculational-approach-to-mathematical-induction,pous}.}

Now that the reader has familiarised with the spirit of this work, we move to its 
formal development.

\section{Mathematical Preliminaries} 
Before going any further, we recall some preliminary notions. 
We assume the reader is familiar with basic category theory \cite{MacLane/Book/1971}. 
We will use standard notation except for: composition of arrows is in diagrammatic order ($f;g$)
and identity is denoted as $\idrel$ (relational notation).

\subsubsection{Initial Algebras}
Given a category $\category$ and a functor $\signature: \category \to \category$ on it,
a $\signature$-algebra consists of an object $A$ (the carrier) and an arrow 
$\alpha: \signature A \to A$ (the algebra map). Such algebras are the objects of a category,  $\signature\text{-}\alg$, 
whose arrows $f: (A, \alpha) \to (B,\beta)$ are $\category$-arrows $f: A \to B$ such that 
$\alpha; f = \signature f; \beta$. The \emph{initial algebra} of $\signature$, if it exists, 
is the initial object in $\signature\text{-}\alg$. Explicitly, it is a 
$\signature$-algebra $(\mu{\signature}, \xi)$  
such that 
for any $\signature$-algebra $(A, \algone)$, there exists a unique $\signature\text{-}\alg$-arrow $\cata{\algone}: \mu \signature \to A$.
% with the following universal property: 
% for any $\signature$-algebra $(A, \algone)$, there exists a unique $\signature\text{-}\alg$-arrow $\cata{\algone}$:
% \[
% \xymatrix{
% \signature(\lfpp{\signature}) 
% \ar@{.>}[r]^{\signature \cata{\algone}}
% \ar[d]_{\xi}
% & 
% \signature A
% \ar[d]^{\algone}
% %%%%%%%%%%%%%%%%%
% \\ 
% %%%%%%%%%%%%%%%%%
% \lfpp{\signature} \ar@{.>}[r]_{\cata{\algone}}
% &
% A
% }
% \]
We denote the carrier of the initial algebra $\signature$ 
by $\mu \signature$ (or $\mu x.\signature x$) and refer to arrows $\cata{\algone}$ as 
\emph{catamorphisms}.
%Intuitively, existence of catamorphisms corresponds to recursion (or definition by induction), 
%whereas their uniqueness gives proofs by induction. 
Being initial objects, initial algebras are unique up-to isomorphism. 
Moreover, the well-known Lambek Lemma~\cite{lambek-1968} states that 
%we can think about initial $\signature$-algebras as least fixed point 
%of $\signature$, in the sense that 
$\xi$ has an inverse and thus 
$\mu{\signature} \cong \signature(\mu{\signature})$.

Initial algebras need not exist, in general. The following result 
\cite{adamek-fixed-point-theorem} gives a sufficient condition on 
functors that guarantees existence of initial algebras. 
%This result has been generalised to the categorical setting of initial algebras by 
%Ad\'amek~\cite{adamek-fixed-point-theorem}. 
%we usually have procedures to incrementally construct 
% such expressions. For instance, defining expressions by means of rules mathematically 
% means giving a monotone map $F: \mathcal{P}(U) \to \mathcal{P}(U)$, with $U$ a universe 
% set. The collection of expression defined this way is nothing but the least fixed ponint 
\begin{theorem}[\cite{adamek-fixed-point-theorem}]
\label{theorem:adamek-theorem}
Let $\catone$ be a category with initial object $0$ and let 
$\signature$ be a finitary endofunctor on it, i.e.
$\signature$ preserves $\omega$-colimits
%.\footnote{
%Sometimes one says that $\signature$ preserves filtered colimits or that it is 
%cocontinuous.} 
Then $\mu{\signature}$ 
exists and coincides with colimit of the chain
$
0 \xrightarrow{!} \signature 0 \xrightarrow{\signature!} \signature^2 0 
\xrightarrow{\signature^2!} \cc
%\signature^3 0 \xrightarrow{\phantom{\signature^3!}} \cc 
%\xrightarrow{\phantom{\signature^n!}}
 \signature^{n}0 \xrightarrow{\signature^{n}!} \cc.
$
\end{theorem}

\subsubsection{Monads}
Recall that a monad on a category $\category$ is a 
triple $(\monad, \unit, \multiplication)$ consisting of a functor 
$\monad: \category \to \category$ and natural transformations 
$\unit_A: A \to \monad A$, $\multiplication_A: \monad \monad A \to \monad A$ 
satisfying suitable coherence conditions~\cite{MacLane/Book/1971}. 
To avoid unnecessary proliferation of notation, 
we denote by $\monad$ both the monad $(\monad, \eta, \multiplication)$ and its 
carrier functor, provided that does not create confusion.

When a functor $\signature$ as above has initial algebra (and 
$\category$ has enough structure), it induces a monad 
$\freemonad$, called the \emph{free monad over $\signature$}. 
Let us assume that $\category$ has finite coproducts
% \footnote{ 
% We denote by $\inl$ and $\inr$ the left and right coproduct 
% injections, by
% $[f,g]: A + B \to C$ the unique arrow obtained from 
% $f: A \to C$ and $g: B \to C$ by universality of coporoduct,
% and by $f+g$ the arrow $[f; \inl, g; \inr]$.} 
% \footnote{
% We recall a couple of useful calculation laws we shall use 
% throughout this work, where $f+g \defeq [f; \inl, g; \inr]$ 
% denotes the action of the coproduct bifunctor on arrows.
% \begin{align*}
% (f; h) + (g; k) &= f;g + h;k
% \\
% [f;h, g;k] &= (f + g); [h, k].
% \end{align*}
% }
and that for any object $A$ the initial algebra 
of the functor $A + \signature(-)$ exists, which it does whenever the one of 
$\signature$ does. 
Then, the assignment $\freemonad A \defeq \mu x.A + \signature x$ 
determines a monad, called the \emph{(algebraically) free monad}
%\footnote{In full generality, 
%free monads over functors $\signature$ are defined as left adjoint to the 
%forgetful functor $U: \signature\text{-}\alg \to \category$ sending a 
%$\signature$-algebra to its carrier. For our purposes, however, the  
%specialised definition through initial algebras and coproducts is enough.
%} 
generated by 
$\signature$ \cite{MacLane/Book/1971}.
%\footnote{To avoid unnecessary proliferation of notation, 
%we denote by $\freemonad$ both the free monad over $\signature$ and its 
%carrier functor.}
The initial algebra map $\xi: A + \signature \freemonad A \to \freemonad A$ can be 
decomposed as $[\unit, \sigma]$, with $\unit: A \to \freemonad A$ and 
$\sigma: \signature \freemonad A \to \freemonad A$. 
Both $\unit$ and $\sigma$ are mono, provided that coproducts injections are monos in 
$\category$, a condition satisfied by any
topos.
%, which we shall use as universes of expressions in this work.
% \footnote{ 
% Indeed if an arrow 
% $[\alpha_1, \alpha_2]: A_1 + A_2 \to A$ has an inverse, say $\beta$, then both 
% $\alpha_1$ and $\alpha_2$ are necessarily monos. For, given $x,y: X \to A_i$ such that 
% $x; \alpha_i = y; \alpha_i$, we have $x; \alpha_i; \beta = y; \alpha_i; \beta$, and thus 
% $x; \mathsf{in}_i = y; \mathsf{in}_i$ which gives $x = y$, since $\mathsf{in}_i$ is a mono.
% \[
% \xymatrix{
% X \ar@<0.8ex>[r]^{x} \ar@<-0.8ex>[r]_{y} & A_i \ar[d]_{\mathsf{in}_i} \ar[rd]^{\alpha_i}
% \\
% & A_1 + A_2  \ar[r]^{[\alpha_1, \alpha_2]} & X \ar@/^1pc/[l]^{\beta} }
% \]
% }
The arrow $\unit$
gives the unit of $\freemonad$, whereas the multiplication $\multiplication$ is defined by initiality as 
$\cata{[\idrel, \sigma]}$.

\section{Outline of a Categorical Theory of Syntax}
Having recalled the notions of initial algebra and free monad, 
in this section we succinctly summarise how these notions can be instantiated 
to give a mathematical theory of syntax~\cite{Goguen/1977,pitts-nominal-sets,pitts-nominal-1,pitts-nominal-2,pitts-nominal-3,pitts-nominal-4,fiore-plotkin-turi-99,fiore-polynomial-functors-simple-type-theories,hamana-free-sigma-monoids,hamana-polymorphic-syntax,Hofmann-higher-order-abstract-syntax,pfenning-higher-order-abstract-syntax,Miller}. 
Although different authors propose different approaches to (different aspects of) abstract syntax,
all such approaches can (perhaps) be understood in the following conceptual framework
(see \autoref{fig:basic-categorical-syntax}).
\begin{varenumerate}
    \item A category (the universe of expressions) 
        $\topos$ capturing the kind of expressions one is interested in 
         is fixed. 
        % Borrowing the vocabulary of logic, $\topos$ is sometimes called the universe of 
        % expressions (or universe of discourse). 
    \item Syntax specification is given by a (signature) functor $\signature: \topos \to \topos$, 
        usually polynomial,
        that specifies how expressions can be combined
        %The functor $\signature$ is usually referred to as the \emph{signature functor} 
        %and it gives a specification of the syntax of expressions. 
    \item The actual syntax of the language is given by the free monad $\freemonad$ 
        generated by $\signature$. 
\end{varenumerate}
 Obviously, the above schema does work only for suitable categories and functors 
 which, in turn, may depend on specific features the framework aims to describe. 
% As we are interested in rewriting, we shall build upon a (as general as possible) 
% theory of abstract syntax. Such a theory will follow the three points in the above list 
% instantiating them in specific ways.
In what follows, we discuss each point in detail and explicitly state 
the axioms of our propaedeutic theory of syntax (upon which we shall 
develop the theory of symbolic maniopulations).

\begin{figure*}[t]
\small
    \centering
    \begin{minipage}{.4\textwidth}
    \textbf{Categorical Syntax}
    \begin{varitemize}
        \item Universe of Expressions: A Grothendieck topos $\topos$
        \item Syntax Specification: A functor $\signature: \topos \to \topos$
            \begin{varitemize}
                %\item Usually $\signature$ is a \emph{polynomial functor}
                \item[-] Finitary syntax means \emph{finitary functor}
                \item[-] If $\signature$ is finitary, then 
                    $\fpinitiall{\signature} = \operatorname{colim}\signature^n(0)$
            \end{varitemize}
        \item Syntax: Free monad $\freemonad: \topos \to \topos$
            \begin{varitemize}
                \item[-] Free monad exists iff $\lfpp{\signature}$ does
                \item[-] $\freemonad A = \lfp{x}{A + \signature x}$
            %    \item[-] $\signature$ polynomial implies $\freemonad$ polynomial 
                    %(polynomial functors are closed under (co)products 
                    %and composition)
            \end{varitemize}
    \end{varitemize}
    \end{minipage}
    \begin{minipage}{.5\textwidth}
    \textbf{Allegorical Syntax}
    \begin{varitemize}
        \item Universe of Relations: The \lcp{} allegory  $\rell{\topos}$
            % \begin{itemize}
            %     \item If $\topos$ Grothendieck topos, then 
            %         $\rell{\allegory}$ is a \lcp{} allegory
            % \end{itemize}
        \item Relational Signature: A relator 
        $\relext{\signature}: \rell{\topos} \to \rell{\topos}$ for $\signature$
        \begin{varitemize}
                \item[-] $\relext{\signature}$ exists unique if 
                $\signature$ preserves strong epis and nearly preserves pbs 
                 %   then $\relext{\signature}$ exists and is unique
               % \item Polynomial functors have relational extensions
              %  \item $\relext{\signature}$ is $\omega$-continuous if 
               %     $\relext{\signature}(\join_n \relone_n) = \join_n \relext{\signature} \relone_n$
                \item[-] $\signature$ is finitary implies 
                    $\relext{\signature}$ $\omega$-continuous: $\relext{\signature}(\join_n \relone_n) = \join_n \relext{\signature} \relone_n$
             %   \item If a relator $\relator$ is finitary, then 
              %      $\lfpp{\relator} = \join_n \relator^n(\bot)$
            \end{varitemize}
        \item Relational Syntax: a relator 
             $\relext{\freemonad}: \rell{\topos} \to \rell{\topos}$ for 
             $\freemonad$
            \begin{varitemize}
                \item[-] $\relext{\freemonad}$ exists if $\relext{\signature}$ does, 
                and $\freemonad \relone = \lfp{x}{\eta\dd; \relone; \eta \vee 
                    \sigma\dd; \relext{\signature}; \sigma}$
                %\item If $\relext{\signature}$ is finitary, then so is 
                  %  $\relext{\signature}_{\relone}(x) \defeq 
                   % \eta\dd; \relone; \eta \vee 
                    % \sigma\dd; \relext{\signature}; \sigma$
                    %(polynomial functors are closed under (co)products 
                    %and composition)
            \end{varitemize}
    \end{varitemize}
    \end{minipage}
    \caption{Basic Notions of Categorical (left) 
    and Allegorical (right) Theory of Syntax}
    \label{fig:basic-categorical-syntax}
\end{figure*}

\subsubsection{Universe of Expressions}
Beginning with point 1, 
i.e. the universe of expressions $\topos$, we have already observed  
that formal expressions come in several flavours 
(first-order, higher-order, typed, two-dimensional, etc.) 
and that each of these alternatives corresponds to a specific 
category (sets, presheaves, nominal sets, spans, hypergraphs, etc.).
The purpose of the category $\topos$ is precisely to formally specify the nature of expressions.
 As the examples mentioned so far share the same structure: the category 
 $\topos$ is a \emph{topos}; even more, it is a Grothendieck topos~\cite{sketches-of-an-elephant}. 
 In light of that, we formulate the first axiom of our theory.
 \begin{axiom} 
 \label{axiom:axiom-of-expression-universe}
 The universe $\topos$ is a Grothendieck topos.
 \end{axiom}

\autoref{axiom:axiom-of-expression-universe} is by no means minimal and we could  
weaken it in many ways (e.g. working with elementary toposes with countable 
colimits, or even weaker structure) 
The advantage of \autoref{axiom:axiom-of-expression-universe} is that 
(i) it covers many interesting examples without requiring the introduction of \emph{ad hoc} 
definitions; (ii) $\topos$ supports an expressive calculus of relations
upon which we shall build a general theory of symbolic manipulation. 
% Grothendieck topoi form a locally complete power 
% allegories~\cite{scedrov-freyd,pitts-quantaloids}, the latter proving a 
% powerful relational framework to study expression dynamics (and being interested in the latter, 
% we will develop our theory 
% taking the latter allegories as primitive and regarding topoi as their induced subcategories). 
% Consequently, we momentary omit the definition of a Grothendieck topos simply remarking 
% that all the examples in \autoref{table:examples} carry such a structure.

\subsubsection{Syntax Specification}
The signature functor $\signature: \topos \to \topos$ specifies how expressions 
can be articulated, 
i.e. combined together to form new expressions. 
We require that $\signature$ captures a crucial features of the kind of syntax 
we are interested in:
syntax is \emph{finitary} and \emph{recursively} defined\footnote{
We leave the investigation 
of infinitary syntax to future work.}.
 By \autoref{theorem:adamek-theorem}, this means that $\signature$ 
 must be finitary. Additionally, we need to be able to manipulate expressions 
 along their syntactic structures so that, for instance, we can apply a 
 syntactic transformation on \emph{parts of} an expression.
 
 \begin{axiom} 
 \label{axiom:axiom-of-signature}
 The signature functor $\signature: \topos \to \topos$:
(i) nearly preserves pullbacks (pbs); 
(ii) preserves strong epimorphisms; (iii) 
is finitary.
 \end{axiom}

Conditions (i) and (ii), which we shall discuss in detail in the next section, ensure
that $\signature$ comes with a well-behaved notion of ``symbol pushing'', the latter being 
obtained via relational extensions of $\signature$. 
Such conditions, due to Carboni et al.~\cite{carboni-kelly-wood}, 
are rather weak: condition (i) is implied by weak pbs preservation, which is in turn implied by 
pbs preservation; condition (ii), instead, is equivalent to regular epimorphisms preservation~\cite{scedrov-freyd},
%(since $\topos$ is a topos and thus, in particular, a regular category~\cite{scedrov-freyd}) 
which is itself implied by epimorphisms preservation. 
In $\set$, the presence of the axiom of choice ensures that any functor preserves epimorphisms, 
but this is not the case in arbitrary topoi. Nonetheless, one can show that in any topos $\topos$ 
simple polynomial functors preserve epimorphisms. 

\subsubsection{Syntax} 
Having axioms on $\topos$ and $\signature$, there is not much to say about 
$\freemonad$. It simply acts as the actual syntax of the language which, given an object 
$A$ representing some collection of basic expressions, builds 
full expressions by recursively combining previously defined expressions 
according to $\signature$. Indeed, since $\topos$ has coproducts, if $\freemonad$ exists it 
maps an object $A$ to $\mu{x}{A + \signature x}$. \autoref{axiom:axiom-of-signature}
ensures such an initial algebra to exist and, additionally, 
to be obtained by \autoref{theorem:adamek-theorem}, since $\freemonad$ is finitary
whenever $\signature$ is. This precisely captures our assumption that syntax is finitary. 
Clearly, $\freemonad$ should also have a relational extension, and thus we may 
ask whether additional requirements have to imposed on $\freemonad$. The answer is in 
the negative, for $\freemonad$ has a relational extension if $\signature$ has, as we shall see 
in next sections.
% $\freemonad$ to nearly preserves pullbacks and covers ???. At the moment, the author 
% does not know whether this follows from the same conditions on $\signature$. 

% \begin{axiom}
% \label{axiom:axiom-preservation-pullbacks-free-monad}
% $\freemonad$ nearly preserves pullbacks and covers.
% \end{axiom}

% All the examples considered in this paper will actually satisfy 
% \autoref{axiom:axiom-preservation-pullbacks-free-monad} for free, since 
% in all of them $\signature$ is polynomial. Consequently, $\freemonad$ is itself polynomial 
% and thus satisfies the required conditions. More succinctly, \autoref{axiom:axiom-of-signature-version-2}
% \emph{de facto} implies \autoref{axiom:axiom-preservation-pullbacks-free-monad}, 
% this way making the latter redundant. 

\subsection{Examples}
We conclude this section by looking at some examples of how specific 
notions of abstract syntax are captured by the general categorical framework. 

\begin{example}
Before moving to concrete examples, we observe that a large family of instances of 
the theory of syntax is obtained throughout \emph{simple polynomial functors}. 
%\subsubsection{Polynomial Functors}
%\autoref{theorem:adamek-theorem} ensures that if a functor 
%$\signature$ is finitary, then it has initial algebra and, 
%consequently, $\freemonad$ exists. 
%There is a well-known class of well-behaved 
%functors satisfying this property, namely the one of 
Recall that a functor on a topos $\topos$ is a \emph{simple polynomial 
functor} \cite{Jacobs/Introduction-to-coalgebra/2016}
if it is built from the identity and constant functors
using composition, finite products, and set-indexed coproducts.\footnote{
More generally, given an arrow $f: B \to A$ in $\topos$, the polynomial 
functor $P_f: \topos \to \topos$ induced by $f$ is defined as 
$P_f(X) = \sum_{a:A} X^{B(a)}$, where the latter expression 
is written using the internal language of $\topos$ 
(equivalently, let us consider the adjoint functors on the slice 
 category $\sum_B \dashv f^* \dashv \prod_B$ with 
 $\sum_B, \prod_B: \topos/B \to \topos/A$ and $f^*: \topos/A \to \topos/B$;
 writing $B^*: \topos \to \topos/B$ for the functor obtained taking 
$A = 1$ (and thus $f: B \to 1$), we have $P_f = B^*; \prod_f; \sum_f$). 
%Polynomial functors are closed under composition and preserve pullbacks~\cite{gambino-kock-polynomical-functors-and-monads,non-wellfounded-trees-in-categories}.
%Moreover, their 
%free monad, if it exists, is itself polynomial. 
%Every polynomial functor has initial algebra, but not all polynomial functors are finitary,
%a condition achieved 
%by requiring the fiber of $f$ to be finite, in suitable sense (notice that simple polynomial functors 
%indeed meet such a condition). 
}
Simple polynomial functors (and suitable extensions thereof) can be thought as abstract notions of syntax (specification)~\cite{fiore-polynomial-functors-simple-type-theories}. 
\end{example}

\begin{example}[First-Order Terms]
We have already seen in \autoref{section:informal-intro} that a first-order 
signature $\signatureop$ induces a (simple polynomial) functor $\signature$ on 
$\set$, and that $\freemonad$ gives the syntax of $\signatureop$-terms.
% Our first example describes (first-order) terms over a (mono-sorted) signature $\signatureop$
% of operation symbols. Being first-order, $\signatureop$-terms are naturally modelled in the category 
% $\set$ of sets and functions. Recall that a signature $\signatureop$ is a set of operation
% symbols $\op$ together with natural numbers representing their arity. 
% Given a set $X$ usually thought as a set of variables or primitive symbols, 
% the collection $\terms{\signatureop}{X}$ is inductively defined thus:
% \[
% \infer{\val{x} \in \terms{\signatureop}{X}}{x \in X}
% \qquad 
% \infer{\op(\termone_1, \hhh, \termone_n) \in \terms{\signatureop}{X}}
% {\termone_1 \in \terms{\signatureop}{X} \cc \termone_n \in \terms{\signatureop}{X}
% &
% (\op, n) \in \signatureop}
% \]
% Each signature $\signatureop$ determines a functor $\signature: \set \to \set$ defined by 
% $\signature A \defeq \coprod_{(\op,n) \in \signatureop} A^{n}$. The functor $\signature$ 
% is a simple polynomnial functor, hence it is finitary 
% and thus has initial algebra. From that follows that 
% $X + \signature-$ is finitary too, meaning that there exists the free monad 
% $\freemonad$ over $\signature$. Moreover, it is easy to see that 
% $\terms{\signatureop}{X} = \freemonad X$, with $\unit$ and $\sigma$ acting as the leftmost 
% and rightmost inductive clause above, respectively. 
\end{example}

\begin{example}[Higher-Order Terms: the $\lambda$-calculus]
\label{example:terms-lambda-calculus}
We now go beyond first-order syntax and introduce  
variable binding. For the sake of exposition, 
instead of defining binding signatures and terms in full generality 
(something that can be easily done~\cite{fiore-plotkin-turi-99}), 
we focus on a single example of such syntax: terms of the $\lambda$-calculus
modulo $\alpha$-conversion~\cite{Barendregt/Book/1984}. 
Following the seminal work by Fiore et al.~\cite{fiore-plotkin-turi-99}, the key insight to 
model terms with binders is to move from $\set$ to categories of expressions in context. 
Let $\nats$ be the category 
of finite cardinals, i.e. the full subcategory of $\set$ with objects 
sets $\numeral{n} \defeq \{0, \hhh, n-1\}$. 
We think of sets $\numeral{n}$ as (indexes of variables of) finite contexts 
and of a function 
$f: \numeral{n} \to \numeral{m}$ as a context renaming.
Accordingly, we consider the presheaf category $\set^{\nats}$ of sets 
(of expressions) in context. 
Fixed a countable collection of variables $\varone_0, \varone_1, \hhh$, 
the presheaf $\Lambda$ of $\lambda$-terms maps 
$\numeral{n}$ to the set 
$\Lambda(\numeral{n})$ of $\lambda$-terms modulo ($\alpha$-)renaming 
 with free variables in 
$\{\varone_0, \hhh, \varone_{n-1}\}$. 
An inductive definition of $\Lambda$ is given thus:
\[
\infer{\varone_i \in \Lambda(\numeral{n})}{i \in \numeral{n}}
\quad 
\infer{\lambda \varone_{n+1}.\termone \in \Lambda(\numeral{n})}{\termone \in \Lambda(\numeral{n+1})}
\quad 
\infer{\termone\;\termtwo \in \Lambda(\numeral{n})}{\termone  \in \Lambda(\numeral{n}) &  \termtwo \in \Lambda(\numeral{n})}
\]
%\begin{varitemize}
%    \item If $i \in \numeral{n}$, then $x_i \in \Lambda(\numeral{n})$;
%    \item If $t \in \Lambda(\numeral{n+1})$, then $\lambda x_{n+1}.t \in \Lambda(\numeral{n})$;
%    \item If $t,s \in \Lambda(\numeral{n})$, then $ts \in \Lambda(\numeral{n})$.
%\end{varitemize}
    % \[
        % \infer{\texttt{x}_i \in \Lambda(\underline{n})}{i \in \underline{n}}
        % \qquad 
        % \infer{\lambda \texttt{x}_{n+1}.\termone \in \Lambda(\underline{n})}{\termone \in \Lambda(\underline{n+1})}
        % \qquad
        % \infer{\termone \termtwo \in \Lambda(\underline{n})}{\termone \in \Lambda(\underline{n}) 
        % &  \termtwo \in \Lambda(\underline{n})}
        % \]
        As it is customary~\cite{Pitts/ATBC/2011}, we oftentimes write 
        $\numeral{n} \vdash \termone$ (or $\vect{\varone} \vdash \termone$, tacitly assuming 
        $\vect{\varone} = \varone_0, \hhh, \varone_{n-1}$) in place of 
        $\termone \in \Lambda(\numeral{n})$.
       Let us consider the signature 
       functor $\signature X \defeq \delta X + (X \times X)$, where 
       $\delta X(\numeral{n}) \defeq X(\numeral{n+1})$, and let us write 
       $V$ for the presheaf of variables mapping 
       $\numeral{n}$ to $\varone_0, \hhh, \varone_{n-1}$.
       Then the free monad $\freemonad$ over $\signature$
       maps $V$ to the presheaf
       $\freemonad V \cong V + \delta(\freemonad V) + (\freemonad V \times \freemonad V)$ of 
       $\lambda$-terms modulo $\alpha$-renaming. 
        The functor $\delta$ is finitary  polynomial)~\cite{fiore-plotkin-turi-99,fiore-polynomial-functors-simple-type-theories} 
       and satisfies the conditions of \autoref{axiom:axiom-of-signature}), so that 
       the whole functor $\signature$ does. In light of that,
       we can extend the class of simple polynomial functors by including
       $\delta$ without altering the `good' properties of simple polynomial functors. 
       The resulting class is sometimes referred to as che class of 
       \emph{binding functors}~\cite{miculan-fusion-calculus}.
\end{example}

\begin{example}[Nominal Sets]
An alternative universe for modelling expressions with variable binding and names is given by 
the category $\nom$ of nominal sets~\cite{pitts-nominal-sets,pitts-nominal-1}.
Due to space constraints, we will not give details about that but simply 
remark that $\nom$ is a Grothendieck topos (it is isomorphic to the 
Schanuel topos~\cite{pitts-nominal-sets}), 
and that syntax for expressions with names and binders 
can be given as free monads over a mild variation of the binding functors
defined in the previous example. 
\end{example}

\begin{example}[Further Examples]
More generally, presheaf categories
% \footnote{Even if we do not cover them in this work, 
% we remark that finer notions of syntax, such as linear or modal syntax, can be given as initial 
% algebras of functors on categories $\mathit{EM}(T)^{\catone^{\mathit{op}}}$, where 
% $\mathcal{EM}(T)$ is the category of algebras of a monad $T$.}
of the form $\set^{\category^{\mathit{op}}}$, with $\category$ small, 
can be used to model many universes of expressions, and several notions of syntax have been 
given as initial algebras of suitable functors on them.
Examples include sorted expressions~\cite{Robinson-initial-algebra-sorted-expression}, 
simply~\cite{fiore-polynomial-functors-simple-type-theories} and polymorphically-typed expressions~\cite{hamana-free-sigma-monoids,hamana-polymorphic-syntax}, graphs and hypergraphs, 
and diagrams~\cite{Selinger2011}. 
For instance, Bonchi et al.~\cite{Bonchi-algebraic-semantics-string-diagrams} 
model string diagram as initial algebra of 
polynomial-like functors 
in the category of spans $\mathbb{N} \leftarrow S \rightarrow \mathbb{N}$ in $\set$ 
(arrows are are span morphisms).Notice that such a category is isomorphic to 
$\set^{\mathbb{N} \times \mathbb{N}}$,  
where $\mathbb{N} \times \mathbb{N}$ is the discrete category with objects pairs of natural numbers.
\end{example}

\subsection{Substitution}
In addition to the aforementioned crucial features of mathematical
syntax (e.g. recursive term formation, structural induction, etc.), 
there is another major syntactic-like structure that characterise  
(many notions of) symbolic syntax: \emph{substitution}. 
Whereas there is a general consensus on the initial algebra approach to term-formation, 
several axiomatics for substitution have been proposed --- such as
monoids on monoidal categories~\cite{fiore-plotkin-turi-99,hamana-free-sigma-monoids}, 
substitution algebras and (heterogeneous) structure~\cite{fiore-plotkin-turi-99,Uustalu-substitution}, 
monads with pointed 
strength~\cite{fiore-pointed-monad}, 
modules~\cite{hirschowitz-andre-1,hirschowitz-andre-2,hirschowitz-syntax-2,hirschowitz-syntax-3}, and monads on nominal sets~\cite{pitts-nominal-1,pitts-nominal-sets},
just to mention but a few --- and a general consensus on them is arguably missing. 

Nonetheless, all these approaches share a common trait; they all
view substitution as a (structurally) recursively-defined syntax-preserving morphism.
When dealing with 
first-order syntax (i.e. signature functors and their free algebras on $\set$), 
a satisfactory account of substitution can be given in terms of monads only: 
being in $\set$, it is possible to internalise the monad structure  
of $\freemonad$, this way
obtaining (monadic binding) maps $\substmap: \freemonad X \times \freemonad X^X \to \freemonad X$, natural in $X$, implementing substitution: given a term $\termone$ and a substitution 
$[\vect{\termtwo}/\vect{x}]$ viewed as 
a map from variables $X$ to $\freemonad X$, the term 
$\substmap(\termone, [\vect{\termtwo}/\vect{x}])$ --- usually written as 
$\termone[\vect{\termtwo}/\vect{x}]$ --- represents the result of 
simultaneously substituting variables 
$x_i$ with $\termtwo_i$ in $\termone$. Monad laws ensures desired equational 
properties of substitution: moreover, since any signature functor 
$\signature$ has a strength $\strength: \signature A \times B \to \signature(A \times B)$, 
we recover substitution via (a suitable instance of) the unique arrow 
$\freemonad A \times B \to \freemonad(A \times B)$ extending $\strength$ \cite{fiore-plotkin-turi-99}. 
This way, we obtain a structurally recursive definition of substitution. 

The aforementioned view of substitution does not scale to 
to richer forms of syntax, such as syntax with variable binding.
To overcome this problem, among the many structures defined, we rely on
\emph{substitution algebras} \cite{fiore-plotkin-turi-99}, 
namely objects $A$ together with 
arrows $\nu: 1 \to A^\vars$ (\emph{generic new variable}) and 
$\substmap: A \times A^\vars \to A$ (\emph{substitution}) subject to 
suitable coherence conditions.
%akin to weakening, extensionality,
%and associativity of substitution (see Definition 3.1 \cite{fiore-plotkin-turi-99}). 
Here, $\vars$ is a suitable object acting as an object of variables. 
Intuitively, the latter is any object ensuring the existence of maps 
for variable manipulation (like duplicating or swapping variables) 
that are necessary to express the aforementioned coherence conditions.\footnote{
For instance, in the presheaf-based framework by Fiore et al. \cite{fiore-plotkin-turi-99}, 
such an object is given by the presheaf of variables, whereas in nominal sets it is given 
by the (nominal) sets of names. A more general analysis of objects of names 
is given by Menni~\cite{menni03}.}

Fixed such an object $\vars$, the (currying of the) map $\eta$ gives us a candidate map $\nu$; 
moreover, assuming $\signature$ to have strength  
$\strength: \signature(\freemonad\vars^\vars) \times \freemonad\vars \to 
\signature(\freemonad\vars^\vars \times \freemonad \vars)$,\footnote{
This point is actually delicate: in fact, it turns out that crucial in
this procedure is the fact that $\freemonad \vars$ is \emph{pointed} on $\vars$, 
meaning that we have an arrow $\vars \to \freemonad \vars$. This observation 
led to the identification of monads with pointed strength as a way to 
internalise notions of substitutions \cite{fiore-pointed-monad}.}
we obtain the map $\substmap$ by initiality. 
These maps are compatible with the $\signature$-algebra structure of 
$\freemonad\vars$, and thus give to $\signature \vars$ the status of 
an initial $\signature$-substitution algebra~\cite{fiore-plotkin-turi-99}.

\begin{definition}
\label{def:substitution-algebra}
\begin{varenumerate}
    \item An object $\vars$ of $\topos$ is an \emph{object of variables}
        if it comes with the maps given in \cite[Definition 3.1]{fiore-plotkin-turi-99}. 
    \item A \emph{substitution algebra} 
        is a triple $(A, \nu, \substmap)$ with $A$ an object of $\topos$, and arrows
        $\nu: 1 \to A^\vars$ and $\substmap: A \times A^\vars \to A$. 
        We require these data to satisfy the compatibility conditions of 
        \cite[Definition 3.1]{fiore-plotkin-turi-99}. 
    \item $\signature$\emph{-substitution algebras} are $\signature$-algebras endowed with 
        a compatible substitution algebra structure as in \cite{fiore-plotkin-turi-99}.
\end{varenumerate}
\end{definition}
Finally, we assume the signature functor $\signature$ to have the aforementioned 
(pointed) strength, this way ensuring $(\freemonad\vars, [\eta, \sigma], \nu, \substmap)$ 
to be $\signature$-substitution algebra. 

\begin{remark}
\autoref{def:substitution-algebra} is deliberately sloppy and
there is no objective reason to rely on substitution algebras rather than on 
other structures.
The reason behind all of that is twofold: on the one hand, 
this choice improves accessibility of the paper by making it digestible to the reader lacking 
the specific categorical background; on the other hand (and most importantly), 
once gone throughout the relational analysis of rewriting, the reader should be convinced that 
the chosen model of substitution is not \emph{operationally} relevant: 
what matters is to have \emph{a} notion of substitution 
inducing an operationally well-behaved relational substitution operator. 
By operationally well-behaved we mean a collection of algebraic laws giving 
an axiomatic definition of relational substitution. 
Such laws 
(\autoref{proposition:lassen-algebra-2} and 
\autoref{lemma:algebra-of-substitution-and-compatible-refinement}) 
are all that matters for rewriting, up 
to the point that we could be completely 
agnostic with respect to the substitution structure used and simply assume to have one inducing 
such a relational operator. Substitution algebras do so, but the reader can easily check 
that many other structures (such as $\signature$-monoids) do that as well. 
As a general (albeit informal) principle, any structure modelling substitution 
as a recursively-defined syntax-preserving map gives raise to a well-behaved relational 
substitution operator (see \autoref{remark:substitution} and the end of \autoref{section:an-allegorical-theory}).
\end{remark}

\section{Allegories: the Theoretical Minimum}

Initial algebras and free monads provide an elegant mathematical description of (abstract) syntax. 
The kind of syntax we are interested in here is the one of symbolic expressions. 
The ``mathematics of syntax approach'' works perfectly for this kind of syntax (as witnessed 
by the many examples previously mentioned), but it does not capture its deep essence, 
that thing that makes symbolic syntax different from the syntax of, e.g., natural language.  
The peculiarity of symbolic expressions, in fact, does not rely in their syntax, but in their 
(operational) ``semantics'': they can be manipulated symbolically 
(cf. \autoref{section:introduction}: \emph{symbolic terms are encapsulated computations}).

In this work, we move 
from the conceptual assumption that symbolic manipulation is an inherently relational notion. 
Such a conceptual point of view is remarkably powerful as it draws a path from 
syntax to semantics: to obtain the symbolic dynamics of expressions, 
we simply take the categorical
theory of abstract syntax and extend it to a relational setting, in a precise sense that 
we are going to define. 
Any universe of expressions $\topos$, in fact, induces a category $\rell{\topos}$ 
of relations describing manipulations between expressions. 
$\rell{\topos}$ is an allegory~\cite{scedrov-freyd}
and thus we can rely on a rich relational framework to study it.

Switching from the categorical to the allegorical point of view one sees that many 
categorical notions have an allegorical counterpart,
and that to define a relational theory of
symbolic manipulation one precisely needs the allegorical extensions of 
the notions defining categorical syntax, namely free monads and initial algebras, 
as summarised in \autoref{fig:basic-categorical-syntax}. 
%to an allegorical setting, hence obtaining \autoref{fig:basic-allegorical-syntax-1}. 
We dedicate this section to study such extensions.

\subsection{Allegories}
Given a universe of expressions $\topos$, the category of its 
relations $\rell{\topos}$ has objects of $\topos$ as objects, whereas 
an arrow from $A$ to $B$ is a subobject $a: A \times B \to \Omega$, 
where $\Omega$ is the subobject classifier of $\topos$. 
Subobjects of the form $a: A \times B \to \Omega$ behave as relations from 
$A$ to $B$: they have converse, compositions, union, etc. The precise sense 
in which they behave relationally has been defined 
through the notion of an \emph{allegory}~\cite{scedrov-freyd}. More precisely, 
$\rell{\topos}$ is a locally-complete power allegory~\cite{scedrov-freyd,algebra-of-programming}. 

\begin{definition}[\cite{scedrov-freyd}]
An allegory $\allegory$ is a category such that each hom-set $\allegory(A,B)$ 
is endowed with: (i) a partial order $\leqq$ and a meet operation $\wedge$ making composition monotone; 
(ii) an order-preserving contravariant involution $-\dd$ 
(so that $\relone\dddd = \relone$, $(\relone; \reltwo)\dd = \reltwo\dd; \relone\dd$, 
and 
$(\relone \wedge \reltwo)\dd = \relone\dd \wedge \reltwo\dd$).
All these data, additionally, have to 
obey the so-called \emph{modular law}:
$
\relone;\reltwo \wedge \relthree \leqq (\relone \wedge \relthree;\reltwo\dd);\reltwo.
$
\end{definition}

Given an allegory $\allegory$, we refer to its arrows as \emph{relations} 
and we call the relation $\relone\dd: B \to A$ the \emph{converse} of $\relone: A \to B$.
As usual, we say that a relation $\relone: A \to A$ is \emph{reflexive}, \emph{symmetric}, and 
\emph{transitive} if $\idrel \leqq \relone$, $\relone\dd \leqq \relone$, and 
$\relone;\relone \leqq \relone$, respectively. Moreover, we say that 
$\relone: A \to B$
is \emph{entire} if $\idrel \leqq \relone;\relone\dd$, \emph{simple} if 
$\relone\dd; \relone \leqq \idrel$, and that it is a \emph{map} if 
it is entire and simple. The subcategory $\mapp{\allegory}$ of an allegory 
$\allegory$ is the category having objects of $\allegory$ as objects and 
\emph{maps} of $\allegory$ as arrows.
% \footnote{ 
% Notice that modularity implies that the partial order $\leqq$ is discrete 
% when restricted to $\mapp{\allegory}$ (meaning that $f \leqq g \implies f = g$); 
%  additionally, if the converse of a map $f$ is itself a map, then $f\dd$ is the inverse of 
%  $f$ (i.e. $f^{-1} = f\dd$) and $f$ is an isomorphism. 
% }

\begin{example}
Maps in $\rell{\topos}$ are precisely arrows in $\topos$, so that 
$\mapp{\rell{\topos}} \simeq \topos$. This means that instead of working with 
$\topos$ as a primitive notion we may (and we will) take a truly relational 
perspective and work with (an axiomatisation) 
of $\rell{\topos}$, this way thinking about $\topos$ as the restriction of $\rell{\topos}$ 
to functional relation.\footnote{Following this perspective, we think about a topos 
as a structure in which relations coincide with set-valued maps~\cite{de-moor-phd-thesis}.}
\autoref{theorem:freyd-equivalence-categories-allegories} 
generalises this correspondence to large classes of allegories and categories.
\end{example}

\begin{remark}
To avoid ambiguities when working both with $\topos$ and $\rell{\topos}$,  
we use the notation $f: A \to B$ and $\relone: A \torel B$ for arrows 
in $\topos$ and in $\rell{\topos}$, respectively. We actually need this convention 
in this section only, since in the next one we will in a full allegorical framework.
\end{remark}

% Modularity implies that the partial order $\leqq$ is discrete 
% when restricted to $\mapp{\allegory}$ (meaning that $f \leqq g \implies f = g$); 
% additionally, if the converse of a map $f$ is itself a map, then $f\dd$ is the inverse of 
% $f$ (i.e. $f^{-1} = f\dd$) and $f$ is an isomorphism. 

\begin{definition}
\label{definition:allegory-power-tabular-etc}
    Given an allegory $\allegory$, we say that $\allegory$ is:
    \begin{varenumerate}
        \item \emph{Tabular} if any relation $\relone: A \to B$ has a 
        (necessary unique, up-to isomorphism)
        tabulation,
        i.e. maps
        $f: R \to A$ and $g: R \to B$ such that $\relone = f\dd; g$ and 
        $f;f\dd \wedge g;g\dd = \idrel$.
        \item \emph{Unitary} if it has a unit $U$, i.e. an object $U$ such that: 
        (i) $\idrel: U \to U$ is the largest relation in $\allegory(U,U)$; 
        (ii) for any object $A$, there is an entire relation (which is necessarily a map) 
        $u: A \to U$. 
        % A unit in $\allegory$ is always a terminal object in $\mapp{\allegory}$. 
        % Conversely, if $\allegory$ is tabular, then a terminal object in $\mapp{\allegory}$ is 
        % a unit in $\allegory$. 
        \item \emph{Locally complete} if it is unitary and tabular and, 
        for all objects $A$, $B$, the set 
        $\allegory(A,B)$ is a complete lattice with composition and finite intersection 
        distributing over arbitrary joins.\footnote{
        In particular, we have 
        $\relone; \join_i \reltwo_i = \join_i \relone; \reltwo_i$ and 
        $(\join_i \relone_i);\reltwo = \join_i \relone_i; \reltwo$.
        % \begin{align*}
        % \relone; \join_i \reltwo_i &= \join_i \relone; \reltwo_i
        %  &
        %  (\join_i \relone_i);\reltwo &= \join_i \relone_i; \reltwo
        %  &
        %  \relone \wedge \join_i \reltwo_i &= \join_i \relone \wedge \reltwo_i
        %  &
        %  (\join_i \relone_i) \wedge \reltwo &= \join_i \relone_i \wedge \reltwo.
        % \end{align*}
        }
       % Notice that the universal property of join (whereby $\join_i \relone \leqq \reltwo \iff 
        %\forall i.\ \relone_i \leqq \reltwo$) implies $(\join_i \relone_i)\dd = \join_i \relone_i\dd$.
        We denote by $\relone \vee \reltwo$ the relation $\join\{\relone, \reltwo\}$ and by 
        $\bot$ the relation $\join \emptyset$. 
    \item A \emph{locally complete power allegory} (\lcp{} allegory) if  it is locally complete and
    for any object $A$ there is a power object 
    $PA$ such that: (i) to any relation $\relone: A \to B$ is associated a map 
    $\Lambda \relone: A \to PB$; (ii) there are relations
    $\ni_A: PA \to A$; (iii) such that
    $
    f = \Lambda \relone$ iff $f; \ni = \relone.
    $
    % If $\allegory$ is not locally complete but still posses finite joins and 
    % left and right adjoints of composition, we say that $\allegory$ is a power 
    % allegory. 
    \end{varenumerate}
\end{definition}
\autoref{definition:allegory-power-tabular-etc} is standard 
in the literature on allegory theory. From a categorical perspective, it 
can be motivated by the following result. 
% but some comments are still in order. 
% First, the definition of a unitary tabular allegory is motivated by the equivalence between 
% such allegories and categories of relations on regular categories. 
% Similarly, the definition of a (locally complete) power allegory is motivated by their equivalence  
% with categories of relations on a (Grothendieck) topos. 
% All of that gives a large array of examples.

\begin{theorem}[\cite{scedrov-freyd,pitts-quantaloids}] 
\label{theorem:freyd-equivalence-categories-allegories}
For any Grothendieck topos $\topos$, 
its category of relations $\rell{\topos}$ 
is a  \lcp{} allegory,
and $\topos \simeq \mapp{\rell{\topos}}$. 
Vice versa, for any \lcp{} allegory
$\allegory$, its subcategory of maps $\mapp{\allegory}$ is a 
Grothendieck topos, and
$\allegory \simeq \rell{\mapp{\allegory}}$. 
% The same results extends to (Grothendieck) topoi and (locally complete) 
% power allegories. In particular, for 
% any (Grothendieck) topos $\mathcal{E}$, $\rell\mathcal{E}$ 
% is a (locally complete) power allegory and $\mathcal{E} \simeq \mapp{\rell\mathcal{E}}$. 
% Vice versa, 
% if $\allegory$ is a (locally complete) power allegory, then 
% $\mapp{\allegory}$ is a (Grothendieck) topos and $\rel(\mapp{\allegory}) \simeq \allegory$. 
\end{theorem}

Consequently, for a universe of expressions $\topos$, 
we see that $\rell{\topos}$ is a \lcp{} allegory.

\begin{example}
Examples are, in principle, not needed (just construct $\rell{\topos}$ for 
the examples of $\topos$ seen in the previous section). Nonetheless, we mention:
%we shortly make explicit relations in a couple of universe of expressions $\topos$.
\begin{varenumerate}
    \item 
        $\rell{\set}$ is $\rel$.
    \item $\rell{\set^{\nats}}$ is the allegory of dependent relations closed under context renaming 
        and weakening. 
        That is, a relation $\relone: A \to B$ is an element 
        $\relone \in \prod_{\numeral{n}}\rel(A(\numeral{n}), B(\numeral{n}))$ such that, 
        for any $\phi: \numeral{n} \to \numeral{m}$, we have:
        $\relone(\numeral{n}); B\phi \subseteq A\phi; \relone(\numeral{m})$.
        \[
        \xymatrix{
        \laxcommuterel
        A(\underline{n}) \ar[r]^{A\phi} \ar[d]_{\relone(\underline{n})} & A(\underline{m}) \ar[d]^{\relone(\underline{m})}
        \\
        B(\underline{n}) \ar[r]_{B\phi}  & B(\underline{m}) 
        }
        \]
        Notice that considering the presheaf of $\lambda$-terms, 
        relations in $\rell{\ctxcat}$ are precisely the so-called term 
        relations~\cite{Lassen/PhDThesis,Gordon/FOSSACS/01,Pitts/ATBC/2011} 
        used in relational reasoning on $\lambda$-terms.
        Accordingly, we employ the notation 
        $\vect{\varone} \vdash \termone \mathrel{\relone} \termtwo$ to state that 
        $\vect{\varone} \vdash \termone, \termtwo$ and 
        $\termone \mathrel{\relone(\vect{\varone})} \termtwo$.
        In particular, notice that we have the following weakening and renaming rule:
        \[
        \infer{\vect{\varone}, \vect{\vartwo} \vdash \termone \mathrel{\relone} \termtwo}
        {\vect{\varone} \vdash \termone \mathrel{\relone} \termtwo}
        \qquad 
        \infer{\vect{\vartwo} \vdash \termone[\vect{\vartwo}/\vect{\varone}] \mathrel{\relone} \termtwo[\vect{\vartwo}/\vect{\varone}]}
        {\vect{\varone} \vdash \termone \mathrel{\relone} \termtwo}
        \]
    \item $\rell{\nom}$ is the allegory of equivariant 
        relations~\cite{pitts-nominal-sets}.
\end{varenumerate}
\end{example}

\subsection{Relators}
The construction of $\rell{\topos}$ from $\topos$ gives a relational counterpart of 
\autoref{axiom:axiom-of-expression-universe}: we need a \lcp{} allegory.
To extend syntax and syntax specification, however, 
we need to understand what are the relational counterparts of initial algebras 
and free monads. The crucial notion to do that is the one of a 
\emph{relator}~\cite{kawahara1973notes,algebra-of-programming,backhouse-polynomial-relators,backhouse-relational-theory-of-datatypes}\footnote{Notions essentially equivalent to the 
one of a relator, such as \emph{relational extensions}~\cite{Barr/LMM/1970}, 
\emph{relation lifting}~\cite{Jacobs/Introduction-to-coalgebra/2016,DBLP:journals/iandc/HermidaJ98}, 
and \emph{lax extensions}~\cite{Hoffman-Seal-Tholem/monoidal-topology/2014,Hoffman/Cottage-industry/2015} have been independently introduced in several fields.}, the relational 
counterpart of functors.

\begin{definition}
A relator on an allegory $\allegory$ is a functor 
$\relator: \allegory \to \allegory$ that is monotone and 
preserves converse.
\end{definition}
In particular, a relator $\relator$ satisfies the law 
$\relator(\relone\dd) = (\relator \relone)\dd$, so that we can unambiguously
write $\relator \relone\dd$.
It is easy to prove that $\relator f$ is a map, whenever $f$ is. Moreover, since 
$\allegory$ is tabular, we see that a functor is a relator if and only if it preserves 
converse \cite[Theorem 5.1]{algebra-of-programming}. 

Since abstract syntax is specified by functors on $\topos$, we are interested 
in the following question: \emph{given a functor 
$\signature$ on $\topos$, 
can we extend it to a relator $\bextt{\signature}$ on $\rell{\topos}$?}
% (or, dually, can we extend $\signature: \mapp{\allegory} \to \mapp{\allegory}$, 
% to a relator on $\allegory$?).

% \begin{definition}
%     A relational extension of a functor $\signature: \mapp{\allegory} \to \mapp{\allegory}$, 
%     is a relator $\bextt{\signature}$ on $\allegory$ such that 
%     $\bextt{\signature} f = \signature f$, for any map $f$.
% \end{definition}

\begin{definition}
    A relational extension of a functor $\signature: \topos \to \topos$, 
    is a relator $\bextt{\signature}$ on $\rell{\topos}$ such that 
    $\bextt{\signature}A = \signature A$ and
    $\bextt{\signature} f = \signature f$, for any object $A$ and map $f$.
\end{definition}

Consequently, we see that if $\relone$ is tabulated as $f\dd;g$, then we must have 
$
\bextt{\signature}\relone = \bextt{\signature}(f\dd;g) = \bextt{\signature}f\dd;\bextt{\signature}g 
= (\signature f)\dd; \signature g.
$
Therefore, a relational extension of $\signature$, if it exists,
it must be unique and defined by 
$\bextt{\signature}(f\dd;g) \defeq (\signature f)\dd; \signature g$ (notice that this definition is 
independent of the choice of the tabulation).
This means that for any $\signature$, we have a candidate relator $\bextt{\signature}$. 
Barr~\cite{Barr/LMM/1970} and Carboni et al.~\cite{carboni-kelly-wood} 
found conditions to answer 
the aforementioned question in the affirmative, this way ensuring that $\bextt{\signature}$ 
is indeed a relator. 

\begin{theorem}[\cite{Barr/LMM/1970},\cite{carboni-kelly-wood}]
\label{theorem:barr-theorem}
$\bextt{\signature}$ is a relator if and only if $\signature$ nearly preserves pullbacks 
and preserves strong epimorphisms. Moreover, $\bextt{\signature}$ is the only 
relational extension of $\signature$.
\end{theorem}

\autoref{theorem:barr-theorem} can also be seen from an allegorical perspective by stating 
that any relator $\relator: \allegory \to \allegory$ gives a functor $\relator: \mapp{\allegory} \to 
\mapp{\allegory}$ that nearly preserves pullbacks 
and preserves strong epimorphisms. It is worth noticing that if a functor preserves 
(weak) pullbacks, then it nearly preserves pullbacks and  
that strong and regular epimorphisms coincide in any topos (cf. \autoref{axiom:axiom-of-signature}).

\begin{example}
\begin{varenumerate}
    \item Since $\set$ 
        satisfies the axiom of choice, any functor on it preserves regular epimorphisms.
        Therefore, if $\signature$ nearly preserves pullbacks, then 
        $\bextt{\signature}$ is a relator.
    \item 
         Simple polynomial functors 
         all extend to relators~\cite{de-moor-phd-thesis}. 
         The resulting class of relators is called the class of 
         \emph{polynomial relators}.  Polynomial relators 
         can be defined explicitly by means of (co)product (bi)relators, 
         constant, and identity relator.\cite{de-moor-phd-thesis}
        %  , as the notion of a relator 
        % generaslises functors to allegories,  one obtains the notions of a birelator and, 
        % more generally, of a $n$-ary relator as relational counterparts of 
        % bifunctors and $n$-ary functors. 
        % Products and coproducts of $\topos$, in particular, extend 
        % to relators. Consequently, we obtain binary
        % operations $\widetilde{+}$ and $\widetilde{\times}$ acting on relators pointwise.
        % For instance, $(\relator \mathbin{\widetilde{+}} \Psi)(\relone) 
        % \defeq \relator\relone \mathbin{\bextt{+}} \Psi\relone$ 
        % (and similarly for arbitrary coproducts).
        % Since any constant functor 
        % gives a constant relator, as well as the identity functor, 
        % and relators are closed under composition, 
        % we can define the class of simple polynomial relators 
        % as the least class of relators containing the identity
        % and constant relators and closed under composition, $\widetilde{\times}$, 
        % and $\widetilde{\coprod}$. 
        % An inductive, explicit definition of polynomial relators on $\set$ can be found in the 
        %  \cite{Jacobs/Introduction-to-coalgebra/2016}.
    \item The functor 
        $\delta$ on $\set^{\nats}$ extends to a relator~\cite{fiore-plotkin-turi-99,miculan-fusion-calculus}, 
        so that binding polynomial functors all extends to relators.
\end{varenumerate}
\end{example}

Summing up, \autoref{axiom:axiom-of-expression-universe} entails 
that $\rell{\topos}$ is a \lcp{} allegory, whereas
\autoref{axiom:axiom-of-signature} implies that 
$\bextt{\signature}$ is a relator. The latter axiom, however, imposes a further condition 
on $\signature$ that is not needed to ensure that $\bextt{\signature}$ is a relator: 
$\signature$ must be finitary. How does that impact on $\bextt{\signature}$? 
Playing a bit with the definition of countable union in $\rell{\topos}$ 
(which uses the coproduct relator~\cite{de-moor-phd-thesis}), 
we see that if $\signature$ is finitary, then 
$\bextt{\signature}$ is $\omega$-continuous, meaning that 
$\bextt{\signature}(\join_n \relone_n) = \join_n \bextt{\signature} \relone_n$, 
for any $\omega$-chain $(\relone_n)_{n \geq 0}$ of relations. 
%As we shall see, 
%$\omega$-continuity will be crucial to reason about least fixed points of relators.

\begin{proposition}
\label{proposition:finitary-implies-continuity}
    If $\signature: \topos \to \topos$ is finitary, then $\bextt{\signature}$ is $\omega$-continuous.
\end{proposition}

In light of \autoref{proposition:finitary-implies-continuity}, we say that 
a relator on $\allegory$ is \emph{finitary} if it is so as a functor on 
$\mapp{\allegory}$. Consequently, $\widehat{\signature}$ is finitary whenever 
$\signature$ is.
What remains to do is to give a relational counterpart to $\freemonad$. 
Since (the carrier of) $\freemonad$ is
a functor, we can simply pick $\bextt{\freemonad}$ as its relational extension. 
Of course, we have to ensure that  $\bextt{\freemonad}$ is a relator. That directly follows 
from the fact that $\bextt{\signature}$ is a relator. 

\begin{proposition}[\cite{algebra-of-programming}]
\label{proposition:existence-relator-free-monad}
If $\bextt{\signature}$ is a relator, then so is $\bextt{\freemonad}$. 
% Moreover, $\bextt{\freemonad}$ is a monadic relator, meaning that 
% $\bextt{\freemonad} \bextt{\freemonad}\relone; \mu = \mu; \bextt{\freemonad}\relone$ 
% and $\relone; \eta = \eta; \bextt{\freemonad}\relone$.
\end{proposition} 

\begin{example}
\label{example:barr-extension-term-monad}
\begin{varenumerate}
\item Let $\signature$ be the functor given by a first-order signature $\signatureop$ on $\set$, 
so that $\freemonad$ is the $\signatureop$-term monad. The relator $\bextt{\freemonad}$ has 
the following inductive characterisation: 
 \[
 \infer{\val{x} \mathrel{\bextt{\freemonad}\relone} \val{y}}{x \mathrel{\relone} y}
 \quad 
 \infer{\op(\termone_1, \hhh, \termone_n) \mathrel{\bextt{\freemonad}\relone} \op(\termtwo_1, \hhh, \termtwo_n)}
 {\termone_1 \mathrel{\bextt{\freemonad}\relone}  \termtwo_1 & \cc & \termone_n\mathrel{\bextt{\freemonad}\relone}  \termtwo_n 
 %& \op \in \signatureop_{n}
 }
 \]
%\begin{varitemize}
%\item If $x \mathrel{\relone} y$, then $\val{x} \mathrel{\bextt{\freemonad}\relone} \val{y}$;
%\item If $t_i \mathrel{\bextt{\freemonad}\relone}  s_i$, for any $i$, then 
%$\op(t_1, \hhh, t_n) \mathrel{\bextt{\freemonad}\relone} \op(s_1, \hhh, s_n)$.
%\end{varitemize}
Equivalently, we say that $\termone \mathrel{\bextt{\freemonad}\relone} \termtwo$ if and only if 
there exists a context $C$ such that
$\termone = C[\varone_1, \hhh, \varone_n]$, $\termtwo = C[\vartwo_1, \hhh, \vartwo_n]$, and 
$x_i \mathrel{\relone} y_i$, for any $i$. 
%In particular, we see that $\termone \mathrel{\relator_{\signatureop}\relone} \termtwo$ if and only if 
%$\termtwo$ is obtained by reducing \emph{all} disjoint redexes of $\termone$ in parallel.
%Notice that $\relator_{\signatureop}$ satisfies \eqref{gamma-comp-rel} precisely because 
%it reduces \emph{all} redexes.
\item A similar inductive characterisation can be given for, e.g.,  
    the monad of $\lambda$-term on $\set^{\nats}$.
%, then $\bextt{\freemonad}$ is inductively 
% 	characterised as follows (we use the notation $\overline{n} | \termone \mathrel{\relone} 
% 	\termtwo$ to indicate that $\termone, \termtwo$ are terms with free variables among 
% 	$\overline{n}$ --- i.e. $\termone, \termtwo \in \Lambda(\overline{n})$ --- 
% 	related by $\relone(\overline{n})$)
% 	\[
% 	\infer{\val{x} \mathrel{\bextt{\freemonad}\relone} \val{y}}{x \mathrel{\relone} y}
% \quad 
% \infer{\op(t_1, \hhh, t_n) \mathrel{\bextt{\freemonad}\relone} \op(s_1, \hhh, s_n)}
% {t_1 \mathrel{\bextt{\freemonad}\relone}  s_1 & \cc & t_n\mathrel{\bextt{\freemonad}\relone}  s_n & \op \in \signatureop_{n}}
% 	\]
\end{varenumerate}
\end{example}

\subsubsection{Relational Initial Algebras} 
At this point a question natural arises. The relationship between a functor 
$\signature$ on $\topos$ and the free monad $\freemonad$ it generates is clear: the latter 
is essentially defined as the (unique) fixed point of a suitable construction on 
$\signature$ (Lambek Lemma). If, additionally, $\signature$ is finitary, 
then $\freemonad$ is the fixed point of iterated application of (a construction on) 
$\signature$ (\autoref{theorem:adamek-theorem}). Does something similar hold for $\bextt{\signature}$ and 
$\bextt{\freemonad}$? The answer is in the affirmative and a fixed point characterisation 
of $\freemonad$ as a suitable fixed point can be elegantly given relying on two 
beautiful results: the already mentioned \emph{Eilenberg-Wright Lemma}~\cite{eilenberg-wright}  and 
the \emph{Hylomorphism Theorem}~\cite{algebra-of-programming}.

\begin{theorem}[Eilenber-Wright Lemma]
\label{eilenberg-wright-lemma}
Given a functor $\signature$ on a topos $\topos$ with relational extension $\bextt{\signature}$, 
initial $\signature$-algebras in $\topos$ coincide with initial 
$\bextt{\signature}$-algebras in $\rell{\topos}$. 
\end{theorem}
Consequently, given an initial algebra 
$\xi: \signature(\lfpp{\signature}) \to \lfpp{\signature}$ (in 
$\topos$) and a relation $\relone: \signature A \torel A$, %in $\rell{\topos}$, 
there is a unique \emph{relation} $\cata{\relone}: \lfpp{\signature} \torel A$ 
such that $\xi; \cata{\relone} = \bextt{\signature}\cata{\relone}; \relone$.
% \footnote{
% Relations obtained by initiality have been extensively studied in 
% the algebra of programming~\cite{algebra-of-programming}, but seem to 
% have received little attention from the program semantics community. 
% Here, we show that relations of the form $\cata{\relone}$ naturally 
% appear in rewriting. Another interesting example come from the field 
% of program equivalence: for a candidate program equivalence $\relone$ 
% on $\lambda$-terms, $\cata{\relone}$ is nothing but the Howe extension of
% $\relone$ \cite{Howe/IC/1996,Pitts/ATBC/2011}.}

% From an allegorical perspective, \autoref{eilenberg-wright-lemma} states that, 
% for any relator $\relator$ on an allegory $\allegory$,
% initial $\relator$-algebras in $\allegory$ coincide with 
% initial $\relator$-algebras in $\mapp{\allegory}$. 

Before stating the hylomorphism theorem, we 
recall that in \lcp{} allegory $\allegory$, by Knaster-Tarski Theorem~\cite{DaveyPriestley/Book/1990},
%for all objects $A$, $B$, the hom-set $\allegory(A,B)$ is a complete lattice. 
%Consequently, 
any monotone (set-theoretic) function of the form
$F: \allegory(A,B) \to \allegory(A,B)$ has a least fixed 
point, denoted by 
$\fpp{F}$ (or $\fp{F}$), which is the least of the pre-fixed points of 
$F$.

\begin{theorem}[Hylomorphism]
\label{thm:hylo}
Given $\relone: \signature A \to A$ and $\reltwo: \signature B \to B$, we have 
$\cata{\reltwo}\dd; \cata{\relone} = \lfp{x}{\reltwo\dd; \bext{\signature}x; \relone}$.
\end{theorem}

\autoref{thm:hylo} gives a powerful proof technique that we shall 
extensively use in subsequent sections. 
We conclude this section showing how \autoref{thm:hylo} 
gives an inductive characterisation of 
$\bext{\freemonad}$. 
 To the best of the author's knowledge, 
 this result, which is folklore on $\set$, is not present 
 in the literature.
% and polynomial functors, 
% but no extension to more general categories and finitary free monad 
% is in literature. 

\begin{proposition}
\label{theorem:inductive-barr-extension}
$\bext{\freemonad}{\relone} 
%= \lfp{x}{[\unit, \sigma]\dd; (\relone + \signature x); [\unit, \sigma]}
=\lfp{x}{(\unit\dd ; \relone; \unit) \vee (\sigma\dd; \bext{\Sigma}x; \sigma)}$.
\end{proposition}

\begin{proof}
Since
$[\unit, \sigma]\dd; (\relone + \functor x); [\unit, \sigma]
= (\unit\dd ; \relone; \unit) \vee (\sigma\dd; \bext{\Sigma}x; \sigma)$, 
it is enough to show 
$\bext{\freemonad}{\relone} 
= \lfp{x}{[\unit, \sigma]\dd; (\relone + \functor x); [\unit, \sigma]}$.
Let us consider a tabulation $A \xleftarrow{f} R \xrightarrow{g} B$ of $\relone$, so that 
$\relone = f\dd;\idrel_R; g$ and
$\bext{\freemonad}\relone = (\freemonad f)\dd; \freemonad g$. 
Since $\freemonad f = \cata{[f;\unit, \sigma]}$ (and similarly for 
$g$), we obtain, 
$\bext{\freemonad}\relone =  \cata{[f;\unit, \sigma]}\dd;  \cata{[g;\unit, \sigma]}$ 
and thus $\bext{\freemonad}\relone = 
\lfp{x}{[f;\unit, \sigma]\dd; (\idrel_R + \bext{\functor}{x});  [g;\unit, \sigma]}$, 
by \autoref{thm:hylo} and the definition of Barr extension of coproduct and constant functors.
We then obtain the desired thesis thus:
% From the identity $[v,u]\dd ; x+y; [s,r] = v\dd ;x ; s \vee u\dd ; y ; r$, 
% we obtain 
\begin{align*}
\bext{\freemonad}\relone 
&=  \cata{[f;\unit, \sigma]}\dd;  \cata{[g;\unit, \sigma]}
\\
&= \lfp{x}{[f;\unit, \sigma]\dd; (\idrel_R + \bext{\functor}{x});  [g;\unit, \sigma]}
\\
&= \lfp{x}{(\unit\dd ; f\dd ; \idrel_R ; g ; \unit) \vee (\sigma\dd ; \bext{\functor}{x} ; \sigma)}
\\
&= \lfp{x}{(\unit\dd ; \relone ; \unit) \vee (\sigma\dd ; \bext{\functor}{x} ; \sigma)}
\end{align*}
%which gives the desired thesis, since $\unit\dd ; f\dd ; \idrel_R ; g ; \unit = 
%\unit\dd ; \relone ; \unit$
% \begin{align*}
% \bext{\freemonad}\relone &=  \cata{[f;\unit, \sigma]}\dd;  \cata{[g;\unit, \sigma]}
% \\
% %&= \lfp{x}{[f;\unit, \sigma]\dd; \bext{K_C + \functor}{x};  [f;\unit, \sigma]}
% &= \lfp{x}{[f;\unit, \sigma]\dd; (\idrel_C + \bext{\functor}{x});  [g;\unit, \sigma]}
% \\
% &= \lfp{x}{f;\unit\dd; \idrel_C;  g;\unit \vee \sigma\dd; \bext{\functor}{x}; \sigma}
% \end{align*}
\end{proof}

% We are now to develop our theory 
% of rewriting.

\subsection{On Fixed Points and Induction}

Before moving to the main subject of this work, namely the allegorical 
theory of symbolic manipulations, we exploit 
 a few fixed point induction principles~\cite{Backhouse-fixed-point-and-galois-connection,DaveyPriestley/Book/1990} that we shall use in 
 proofs of theorem about such a theory
notions about fixed points. 
In the following, we tacitly assume that functions are of the form 
of the form
$F: \allegory(A,B) \to \allegory(A,B)$. 
The first induction principle we state is the so-called fixed point induction principle, 
which is an immediate consequence of Knaster-Tarski Theorem.
\begin{proposition}[Fixed Point Induction]
If $F$ is monotone, then to prove $\mu F \leq \relone$, it is sufficient to prove 
$F(\relone) \leq \relone$.
\end{proposition}
Almost all the relational operators we will define in next sections 
are not just monotone, but $\omega$-continuous --- 
recall that $F$ is $\omega$-continuous if preserves 
joins of $\omega$-chains: $F(\join_n x_n) = \join_n F(x_n)$.
By Kleene Fixed Point Theorem~\cite{DaveyPriestley/Book/1990}, if $F$ is 
$\omega$-continuous, then we can give an 
iterative characterisation of $\mu F$, namely: $\mu F = \join_n F^n(\bot)$, 
where $F^n$ is the $n$-th iteration of $F$.
As for Knaster-Tarski, also Kleene Fixed Point Theorem comes with an 
associated induction principle, to which we refer to 
as \emph{$\omega$-continuous fixed point induction}.\footnote{
Even if straightforward to prove, the author was unable to find 
the induction principle of \autoref{prop:omega-continuous-fixed-point-induction} 
in the literature. For the sake of completeness, we thus give a proof of it.}
\begin{proposition}[$\omega$-Continuous Fixed Point Induction]
\label{prop:omega-continuous-fixed-point-induction}
If $F$ is $\omega$-continuous, then to prove
$\mu F \leq \relone$, it is sufficient to prove 
$x \leq \mu F \wedge \relone \implies F(x) \leq \relone$, for any $x$.
\end{proposition}
\begin{proof}
 Let us assume $\forall x.\ x \leq \mu F \wedge \relone \implies F(x) \leq \relone$
 (to which we refer to as the induction hypothesis). 
 Since $F$ is $\omega$-continuous, proving $\mu F \leq \relone$ 
 means proving $\join_n F^n(\bot) \leq \relone$. We proceed by 
 induction on $n$. 
 The base is trivial, since $F^0(\bot) = \bot \leq \relone$. 
 Assuming now $F^k(\bot) \leq \relone$, we show 
 $F(F^k(\bot)) \leq \relone$. Since $F^k(\bot) \leq \join_n F^n(\bot) = \mu F$, 
 from $F^k(\bot) \leq \relone$ we infer $F^k(\bot) \leq \mu F \wedge \relone$. 
 We can thus use the induction hypothesis to conclude 
 $F(F^k(\bot)) \leq \relone$.
\end{proof}

Finally, we mention an enhancement of \autoref{prop:omega-continuous-fixed-point-induction} 
whereby we can perform induction insider 
an $\omega$-continuous \emph{strict} function (recall that $F$ is strict if 
$F(\bot) = \bot$).
\begin{proposition}[Enhanced $\omega$-Continuous Fixed Point Induction]
\label{prop:omega-continuous-fixed-point-induction-up-to}
    Let $F,G$ be $\omega$-continuous functions. 
    Assume also that $G$ is strict.
    Then, to prove $G(\mu F) \leq \relone$, 
     it is sufficient to show 
that for any $x$ such that $x \leq \mu{F}$ and 
$G(x) \leq \relone$, we have $G(F(x)) \leq x$.
\end{proposition}
\begin{proof}
Let us assume $G(x) \leq \relone$ implies $G(F(x)) \leq x$, 
for any $x \leq \mu{F}$. We call this implication the induction hypothesis.
 Proving $G(F(x)) \leq \relone$ means proving 
 $G(\join_n F^n(\bot)) \leq \relone$, i.e. 
 $\join_n G(F^n(\bot)) \leq \relone$, since $G$ is $\omega$-continuous. 
 We proceed by induction on $n$. The base case amounts to prove 
 $G(\bot) \leq \relone$, which holds since $G$ is strict (hence $G(\bot) = \bot$). 
 For the inductive, we assume $G(F^k(\bot)) \leq \relone$ and notice that we can 
 appeal to the induction hypothesis, since $F^k(\bot) \leq \mu F$.
\end{proof}

We will use \autoref{prop:omega-continuous-fixed-point-induction-up-to} 
for functions $G$ of the form 
$G(x) \defeq \relone;x;\reltwo$, for given relations $\relone$, $\reltwo$. 
Notice that $G$ is indeed strict and $\omega$-continuous (this follows from 
distributivity of composition over join).

% Another induction principle we 
% Using distributivity of composition over join, 
% we also infer the following strengthening of fixed point induction:
% for an $\omega$-continuous function $G$, 
% to prove $G(\mu{F}) \leq \relone$ it is sufficient to show 
% that for any $x$ such that $x \leq \mu{F}$ and 
% $G(x) \leq \relone$, we have $G(F(x)) \leq x$.
% %to prove $\relone; \mu F; \reltwo \leq y$ it is sufficient to show 
% %that for any $x$ such that $x \leq \fpp{F}$ and 
% %$\relone; x; \reltwo \leq y$, we have $\relone; F(x); \reltwo \leq y$.

Finally, we observe that using fixed points, we can easily generalise relational notions 
useful in rewriting to any \lcp{} allegory. For instance, 
the reflexive and transitive closure $\relone^*$ of  
$\relone: A \to A$ is defined as $\fp{\idrel \vee \relone;x}$.
Moreover, any relator has least fixed point, and on finitary
ones we can apply $\omega$-continuous fixed point induction principle just stated. 
Notice also  that if $\relator$ is finitary and $f,g$ are maps, then 
$f\dd; \relator(-); g$ is finitary too.
This also entails that both $\bext{\signature}$ and 
$\bext{\freemonad}$ are finitary whenever $\signature$ is.

\section{An Allegorical Theory of Symbolic Manipulations}
\label{section:an-allegorical-theory}

We are now ready to put the allegorical machinery to work.
In this section, we formalise the main contribution of the paper, this 
way beginning to develop an allegorical theory of symbolic manipulations.
The main structure studied by such a theory is the one of an (abstract) \emph{expression system} 
(\expr-system, for short),\footnote{For the sake of readaability, we depart 
from standard rewriting nomenclature and follow Aczel's terminology~\cite{aczel-general-church-rosser}.}
namely a triple $(\signature, \vars, \relone)$ consisting of 
a signature functor $\signature: \topos \to \topos$, an object of variables $\vars$, 
and a (ground) reduction relation $\relone: \freemonad \vars \torel \freemonad \vars$ in 
$\rell{\topos}$. 
As in concrete systems one assumes the collection variables and the signature to be 
disjoint, we assume $\eta;\sigma\dd = \bot$. 

\begin{example}
Standard examples of \expr-systems include \trs{s} $(\signature, X, \stepto)$ on $\set$ and 
higher-order rewriting systems \cite{hamana-rewriting} in presheaves. As a paradigmatic example 
of the latter, we consider the \expr-system of $\lambda$-terms 
$(\signature, V, \beta)$, where $\signature$ is the signature of 
$\lambda$-terms and $\vec{\varone} \vdash (\lambda \varone.\termone)\termtwo 
\mathrel{\beta} \termone[\termtwo/\varone]$.
\end{example}

%$\relone$ propagates throughout arbitrary expressions. 
%This is formalised by defining suitable maps $\rel(\freemonad A, \freemonad A)$ which, 
%without much of a surprise, builds upon the relational extension of $\freemonad$. 
%As we are going to see, there are two natural such maps induced by $\freemonad$, one obtained by 
%looking at $\freemonad$ as a \emph{monad} --- which gives a form of parallel reduction --- 
%and one obtained by 
%looking at $\freemonad$ as a constructing an initial algebra --- and this gives a form of nested reduction.

\begin{remark}
To facilitate the development of the theory of \expr-systems, it is convenient
to work within an allegorical setting right from the beginning. 
Consequently, instead of starting with a topos $\topos$ and a signature functor 
$\signature$ 
(from which one constructs $\rell{\topos}$ and $\bextt{\signature}$), 
from now on we assume to have fixed (i) 
a \lcp{} allegory $\allegory$ and (ii)
a finitary signature relator $\signature: \allegory \to \allegory$. Consequently, we think about 
$\topos$ as $\mapp{\allegory}$ and as the signature functor as the relator $\signature$ on 
$\mapp{\allegory}$. This way we also obtain the syntax relator 
$\freemonad$ which indeed gives the free monad over $\signature$ when 
restricted to $\mapp{\allegory}$. In light of that, we use the notation 
$\relone: A \to B$ for arrows (hence relations) in $\allegory$, this 
way dropping the distinction between $\to$ and $\torel$ 
(but we still reserve letters $f,g, \hhh$ for maps in $\allegory$). 
Finally, recall that by 
\autoref{eilenberg-wright-lemma}, 
$[\eta, \sigma]: \vars + \signature(\freemonad \vars) \to \freemonad \vars$  
is the initial $(\vars + \signature(-))$-algebra in $\allegory$.
\end{remark}

% \begin{definition}
% A \emph{structured abstract reduction system} (\sars{)} consists of an object $A$ and a 
% \emph{reflexive}
% relation $\relone: \freemonad A \to \freemonad A$.
% \end{definition}

% \begin{example}
% \begin{enumerate}
%     \item Given a signature $\signatureop$ and a set of variables $X$, 
%         a term rewriting system (TRS) is given by a relation 
%         ${\mapsto}$ on $\signatureop$-terms. 
% \end{enumerate}
% \end{example}

%Standard examples of \rs{s} include term and higher-order rewriting systems
%a notable example of the former being the system of (the reflexive closure of) $\beta$-reduction in the 
%$\lambda$-calculus given by the relation $\beta \vee \idrel$, where
%$(\lambda \texttt{x}.\termone)\termtwo \mathrel{\beta(\overline{n})} 
%\termone[\termtwo/\texttt{x}]$ (here
%$(\lambda \texttt{x}.\termone)\termtwo$ and, consequently, 
%$\termone[\termtwo/\texttt{x}]$ belong to $\Lambda(\overline{n})$). 

Given an \expr-system $(\signature, \vars, \relone)$, the relation $\relone$ is meant to 
model \emph{ground} reduction. Actual reduction relations shall be then obtained 
by extending $\relone$ to account for the substitution structure of syntax --- this 
way allowing to consider \emph{substitution instances} of $\relone$ --- and by 
propagating reductions along the syntactic structure of terms. 
Both these (families of) operations, which are generally defined in a syntactic fashion, 
can be elegantly recovered in a purely relational fashion. In what follows, 
we introduce the powerful operations of \emph{relational substitution} 
and \emph{compatible refinement} \cite{Lassen/PhDThesis,Gordon/FOSSACS/01,Gordon-1995} 
which will be crucial to define the aforementioned 
extensions of ground reduction.

\subsubsection{Compatible Refinement} 
% The next step is to propagate along the syntactic constructs. 
% As noticed in \autoref{section:informal-intro}, there are several ways 
% to do so, depending on `how much' one wants to reduce a term. 
% Among such ways, it is natural to distinguish between \emph{parallel} 
% and \emph{sequential} reductions. In this paper, we focus on the former, 
% although we shall discuss sequential reduction in the last part of the work. 
The compatible refinement of \cite{Gordon-1995} relation $\relone$ 
relates expressions that have the same outermost syntactic construct and 
$\relone$-related arguments, and thus 
plays a crucial role 
in the definition and analysis of many forms of (parallel) reduction.

\begin{definition}
    Given an \expr-system $(\signature, \vars, \relone)$, the \emph{compatible 
    refinement} of $\relone$ is the relation 
    $\compref{\relone}: \freemonad \vars \to \freemonad \vars$ defined as 
    $[\unit, \sigma]\dd; (\idrel_{\vars} + \signature \relone); [\unit, \sigma]$. 
    Notice that $\compref{\relone} = \unit\dd; \unit \vee \sigma\dd; \signature \relone; \sigma$.
\end{definition}

Since 
$\compref{\relone} = \unit\dd; \unit \vee \sigma\dd; \signature \relone; \sigma$, 
we can clean up the definition of $\compref{-}$ by defining the operator
$\compreff{\relone} \defeq \sigma\dd;\signature\relone;\sigma$, hence recovering 
$\compref{\relone}$ by joining $\compreff{\relone}$ with
the relation $\ideta \defeq \eta\dd;\eta$ (viz. 
 $\compref{\relone} = \ideta \vee \compreff{\relone}$). 
 The latter is a so-called \emph{coreflexive}~\cite{scedrov-freyd,algebra-of-programming}, namely relation 
$\relone$ such that $\relone \leq \idrel$, 
and it can be regarded as the property of being a variable (in $\set$, for instance, 
$\ideta$ states that a term is actually a variable). 
%Consequently, we see that $\compref{\relone} = \ideta \vee \compreff{\relone}$ 
% Notice that since 
% we assume that variables and signatures are disjoint --- i.e. $\eta;\sigma\dd = \relbot$ --- 
% we have $\compreff{\relone} \wedge \ideta = \bot$, from which follows 
% $\ideta;\compreff{\relone} = \relbot$ ($= \compreff{\relone}; \ideta$), since 
% $\ideta \leq \idrel$.
 Moreover, since 
we assume that variables and signatures are disjoint --- i.e. $\eta;\sigma\dd = \relbot$ --- 
we have $\compreff{\relone} \wedge \ideta = \bot$, from which follows 
$\ideta;\compreff{\relone} = \relbot$ ($= \compreff{\relone}; \ideta$), since 
$\ideta \leq \idrel$.

\begin{proposition}
\label{proposition:lassen-algebra-1}
Both $\compreff{-}$ and $\compref{-}$ are $\omega$-continuous relators.
\end{proposition}

When instantiated on a first-order system $(\signatureop, X, \mapsto)$, 
we see that $\compref{\relone}$ is defined by the following rules: 
\[
\infer{\val{x} \mathrel{\compref{\mapsto}} \val{x}}{x \in X}
\qquad 
\infer{o(\termone_1, \hhh, \termone_n) \mathrel{\compref{\mapsto}} o(\termtwo_1, \hhh, \termtwo_n)}
{\termone_1 \mathrel{\mapsto} \termtwo_1 & \cc & \termone_n \mathrel{\mapsto} \termtwo_n 
& 
o \in \signatureop}
\]
If we consider the second rule only, we obtain $\compreff{\stepto}$.

% The compatible refinement operator is a relator, meaning that it 
% is monotone and preserves identity, composition, and converse. 
% Moreover, since $\signature$ is $\omega$-continuous, so is $\compref{-}$.
Having the notion of a compatible refinement, the natural next step 
is to define the notion of \emph{compatibility} and 
the associated \emph{context closure} operator.

\begin{definition}
\begin{varenumerate}
    \item A relation $\relone$ is \emph{compatible} if $\compref{\relone} \leq \relone$.
    \item The \emph{context closure} of $\relone$ is defined thus: 
        $\relone\ccc \defeq \lfp{x}{\relone \vee \compref{x}}$. 
\end{varenumerate}
\end{definition}
% Having the notion of a compatible refinement, the natural next step 
% is to define the \emph{context closure} of $\relone$ as 
% the least compatible relation containing $\relone$. 

% \begin{definition}
% The \emph{context closure} of $\relone$ is defined thus: 
% $\relone\ccc \defeq \lfp{x}{\relone \vee \compref{x}}$. 
% \end{definition}
Continuing the example of first-order systems $(\signatureop, X, \mapsto)$, 
we see that $\mapsto\ccc$ is inductively defined as follows: 
\[
\infer{\termone \mathrel{\mapsto\ccc} \termtwo}{\termone \mathrel{\mapsto} \termtwo}
\quad
\infer{\val{x} \mathrel{\mapsto\ccc} \val{x}}{x \in X}
\quad 
\infer{o(\termone_1, \hhh, \termone_n) \mathrel{\mapsto\ccc} o(\termtwo_1, \hhh, \termtwo_n)}
{\termone_1 \mathrel{\mapsto\ccc} \termtwo_1 & \cc & \termone_n \mathrel{\mapsto\ccc} \termtwo_n 
& 
o \in \signatureop}
\]

It is easy to see that $-\ccc$ is monotone and idempotent, and that 
if $\compref{\relone} \leq \relone$, then $\relone\ccc = \relone$. 
Since 
$\relone\ccc = \relone \vee \compref{\relone\ccc}$, we observe that $\relone\ccc$ is 
compatible and extends $\relone$, and 
thus $\relone\ccc$ is 
the least compatible relation containing $\relone$. 
Moreover, as 
$\compref{\relone} = \eta\dd;\eta \vee \sigma\dd; \signature \relone; \sigma$, 
we have that $\relone\ccc$ is reflexive on variables.
Given the the inductive nature of $\relone\ccc$, it is then natural to 
expect to have full reflexivity of $\relone\ccc$. This is indeed the case
(cf. binary induction principle \cite{Jacobs1997ATO}). 

\begin{proposition}
    The identity relation is the least compatible relation --- i.e. 
    $\idrel = \lfp{x}{\compref{x}}$ --- and thus any compatible relation 
    is reflexive.
\end{proposition}
\begin{proof}
We already know $\compref{\idrel} \leq \idrel$. We prove that it is 
the least such a relation. Given a compatible relation $\relone$, 
by initiality we have $\idrel = \cata{[\unit, \sigma]}$, so that to 
prove $\idrel \leq \relone$ we can rely on \autoref{thm:hylo} 
and prove $\unit\dd; \unit \vee \sigma\dd; \signature \relone; \sigma \leq \relone$,
but this is nothing but $\compref{\relone} \leq \relone$.
\end{proof}

\subsubsection{Relational Substitution} 
Having compatible refinement and context closure --- hence ways to propagate 
reductions along syntactic constructs --- we now need an operator extending substitution 
to (reduction) relations. For that, we rely on an extension of 
Lassen's \emph{relation substitution}~\cite{Lassen/PhDThesis}.
\begin{definition}
Given a substitution algebra $A \times \nameabs{\vars}{A} \xrightarrow{\substmap} A \xleftarrow{\new} \vars$ and relations $\relone, \reltwo : A \to A$, we define the substitution of 
$\reltwo$ into $\relone$ as the relation $\relone[\reltwo]: A \to A$ 
defined by $\dual{\substmap}; (\relone \times \nameabs{\vars}{\reltwo}); \substmap$, 
where we recall that both $\times$ and $\nameabs{-}{\vars}$ are the (bi)relator 
associated with the corresponding functor. 
\end{definition}
When instantiated on, e.g., first- or second-order syntax, the relation $\relone[\reltwo]$ 
relates all terms $\termone[\vect{\termtwo}/\vect{x}]$, $\termone'[\vect{\termtwo}'/\vect{x}]$ such that 
$\termone, \termone'$ are related by $\relone$ and $\termtwo_i$, $\termtwo_i'$ are related 
by $\reltwo$, for any $i$. 

\begin{proposition}[\cite{Lassen/PhDThesis,Levy/ENTCS/2006}]
\label{proposition:lassen-algebra-2}
\begin{enumerate}
    \item $-[=]$ is a (bi-)relator.
    \item  $-[=]$ is $\omega$-continuous in the first argument.
    \item $-[=]$ is associative: $\relone[\reltwo][\relthree] = \relone[\reltwo[\relthree]]$.
\end{enumerate}    
\end{proposition}
%$(\relone;\relone')[\reltwo;\reltwo'] \leq \relone[\reltwo]; \relone'[\reltwo']$. 
When it comes to calculate with $-[=]$, it is useful to notice that 
it has a right adjoint~\cite{Lassen/PhDThesis}:
$\relone[\reltwo] \leq x$ iff $\relone \leq \reltwo {\gg} x$. Explicitly, 
$\relone {\gg} \reltwo = \join\{x\mid x[\relone] \leq \reltwo\}$. 
Notice that $\gg$ is lax functorial 
(i.e. $(\relone \gg \reltwo);(\relone' \gg \reltwo') \leq 
(\relone;\relone') \gg (\reltwo; \reltwo')$), antitone in the first argument, and 
monotone in the second one. 

% The operation $-[=]$ is monotone in both arguments and lax functorial, in the sense that 
% $(\relone;\relone')[\reltwo;\reltwo'] \leq \relone[\reltwo]; \relone'[\reltwo']$. 
% Moreover, it has a right adjoint $\gg$ \cite{Lassen/PhDThesis} which is useful in calculation:
% $\relone[\reltwo] \leq x$ iff $\relone \leq \reltwo {\gg} x$. Explicitly, 
% $\relone {\gg} \reltwo = \join\{x\mid x[\relone] \leq \reltwo\}$. 
% Notice that $\gg$ is lax functorial 
% (i.e. $(\relone \gg \reltwo);(\relone' \gg \reltwo') \leq 
% (\relone;\relone') \gg (\reltwo; \reltwo')$), antitone in the first argument, and 
% monotone in the second one. 

\begin{definition}
We say that a relation $\relone$ is \emph{closed under substitution} if 
$\relone[\idrel] \leq \relone$, and that it is \emph{substitutive} 
if $\relone[\relone] \leq \relone$. 
\end{definition}
In particular,
we think about $\relone[\idrel]$ as the reduction obtained by taking substitution instances 
of $\relone$. For instance, in a \trs{} $(\signatureop, X, \mapsto)$, the relation 
$\rightarrowtriangle$ seen in \autoref{section:informal-intro} 
is precisely ${\mapsto}{[\idrel]}$.
Notice that since $-[=]$ is $\omega$-continuous in the first argument, 
the (unary) operator $-[\idrel]$ is
$\omega$-continuous itself. 

At this point we have introduced some new relational operators that compactly 
describe rewriting notions (as we shall better see in forthcoming sections). 
To make them really useful, we also have to provide algebraic laws for 
calculating with them.\footnote{Laws in 
\autoref{lemma:algebra-of-substitution-and-compatible-refinement} have been 
first proved by Lassen~\cite{Lassen/PhDThesis}, and then extended by 
Levy~\cite{Levy/ENTCS/2006}, in the context of specific $\lambda$-calculi. 
Our rule $\compref{\relone}[\reltwo] \leq \compref{\relone[\reltwo]} \vee \reltwo$ 
differs from Lassen's one --- namely 
$\compref{\relone}[\reltwo] \leq \compref{\relone[\reltwo] \vee \reltwo}$ --- 
which seems wrong. For suppose $\varone \mathrel{\compref{\relone}} \varone$ and 
$\lambda \vartwo.\vartwo \mathrel{\reltwo} \vartwo\vartwo$. Then, taking the two
substitutions  $[\lambda \vartwo.\vartwo/\varone]$ and 
$\vartwo\vartwo/\varone]$, we obtain $\lambda \vartwo.\vartwo \mathrel{\compref{\relone}[\reltwo]} 
\vartwo\vartwo$. The latter terms, however, cannot be related by 
the compatible refinement of any relation, as they are not variables and have different 
outermost syntactic constructs (viz. abstraction, for the first, and application, for the second).
Notice that this inequality plays a crucial role in Lassen's proof of substitutivity of 
the Howe extension of a relation. Contrary to usual presentations of the same results, Lassen 
does not require the relation to be transitive, a condition which is instead needed 
if one reviews the proof using the correct inequality as stated in 
\autoref{lemma:algebra-of-substitution-and-compatible-refinement}.
}
\begin{proposition}
\label{lemma:algebra-of-substitution-and-compatible-refinement}
We have the following laws~\cite{Lassen/PhDThesis,Levy/ENTCS/2006}:
% \begin{align*}
%     \compreff{\relone}[\reltwo] &\leq \compreff{\relone[\reltwo]}
%     \\
%     \ideta[\reltwo] &\leq \reltwo
%     \\
%     \compref{\relone}[\reltwo] &\leq \compref{\relone[\reltwo]} \vee \reltwo.
% \end{align*}
\begin{enumerate}
    \item $\compreff{\relone}[\reltwo] \leq \compreff{\relone[\reltwo]}$
    \item $\ideta[\reltwo] \leq \reltwo$
    \item $\compref{\relone}[\reltwo] \leq \compref{\relone[\reltwo]} \vee \reltwo$.
\end{enumerate}
% \begin{align*}
%     \compreff{\relone}[\reltwo] &\leq \compreff{\relone[\reltwo]},
%     &
%     \ideta[\reltwo] &\leq \reltwo,
%     &
%     \compref{\relone}[\reltwo] &\leq \compref{\relone[\reltwo]} \vee \reltwo.
% \end{align*}
\end{proposition}

\begin{remark}
\label{remark:substitution}
The algebraic laws in
\autoref{lemma:algebra-of-substitution-and-compatible-refinement} and 
\autoref{proposition:lassen-algebra-2} constitute the \emph{operational} definition of 
substitution at a relational level. 
Even if built upon a specific definition of substitution structure (viz. substitution algebra), 
what truly matters when it comes to (operational) reasoning and symbolic manipulation 
is to have an operator $-[=]$ obeying the aforementioned algebraic laws. 
For instance, $\omega$-continuity of $-[=]$ in the first argument 
says that substitution is defined by structural recursion, whereas the rule
$\compreff{\relone}[\reltwo] \leq \compreff{\relone[\reltwo]}$
states that substitution behaves as a syntax-preserving morphism. 
The actual structure used to model substitution is irrelevant: any 
`good' structure will induce a relational substitution operator satisfying the aforementioned laws.
For instance, readers can convince themselves that replacing substitution algebras with, e.g., 
$\signature$-monoid, leads to essentially (i.e. operationally) the same operator $-[=]$. 
Moreover, as soon as a `substitution' operator satisfying 
\autoref{lemma:algebra-of-substitution-and-compatible-refinement} and 
\autoref{proposition:lassen-algebra-2} is available, one can study rewriting properties of 
substitution, regardless of the actual definition of the latter. 
This observation can be pushed even further by completely forgetting 
the actual syntax of a system and working axiomatically within an augmented calculus of relations, 
viz. a traditional calculus of relations enriched with operators and laws 
as described in this section: whenever there is a model of syntax admitting an instance of 
such a relational calculus, rewriting is obtained for free. 
We will comment further on the axiomatic approach at the end of this section.
\end{remark}

Finally, we can merge the definition of substitution and context closure, 
this way obtaining the \emph{substitutive context closure} of $\relone$. 

\begin{definition}
The \emph{substitutive context closure} of a relation $\relone$ 
is the relation $\relone\scc \defeq \relone[\idrel]\ccc = \lfp{x}{\relone[\idrel] \vee \compref{x}}$.
\end{definition}
When instantiated on a first-order system $(\signatureop, X, \relone)$, 
we obtain the following inductive characterisation of 
$\relone\scc$:
\[
\infer{\termone[\vect{\valone}/\vect{x}] \stepto\scc \termtwo[\vect{\valone}/\vect{x}]}{t \stepto s}
\hspace{0.2cm}
\infer{\val{x} \stepto\scc \val{x}}{x \in X}
\hspace{0.2cm}
\infer{\op(\termone_1, \hhh, \termone_n) \stepto\scc \op(\termtwo_1, \hhh, \termtwo_n)}
{\forall i \leq n.\ \termone_i \stepto\scc \termtwo_i 
& 
o \in \signatureop}
\]
% Before 
% moving into that, we notice that relying, among others, 
% the equational laws of a substitution algebra, 
% we can prove algebraic laws exploiting the interaction between 
% compatible refinement and relational substitution \cite{Lassen/PhDThesis}. For instance, we have
% $\compref{\relone}[\reltwo] \leq \compref{\relone[\reltwo] \vee \reltwo}$.
We conclude this section with a methodological (and perhaps conceptual) 
consideration.

\subsubsection{The Augmented Calculus of Relations} 
Let us have a look at the relational apparatus developed so far 
from an operational perspective. 
Accordingly, can think about our framework as an \emph{augmented calculus of relations} where, 
in addition to the classic operations on relations (such as composition, meet, join, etc), 
we have the operations $\compreff{-}$, $\ideta$ (hence $\compref{-}$), 
and $-[=]$ together with suitable
equational laws 
(viz. those in \autoref{proposition:lassen-algebra-1}, \autoref{proposition:lassen-algebra-2}, and \autoref{lemma:algebra-of-substitution-and-compatible-refinement}) 
and proof principles (viz. fixed point induction).
Assuming to have fixed point operators, one can then define inside such an augmented calculus 
the operators $-\scc$ and $-\ccc$ (otherwise, one can add them as primitive operators, together 
with suitable equational laws and proof principles). 

As we are going to see, this augmented calculus of relation is expressive enough to 
define interesting notions of reductions and to prove nontrivial properties 
about them. Moving from this observation, we could make the whole relational framework 
developed completely axiomatic, abstracting over syntax and simply working with 
the aforementioned augmented relational calculus. The fact that we have extracted such a 
calculus out of a syntactic system $(\signature, \vars)$ can be then read as a way to build a model 
of the augmented calculus.

The remaining part of the paper is devoted to the definition and analyisis 
of specific notions of reduction: we shall do so first relying on 
the algebra of syntax (this way showing how they indeed correspond to those given 
in the rewriting literature), and then showing how syntax-dependency can be avoided 
by giving equivalent definitions in the augmented calculus of relations. Moreover, 
to prove our confluence theorems --- the main result proved --- 
we will use the laws of the augmented calculus only, hence witnessing 
the effectiveness of the axiomatic approach (see also \autoref{section:conclusion} 
for a more general discussion on the impact of such an approach in operational reasoning).

\section{Parallel Reduction} 
\label{section:parallel-reduction}
Having defined \expr-systems and a (augmented) relational calculus to 
reason about them, it is time to introduce extensions of ground reductions, 
such extensions giving actual (operational) reduction.
We shall focus on two such extensions --- namely \emph{parallel} and \emph{full} 
reduction --- confining ourselves to just few observations on \emph{sequential} reduction 
in the last part of the work. 
Besides introducing such notions and proving basic facts about them, 
the main results we proved are confluence theorems for a generalisation of
the so-called \emph{orthogonal} systems~\cite{Huet80}.
We begin with \emph{parallel reduction}. 

\begin{remark}
\label{remark:variables-not-left-rule}
In the remaining part of the paper, we assume that in an \expr-system $(\signature, \vars, \relone)$, 
we have $\ideta;\relone = \bot$. This corresponds to the usual assumption that the 
left-hand side of reduction rule cannot be a variable~\cite{terese}.
\end{remark}

Let us ignore substitution for the moment 
and recall that, in concrete reduction systems (such as term and higher-order systems),
parallel reduction applies ground reduction on arbitrarily chosen 
set of disjoint redexes in parallel. Abstractly, we obtain parallel reduction 
relying on the monad structure of syntax (as we will see, parallel reduction corresponds 
to looking at syntax as a \emph{monad}, whereas full reduction corresponds to 
looking at syntax as a \emph{free algebra}).

\begin{definition}
\label{definition:parallel-reduction}
Given an \expr-system $(\signature, \vars, \relone)$, we define \emph{parallel reduction} 
(without substitution) $\parr{\relone}: \freemonad \vars \to \freemonad \vars$ 
 as 
$\multiplication\dd; \freemonad\relone; \multiplication$.\footnote{Recall that 
working naively within allegories, $\freemonad$ is a relator (which is 
necessarily the Barr extension of its restriction to $\mapp{\allegory}$).}
\emph{Substitutive parallel extension} is defined as 
$\relone\spp \defeq \relone[\idrel]\pp$.
%Diagramatically:
%    \[
%    \xymatrix{
%    \freemonad \freemonad A 
%    \ar[r]^{\mu}
%    \ar[d]_{\bext{\freemonad} \relone}
%    & 
%    \freemonad A 
%    \ar@{.>}[d]^{\widehat{\relone}}
%    \\
%    \freemonad \freemonad A \ar[r]_{\mu} 
%    & \freemonad A  
%    }
%    \]
\end{definition}
The relation $\relone\spp$ 
generalises the context-based definition of parallel reduction in 
syntax-based rewriting systems. 
\begin{example}
When instantiated on a \trs{} 
$(\signatureop, X, \mapsto)$,
we see that $\termone \Rightarrow \termtwo$, where 
${\Rightarrow} = {\mapsto\spp}$, if and 
only if there exist a context $C$, terms $\termone_1, \hhh, \termone_n$, 
$\termtwo_1, \hhh, \termtwo_n$, and substitutions $\gamma_i$ % = [\vect{v}^i/\vect{x}^i]$ 
such that: 
(i) $\termone_i \mapsto \termtwo_i$, for each $i$;
(ii) $t = C[\termone_1\gamma_1, \hhh, \termone_n\gamma_n]$; 
(iii) $s = C[\termtwo_1\gamma_1, \hhh, \termtwo_n\gamma_n]$;
(iv) all $\termone_i$ and $\termtwo_j$ are pairwise disjoint in $t$ and $s$, respectively. 
% Writing can summarise clauses (i), (ii), (iii) as the rule:
% \[
% \infer{C[\termone_1\gamma_1, \hhh, \termone_n\gamma_n] 
% \Rightarrow C[\termtwo_1\gamma_1, \hhh, \termtwo_n\gamma_n]}
% {\termone_1 \mapsto \termtwo_1 & \cc & \termone_n \mapsto \termtwo_n}
% \]
%Similar definitions are recovered by instantiating $\widehat{-}$ on other examples. 
\end{example}
Since $\freemonad$ is finitary, we immediately notice that 
$\parr{-}$ inherits many of the structural properties of $\freemonad$: 
it is $\omega$-continuous (and thus monotone)
and commutes with converse ($\relone\ppdd = \relone\ddpp$). 
Moreover, it indeed extends $\relone$, i.e. 
$\relone \leqq \relone\pp$.

The definition of $\relone\pp$ (and thus of $\relone\spp$) relies on the monad 
multiplication of $\freemonad$, and thus goes beyond the augmented calculus of relations 
outlined in the previous section. 
As it happens in concrete syntax-based systems, $\relone\pp$ can be characterised 
inductively and such a characterisation precisely shows us how to 
define parallel extension in the relational calculus.

% of parallel reduction; indeed, the 
% reader may have noticed that the definition of substitutive context closure 
% resemble the \emph{inductive} definition of parallel reduction in 
% concrete syntax-based systems. 

%\begin{definition}
%The Barr extension of a relation $\relone: A \torel B$ is defined by 
%$\bext{\freemonad}\relone \defeq (\freemonad f)\dd; \freemonad g$, where 
%$(f,g)$ tabulates $\relone$.
%\end{definition}

% \begin{proposition}
%     $\bext{\freemonad}$ is a relational extension. Moreover, the definition of 
%     $\bext{\freemonad}\relone$ is independent of the choice of the tabulation.
% \end{proposition}

% From now on, we assume that $\bext{\signature}$ is $\omega$-continuous, meaning that 
% for any $\omega$-chain $(\relone_n)_{n \in \omega}$, we have 
% $\bext{\signature}(\join_n \relone_n) = \join_n \bext{\signature} \relone_n$. 

% Notice that if $\bext{\signature}$ is $\omega$-continuous, then so is 
% $\bext{\freemonad}$. 

\begin{proposition}
\label{proposition:parallel-reduction-inductively}
$\relone\pp = \relone\ccc$ ($= \lfp{x}{\relone \vee \compref{x}}$), and thus 
$\relone\spp = \relone \scc$.
\end{proposition}

\begin{proof}
        First, we notice that $\relone\pp = 
    \relone \vee \compref{\relone\pp}$, 
    and thus $\relone\ccc \leq \relone\pp$.
    To see that, consider the following diagram and calculation 
    (where we use the inductive characterisation of 
    $\freemonad$ provided by 
    \autoref{theorem:inductive-barr-extension}):
    \[
    \xymatrix{
    & \freemonad A + \signature \freemonad A \ar[rd]^{[\idrel, \sigma]} &
    \\
    \freemonad A + \signature\freemonad \freemonad A
    \ar[r]^{[\unit, \sigma]}
    \ar[ru]^{\idrel + \signature \mu}
    \ar[d]_{\relone + \signature \freemonad \relone}
    & 
    \freemonad \freemonad A 
    \ar[r]^{\mu}
    \ar[d]_{\bext{\freemonad} \relone}
    & 
    \freemonad A 
    \ar[d]^{\widehat{\relone}}
    \\
    \freemonad A + \signature\freemonad \freemonad A
    \ar[r]_{[\unit, \sigma]}
    \ar[rd]_{\idrel + \signature \mu}
    & 
    \freemonad \freemonad A \ar[r]_{\mu} 
    & \freemonad A  
    \\
    & \freemonad A + \signature \freemonad A \ar[ru]_{[\idrel, \sigma]} &
    }
    \]
%    We have:
    \begin{align*}
        \relone\pp &=
        \multiplication\dd; \freemonad\relone; \multiplication 
        \\
        &= ((\idrel + \signature \multiplication); [\idrel, \sigma])\dd; (\relone + \signature \freemonad \relone); 
        (\idrel + \signature \multiplication); [\idrel, \sigma]
        \\
        &= [\idrel, \signature \multiplication; \sigma]\dd; (\relone + \signature \freemonad \relone); 
        [\idrel, \signature \multiplication; \sigma]
        \\
        &=\relone \vee  \sigma\dd; \signature \multiplication\dd; \signature \freemonad \relone; \signature \multiplication; \sigma
        \\
        &= \relone \vee  \sigma\dd; \signature (\multiplication\dd; \freemonad \relone; \multiplication); \sigma
        \\
        &= \relone \vee  \sigma\dd; \signature \relone\pp; \sigma
        \\
        &= \relone \vee  \compref{\relone\pp}
    \end{align*}
 Let us now prove $\relone\pp \leq \relone\ccc$. 
 Since by \autoref{theorem:inductive-barr-extension} we have 
 $\freemonad\relone = \mu x. (\unit\dd ; \relone; \unit) \vee (\sigma\dd; \signature x; \sigma)$, 
we can proceed by $\omega$-continuous 
fixed point induction on. Assuming $\multiplication\dd;x;\multiplication \leq \relone\ccc$, we have:
\begin{align*}
    & \multiplication\dd;((\unit\dd ; \relone; \unit) \vee (\sigma\dd; \signature x; \sigma));\multiplication 
    \\
    &=  (\multiplication\dd;\unit\dd ; \relone; \unit;\multiplication) \vee (\multiplication\dd; \sigma\dd; \signature x; \sigma;\multiplication)
    \\
    &= \relone \vee (\multiplication\dd; \sigma\dd; \signature x; \sigma;\multiplication)
    \\
    &= \relone \vee (\sigma\dd; \signature \multiplication\dd; \signature x; \signature \multiplication; \sigma;)
    \\
    &= \relone \vee (\sigma\dd; \signature(\multiplication\dd; x; \multiplication); \sigma;)
    \\
    &= \relone \vee (\sigma\dd; \signature\relone\ccc; \sigma;)
    \\
    &= \relone \vee \compref{\relone\ccc}
    \\
    &= \relone\ccc.
\end{align*}
\end{proof}

By \autoref{proposition:parallel-reduction-inductively}, 
$\relone\pp$ and $\relone\spp$  are compatible, 
and since compatible relations are reflexive, 
they are reflexive too. Therefore, we see that indeed
$\relone\pp$ reduces disjoint redexes \emph{at will}.  
Before moving to confluence properties of $\relone\spp$, we 
observe that $\relone\spp$ is closed under substitution. 

\begin{proposition}
    $\relone\spp[\idrel] \leq \relone\spp$.
\end{proposition}
\begin{proof}
 It is sufficient to prove $\relone\spp \leq \idrel \gg \relone\spp$, 
 and we do so by fixed point induction. 
 The interesting case is showing $\compref{\idrel \gg \relone\spp} \leq \idrel \gg \relone\spp$. 
 For that, we notice that we have the general law 
 $\compref{x \gg y} \leq x \gg (\compref{y} \vee x)$,\footnote{
It is sufficient to prove $\compref{x \gg y}[x] \leq \compref{y} \vee x$. 
The latter follows from \autoref{proposition:lassen-algebra-2} thus: 
$\compref{x \gg y}[x] \leq \compref{(x \gg y)[x]} \vee x 
\leq \compref{y} \vee x.
$
} which gives 
 $\compref{\idrel \gg \relone\spp} \leq \idrel \gg (\compref{\relone\spp} \vee \idrel) 
 \leq \idrel \gg \relone\spp$, since $\gg$ is monotone in the second argument and $\relone\spp$ is 
 reflexive and compatible.
\end{proof}

\subsection{The Relational Parallel Moves Technique}
Having defined parallel reduction, we now give evidences of its 
effectiveness.  
We do so by looking at one of the most important property of 
symbolic systems: \emph{confluence}~\cite{church_rosser_1936}.
In full generality: (i) a relation $\relone: A \to A$ has the diamond 
property if $\relone\dd; \relone \leqq \relone; \relone\dd$,
(ii) is \emph{confluent} if $\relone^*$ has the \emph{diamond property}, 
(iii) is \emph{weakly confluent} if 
$\relone\dd; \relone \leqq \relone^*; \relone\stardd$ (notice that 
$\relone\ddstar = \relone\stardd$).

When it comes to prove confluence of parallel reduction $\relone\spp$ (and reductions alike), 
it is desirable to have \emph{local} proof 
techniques at disposal. Confluence, in fact, is a non-local property in two ways: 
(i) it refers to reduction sequences (viz. $\relone\sppstar$) rather than 
single reduction steps; (ii) it performs reduction inside complex expressions. 
The proof techniques we shall prove in this paper are mostly given in 
the form of diamond-like properties (hence obtaining locality of the first type above)
with hypotheses formulated on ground reduction $\relone$ or substitution 
instances thereof, viz. $\relone[\idrel]$,
hence obtaining semi-locality of the second type above. 
% As a side-effect of our 
% abstract theory, we see that this patter of semi-local proof 
% techniques appears everywhere in syntax-specific rewriting (e.g. critical pairs and 
% orthogonality).

The first main result we prove here is an abstract and relational version 
of the so-called parallel moves lemma~\cite{klop}, a well-known 
technique to prove confluence of \emph{orthogonal} systems~\cite{Huet80}. 
When specialised to first-order systems, orthogonality of a \trs{} means that 
the system has no critical pair~\cite{Huet80} and all ground reductions 
are left-linear (i.e. variables in a ground redexes occur at most once). 
These conditions are heavily syntax-dependent and they hardly generalise to 
arbitrarily syntax-based systems. 

We overcome this issue by isolating an \emph{operational} notion 
of orthogonality which, when instantiated to \trs{s}, is indeed implied by the 
aforementioned syntactic definition of orthogonality. Such an operational 
notion of orthogonality directly translates into the relational framework the 
informal intuition behind orthogonal systems, namely that redexes remains so 
whenever their subterms are reduced (stated otherwise: reduction is local, 
in the sense that reducing a redex does not affect other redexes). 

\begin{definition}
\label{def:orthogonality}
An \expr-system $(\signature, \vars, \relone)$ is \emph{orthogonal} 
if:
\begin{align*}
    \relone[\idrel]\dd; \relone[\idrel] &\leq \idrel
    & 
    \relone[\idrel]\dd; \compreff{\relone\spp} &\leq \relone\dd[\relone\spp].
\end{align*}
% $\relone[\idrel]\dd; \relone[\idrel] \leq \idrel$ and 
% $\relone[\idrel]\dd; \compref{\relone\spp} \leq \relone\dd[\relone\spp]$.
\end{definition}
The first condition in \autoref{def:orthogonality} is clear: 
ground reduction rules are essentially unique.\footnote{Notice that 
such a rule actually gives \emph{weak} orthogonality \cite{terese}.} 
The second condition 
states that a redex (instance) remains such when we reduce its subterms. 
In the case of \trs{s}, we can spell out such a condition as follows: 
for any ground reduction $\ell \mapsto r$, if we reduce a proper subterm 
$\termtwo$ of an instance $\ell[\vect{\valone}/\vect{x}]$ of $\ell$, 
say $\termtwo \Rightarrow \termone$, then 
$\termone = \ell[\vect{\valtwo}/\vect{x}]$ and $\valone_i \Rightarrow \valtwo_i$. 
Notice that using the compatible refinement operator, we can indeed express that 
the reduction $\termtwo \Rightarrow \termone$ happens on a proper subterm of 
$\ell[\vect{\valone}/\vect{x}]$.

Our goal now is to prove that orthogonal systems are confluent by showing that 
parallel reduction has the diamond property: this technique goes under the name of 
\emph{parallel moves} \cite{klop,terese}. Before going any further, we remark 
that the latter statement is not true in general; it fails, for instance, for
syntax involving variable binding. In those cases, parallel reduction is 
only weakly confluent. Working relationally, we see clearly the point 
where the parallel moves technique breaks and thus we can isolate 
the operational condition needed to make it work, such a condition holding 
for, e.g., first-order syntax. Moreover, by weakening the orthogonality condition, 
we obtain (perhaps) novel techniques to prove weak confluence of non-orthogonal 
systems.

As a first step towards confluence, we notice that the parallel operator 
enjoys a Kleisli-like lifting property.

\begin{lemma}
\label{theorem:parallel-moves}
If $\relone\dd; \relone\pp \leq \relone\pp; \relone\ppdd$, then 
$\relone\ppdd; \relone\pp\leq \relone\pp; \relone\ppdd$
\end{lemma}

\begin{proof}
    We assume $\relone\dd; \relone\pp \leq \relone\pp; \relone\ppdd$ and notice that, dualising, 
    we also obtain $\relone\ppdd; \relone \leq \relone\pp; \relone\ppdd$. 
    Since $\relone\ppdd = \relone\ddpp =\mu x. \relone\dd \vee \compref{x}$, we 
    can prove the thesis by $\omega$-continuous induction 
    (notice that indeed $\phi(x) \defeq \relone\dd \vee \compref{x}$ is $\omega$-continuous --- 
    this also follows from \autoref{proposition:parallel-reduction-inductively} --- 
    and that so is $\phi(x); \relone\pp$).
    Let us assume $x \leq \relone\ppdd$ and $x;\relone\pp \leq \relone\pp; \relone\ppdd$. 
    We show $(\relone\dd \vee \compref{x});\relone\pp \leq \relone\pp; \relone\ppdd$. 
    The latter amounts to prove 
    $\relone\dd;\relone\pp \vee \compref{x};\relone\pp \leq \relone\pp; \relone\ppdd$, which 
    in turn follows from 
    $\relone\dd;\relone\pp \leq \relone\pp; \relone\ppdd$
    and 
    $\compref{x};\relone\pp \leq \relone\pp; \relone\ppdd$. 
    The former follows by hypothesis. For the latter, since 
    $\relone\pp = \relone \vee \compref{\relone\pp}$, it is sufficient to prove 
     $\compref{x};\relone \leq \relone\pp; \relone\ppdd$ and 
      $\compref{x};\compref{\relone\pp} \leq \relone\pp; \relone\ppdd$. 
      The first follows from (the dualised version of the) hypothesis, since 
      $x \leq \relone\ppdd$ gives 
      $$
      \compref{x};\relone
      \leq \compref{\relone\ppdd};\relone
      \leq \compref{\relone\pp}\dd; \relone
      \leq \relone\ppdd; \relone.
      $$
      For the second inequality, we first use the induction hypothesis 
      as follows
      $$
      \compref{x};\compref{\relone\pp} 
      = \compref{x;\relone\pp}
      \leq \compref{\relone\pp; \relone\ppdd}
      $$
      and then notice that $\relone\pp; \relone\ppdd$ is indeed compatible 
      (from which the thesis follows).
      For
      \begin{align*}
          \compref{\relone\pp; \relone\ppdd} 
          &=  \compref{\relone\pp}; \compref{\relone\ppdd}
          \\
          &=  \compref{\relone\pp}; \compref{\relone\ddpp}
          \\
          &\leq (\relone \vee \compref{\relone\pp}); (\relone\dd \vee \compref{\relone\ddpp})
          \\
          &= \relone\pp; \relone\ddpp
          \\
          &= \relone\pp; \relone\ppdd.
      \end{align*}
\end{proof}

Since $\relone\spp = \relone[\idrel]\pp$, 
we can prove that the latter has the diamond property 
relying on \autoref{theorem:parallel-moves} by showing 
$\relone[\idrel]\dd; \relone[\idrel]\pp \leq \relone[\idrel]\pp; \relone[\idrel]\ppdd$.
To do so, we need two auxiliary results. The first stating 
that parallel reduction extends to substitutions.\footnote{
On a \trs{} we have: $\valone_i \Rightarrow \valtwo_i$ implies 
$\termone[\vect{\valone}/\vect{x}] \Rightarrow \termone[\vect{\valtwo}/\vect{x}]$.}

\begin{lemma}
\label{lemma:parallel-reduction-closed-under-substitution}
$\idrel[\relone\spp] \leq \relone\spp$.
\end{lemma}

\begin{proof}
 It is sufficient to prove $\idrel \leq \relone\spp \gg \relone\spp$. 
 Since $\idrel = \lfp{x}{\compref{x}}$, we proceed by fixed point induction 
 showing that $\relone\spp \gg \relone\spp$ is compatible. 
 Since $\relone\spp$ is compatible (and $\gg$ is monotone in the second argument), we have:
$
     \compref{\relone\spp \gg \relone\spp} 
        \leq \relone\spp \gg (\compref{\relone\spp} \vee \relone\spp)
        \leq \relone\spp \gg \relone\spp.
$
\end{proof}

The second property needed states that a reduction cannot produce 
\emph{nested} redexes. 
\begin{definition}
\label{def:nesting-property}
An \expr-system $(\signature, \vars, \relone)$ has the \emph{nesting property} if
$\relone\dd[\relone\spp] \leq \relone\spp; \relone\dd[\idrel].$
\end{definition} 
 
The nesting property holds for \trs{s}, as well as for 
higher-order systems without binders~\cite{yamada-orthogonal}. 
However, as already mentioned, it fails on syntax with variable binding. 
For instance, in the case of the $\lambda$-calculus, it states 
that whenever we have 
$$
\termone[\vect{\valone}/\vect{x}][\termtwo[\vect{\valone}/\vect{x}]/x]
\leftarrow_{\beta} 
(\lambda x.\termone[\vect{\valone}/\vect{x}])\termtwo[\vect{\valone}/\vect{x}] 
\Rightarrow_{\beta} 
(\lambda x.\termone[\vect{\valtwo}/\vect{x}])\termtwo[\vect{\valtwo}/\vect{x}],
$$ 
then $\termone[\vect{\valone}/\vect{x}][\termtwo[\vect{\valone}/\vect{x}]/x] \Rightarrow 
\termone[\vect{\valtwo}/\vect{x}][\termtwo[\vect{\valtwo}/\vect{x}]/x]
$. The latter reduction does not hold, as there could \emph{nested} redexes $\valone_i$s. 
As we shall see in the next section, such a property is implied by 
substitutivity, a property not enjoyed by $\relone\spp$.

\begin{theorem}[Parallel Moves]
\label{theorem:relational-parallel-moves}
Let $(\signature, \vars, \relone)$ be an orthogonal system satisfying the nesting property. 
Then $\relone\spp$ has the diamond property.
\end{theorem}

% \begin{remark}
% \label{remark:no-left-hand-side-variable-rule}
% Stated as it is, \autoref{theorem:relational-parallel-moves} does not hold. 
% To hold, one actually needs a stronger orthogonality condition that replaces 
% $\compreff{-}$ with $\compref{-}$ in \autoref{def:orthogonality}. 
% To prove the theorem with the current definition of orthogonality, 
% we need to exploit a further condition on \expr-system which is often (more or less tacitly) 
% assumed in the definition of \trs{s}, namely that the left-hand side of ground reduction rules 
% cannot be a variable. In the relational formalism, that becomes 
% $\relone\dd; \ideta \leq \relbot$ (from which follows $\relone[\idrel]\dd; \ideta \leq \relbot$ too).
% \end{remark}

\begin{proof}
By \autoref{theorem:parallel-moves} it is sufficient to prove
$\relone[\idrel]\dd; \relone[\idrel]\pp \leq \relone[\idrel]\pp; \relone[\idrel]\ppdd$. 
Let  $\reltwo$ be $\relone[\idrel]\pp; \relone[\idrel]\ppdd$. 
Since $\relone[\idrel]\pp = \relone[\idrel] \vee \compref{\relone[\idrel]\pp}$, 
it is sufficient to show 
$\relone[\idrel]\dd; \relone[\idrel] \leq \reltwo$ and 
$\relone[\idrel]\dd; \compref{\relone[\idrel]\pp} \leq \reltwo$, i.e. 
$\relone[\idrel]\dd; \compref{\relone\spp} \leq \reltwo$.
The former directly follows from orthogonality, since $\idrel \leq \reltwo$. 
% PROOF WITHOUT CONDITON ON VARIABLES
% For the latter, by orthogonality and the nesting property we have 
% $$\relone[\idrel]\dd; \compref{\relone\spp} \leq \relone\dd[\relone\spp] 
% \leq \relone\spp; \relone\dd[\idrel].$$
% We conclude the thesis since $\relone\spp$ is closed under substitution 
% (\autoref{lemma:parallel-reduction-closed-under-substitution})
% (and thus $\relone\dd[\idrel] \leq \relone\sppdd[\idrel]\leq \relone\sppdd$).
For the latter, it is sufficient to show 
$\relone[\idrel]\dd; \ideta \leq \reltwo$ and $\relone[\idrel]\dd; \compreff{\relone\spp} \leq \reltwo$. 
The first trivially follows by \autoref{remark:variables-not-left-rule}, 
whereas for the second 
by orthogonality and the nesting property, we have 
$$\relone[\idrel]\dd; \compreff{\relone\spp} \leq \relone\dd[\relone\spp] 
\leq \relone\spp; \relone\dd[\idrel].$$
We conclude the thesis since $\relone\spp$ is closed under substitution 
(\autoref{lemma:parallel-reduction-closed-under-substitution})
(and thus $\relone\dd[\idrel] \leq \relone\sppdd[\idrel]\leq \relone\sppdd$).
 \end{proof}
 
 \subsubsection{Beyond Confluence} 
 We have seen that in presence of the nesting property orthogonal 
 systems are confluent. But what happens if such a property fails, as 
 in the case of the $\lambda$-calculus? In the next section, 
 we shall see another route to achieve confluence. Here, we notice 
 that our relational parallel moves technique outlines a blueprint 
 that can be used to prove weaker forms of confluence in presence of 
 relaxed conditions. For instance, 
 we can massage \autoref{theorem:parallel-moves} to deal with 
 weak confluence.
 
 \begin{lemma}
\label{lemma:parallel-moves-weak-helper}
$\compref{\relone^*} = \compref{\relone}^*$.
\end{lemma}
\begin{proof}
By \autoref{proposition:lassen-algebra-1} 
using that $\relone^* = \mu x. \idrel \vee \relone;x$.
\end{proof}
 
 \begin{proposition}
\label{lemma:parallel-moves-weak}
  If $\relone\dd; \relone\pp \leq \relone\ppstar; \relone\ppstardd$, then 
        $\relone\ppdd; \relone\pp\leq \relone\ppstar; \relone\ppstardd$
\end{proposition}
\begin{proof}[Proof Sketch]
We proceed as in the proof of 
\autoref{theorem:parallel-moves} by
noticing that: \emph{(i)} $\compref{\relone\ppstar} \leq \relone\ppstar$ 
(this follows from \autoref{lemma:parallel-moves-weak-helper}); and \emph{(ii)} 
$\compref{\relone\ppstar; \relone\ppstardd} \leq \relone\ppstar; \relone\ppstardd$.
\end{proof}

Taking advantage of \autoref{lemma:parallel-moves-weak}, we can now 
mimic the proof of \autoref{theorem:relational-parallel-moves} but with 
weaker assumptions. For instance, we can weaken the nesting property
by allowing multiple steps of parallel reductions, i.e. 
$\relone\dd[\relone\spp] \leq \relone[\idrel]; \idrel[\relone\sppstardd]$. 
Such a property, that holds also on syntax with binders, allows us to 
conclude that orthogonal systems are weakly confluent. 
This may not be that interesting, as we are going to see that 
using deep reduction we can prove confluence of orthogonal systems, but observe 
that the result can be further weakened by, e.g., requiring 
$\relone\dd[\idrel]; \relone[\idrel] \leq \relone\sppstar; \relone\sppstardd$, 
hence going beyond orthogonality.

\section{Full Reduction}
Looking at syntax as the (free) \emph{monad} $\freemonad$, we can qualify parallel
reduction --- which is defined relying on the monad multiplication and 
relational extension ---
as the canonical notion of reduction induced 
by the syntax $\freemonad$. We can also look at
$\freemonad$ as an \emph{initial algebra} and rely on 
\autoref{eilenberg-wright-lemma} to exploit 
initiality at a relational level. In fact, 
any ground relation $\relone: \freemonad \vars \to \freemonad \vars$ 
(for reasons that will become clear soon, we will actually consider 
$\relone \vee \idrel$)
induces a relatonal $(\vars + \signature)$-algebra on $\freemonad \vars$ via 
post-composition with the algebra map 
$[\eta, \sigma]: \vars + \signature(\freemonad \vars) \to \freemonad \vars$.

\begin{definition}
\label{def:full-reduction}
Given an \expr-system $(\signature, \vars, \relone)$, 
we define the \emph{full reduction} relation as 
$\relone\ff \defeq \cata{[\eta,\sigma]; (\relone \vee \idrel)}$. 
% \[  
% \xymatrixcolsep{3.5pc}\xymatrix{
% A + \signature(\freemonad A)  
% \ar[d]_{[\eta, \sigma]}
% \ar[r]^{\idrel + \bext{\signature}\cata{[\eta, \sigma]; \relone}}
% & 
% A + \signature(\freemonad A) \ar[d]^{[\eta, \sigma]; \relone}
% \\
% \freemonad A \ar[r]_{\cata{[\eta, \sigma]; \relone}} & \freemonad A
% }
% \]
\end{definition} 

Specific instances of deep reduction have been extensively employed both 
in first- and higher-order rewriting to prove confluence. 
For instance,  for a \trs{} $(X, \signatureop, \stepto)$, 
we see that $\stepto\ff$ is the relation inductively defined thus:
\[
\infer{\val{x} \mathrel{\stepto\ff} \val{x}}{}
\qquad 
\infer{\val{x} \mathrel{\stepto\ff} \termone}{\val{x} \mathrel{\stepto} \termone}
\qquad
\infer{\op(\termone_1, \hhh, \termone_n) \mathrel{\stepto\ff} \op(\termtwo_1, \hhh, \termtwo_n)}
{\termone_1 \mathrel{\stepto\ff} \termtwo_1 \text{ }\cdots \text{ }  \termone_n \mathrel{\stepto\ff} \termtwo_n}
\]
\vspace{-0.2cm}
\[
\infer{o(\termone_1, \hhh, \termone_n) \mathrel{\stepto\ff} \termthree}
{\termone_1 \mathrel{\stepto\ff} \termtwo_1 \text{ }\cdots \text{ }  \termone_n \mathrel{\stepto\ff} \termtwo_n
&
o(\termtwo_1, \hhh, \termtwo_n) \mathrel{\stepto} \termthree}
\]

Notice that since in \trs{s} variables cannot be redexes, 
the second clause above never applies. 
By taking $\relone[\idrel]\ff$, we recover the traditional notion of 
full reduction $\Rrightarrow$.

The relation $\relone\ff$ recursively applies $\relone \vee \idrel$ 
on the whole expression, hence reducing in parallel possibly nested 
redexes at will. This ensures that $\relone\ff$ extends $\relone$. 
Moreover, the presence of $\idrel$ allows one to stop reducing 
at any time. 

\begin{lemma} 
\begin{enumerate}
    \item $\idrel \leq \relone\ff$,
    \item $\relone \leq \relone\ff$.
\end{enumerate} 
\end{lemma}

Using \autoref{thm:hylo} we can exploit the inductive nature of 
$\relone\ff$.

\begin{proposition}
\label{proposition:deep-reduction-inductive}
    $\relone\ff =\lfp{x}{\ideta \vee \ideta;\relone \vee 
    \compreff{x} \vee  \compreff{x}; \relone}$.
\end{proposition}

\begin{remark}
Actually, if we exploit the assumption in \autoref{remark:variables-not-left-rule}, 
we further simplify \autoref{proposition:deep-reduction-inductive} obtaining 
 $\relone\ff =\lfp{x}{\ideta \vee \ideta;\relone \vee 
    \compreff{x} \vee  \compreff{x}; \relone}$.
\end{remark}

Before studying rewriting properties of $\relone\ff$, it is natural to ask 
how $\relone\ff$ relates to $\relone\pp$. Intuitively, the latter is the subrelation 
of the former obtained by reducing non-nested redexes only. Consequently, 
one expects $\relone\pp \leq \relone\ff$. Moreover, 
the same kind of argument suggests that $\relone\ff$ can be recovered by 
possibly many steps of $\relone\pp$. 
This is indeed the case: actually, $\relone\ff$ and $\relone\pp$ 
determine the same reduction sequences.

\begin{lemma}
    $\relone\pp \leqq \relone\ff \leqq \relone\ppstar = \relone\ffstar$.
\end{lemma}

\begin{proof}[Proof Sketch.]
The proof is a straightforward (fixed point) induction. The only (perhaps) non-immediately trivial passage is 
observing that $\compreff{\relone\ppstar};\relone \leq \relone\ppstar$. 
This follows from \autoref{lemma:parallel-moves-weak-helper} thus:
$
\compreff{\relone\ppstar};\relone 
\leq 
\compref{\relone\ppstar};\relone 
\leq 
\relone\ppstar;\relone 
\leq 
\relone\ppstar;\relone\pp 
\leq \relone\ppstar.
$
\end{proof}
\noindent Writing $\relone\sff$ for $\relone[\idrel]\ff$, we see that all the above 
results extend to $\relone\sff$ and $\relone\spp$. 

Let us now move to confluence of full reduction. Since $\relone\sff$ 
reduces also nested redexes, we expect such a relation to 
satisfy the nesting property of \autoref{def:nesting-property} 
(properly reformulated replacing $\relone\spp$ with $\relone\sff$), at least on 
orthogonal systems, where now orthogonality is defined as in \autoref{def:orthogonality} 
but with $\relone\sff$ in place of $\relone\spp$. 
To prove the nesting property for $\relone\sff$ we first observe that the latter 
is implied by substitutivity.

\begin{lemma}
\label{lemma:substitutivity-implies-nesting}
If 
$\relone\sff$ is substitutive (i.e. 
$\relone\sff[\relone\sff] \leq \relone\sff]$), then 
it has the nesting property: that is, 
$\relone\dd[\relone\sff] \leq \relone\sff; \relone\dd[\idrel]$.
\end{lemma}

\begin{proof}
 Recall that $-[=]$ is functorial
%$(x_1;x_2)[y_1;y_2] \leq x_1[y_1]; x_2[y_2]$, 
 and that 
 $\idrel \leq \relone\sff$. We have: 
 $\relone\dd;[\relone\sff] = (\idrel;\relone\dd)[\relone\sff; \idrel] 
\leq (\relone\sff; \relone\dd)[\relone\sff;\idrel] =
 \relone\sff[\relone\sff]; \relone\dd[\idrel] 
 \leq \relone\sff; \relone\dd[\idrel]$, where the last inequality 
 follows from substitutivity. 
\end{proof}

We now aim to prove substitutivity of $\relone\sff$. We do so by exploiting an 
unexpected connection between $\relone\sff$ and a well-known relational technique in program 
equivalence: \emph{Howe's method} \cite{Howe/IC/1996,Pitts/ATBC/2011}.

\subsection{Full Reduction and Howe's Method}

Howe's method is a powerful operational technique to prove congruence of 
applicative (bi)similarity originally developed in the context of 
the pure $\lambda$-calculus. Howe's method has been extended to a variety 
of concrete formalism --- such as calculi with computational effects~\cite{Lassen/PhDThesis,Ong/LICS/1993,DalLagoSangiorgiAlberti/POPL/2014,Levy/ENTCS/2006,DalLagoGavazzoLevy/LICS/2017,dal-lago/gavazzo-mfps-2019,DBLP:phd/basesearch/Gavazzo19,Gavazzo/ICTCS/2017,CrubilleDalLago/ESOP/2014,DBLP:journals/entcs/BiernackiL14,DBLP:journals/tcs/LagoGT20,DBLP:journals/pacmpl/LagoG22a} ---
and categorical semantics~\cite{hirschowitz-howe-1,hirschowitz-howe-2,goncharov-howe}. 
Gordon~\cite{Gordon-1995,Gordon/FOSSACS/01} and 
Lassen~\cite{Lassen/PhDThesis,Lassen/RelationalReasoning} have developed an 
elegant relational account of Howe's method on specific $\lambda$-calculi;
such an account can be made completely general 
by abstracting over the concrete syntax, along the lines of \autoref{section:an-allegorical-theory}. 

We are going to show that the Howe extension of a relation $\relone$ 
coincides with $\relone\ff$. 
The advantage of such an equality --- which is, in spite of its simplicity, new (
at least to the best of the author's knowledge) --- is twofold: on the one hand, we obtain a novel understanding of 
Howe's method in terms of initial relation algebras as in \autoref{def:full-reduction}; 
on the other hand, we obtain powerful proof techniques for reasoning about full reduction. 

\begin{definition}
For a relation $\relone: \freemonad \vars \to \freemonad \vars$, define 
its \emph{Howe extension}\footnote{Usually one defines 
$\relone\hh$ as $\lfp{x}{\compref{x}; \relone}$ --- hence without forcing 
reflexivity on $\relone$ --- and then restricts the analysis to reflexive relations 
(program approximations and equivalences being such). For the ease of exposition, 
we force reflexivity into the very definition of $\relone\hh$, much in the same way as we 
did with $\relone\ff$. Of course, it is possible to remove reflexivity from 
both such definitions and obtain the same results we prove in this section 
\emph{mutatis mutandis}.}
as $\relone\hh \defeq \lfp{x}{\compref{x}; (\relone \vee \idrel)}$. 
We define $\relone\shh$ as $\relone[\idrel]\hh$.
\end{definition}
We now show that $\relone\hh$ and $\relone\ff$ coincides, and infer from that 
an inductive characterisation of $\relone\sff$. 
For the remaining part of this section, we use the notation 
$y\eqq$ 
for the reflexive closure of $y$, i.e. $y\eqq \defeq y \vee \idrel$.
\begin{lemma}
\label{lemma:howe-equals-full-reduction}
    $\relone\hh= \relone\ff$.
\end{lemma}
\begin{proof}
By \autoref{thm:hylo}, we have:
\begin{align*}
    \relone\ff 
    &= \cata{[\eta,\sigma];\relone\eqq}
    \\
    &= \mu x. [\eta,\sigma]\dd; (\idrel_{\vars} + \signature x); [\eta,\sigma];\relone\eqq
    \\
    &= \mu x. \eta\dd;\eta;\relone\eqq \vee \sigma\dd;\signature x;\sigma; \relone\eqq 
    \\
    &= \mu x. \ideta;\relone\eqq \vee \compreff{x};\relone\eqq 
    \\
    &= \mu x. (\ideta \vee \compreff{x}); \relone\eqq 
    \\
    &= \mu x. \compref{x};\relone\eqq 
    \\
    &= \relone\hh.
\end{align*}
\end{proof}

To exploit the consequences of \autoref{lemma:howe-equals-full-reduction}, 
we first make explicit a (straightforward) 
property of the reflexive closure operator.

\begin{lemma}
\label{lemma:reflexive-closure-substitutive}
$\relone[\idrel]\eqq = \relone\eqq[\idrel]$.
\end{lemma}
\begin{proof}
Unfolding the definition of $(-)\eqq$, we see that 
we have to prove $\relone[\idrel] \vee \idrel = (\relone \vee \idrel)[\idrel]$.
First, we notice that 
$\relone[\idrel] \vee \idrel \leq (\relone \vee \idrel)[\idrel]$
follows from
$\relone[\idrel]  \leq (\relone \vee \idrel)[\idrel]$ 
((which follows from \autoref{proposition:lassen-algebra-2}, since 
$\relone \leq \relone \vee \idrel$))
and 
$\idrel \leq (\relone \vee \idrel)[\idrel]$. 
(again, by \autoref{proposition:lassen-algebra-2} we have 
$\idrel = \idrel[\idrel] \leq (\relone \vee \idrel)[\idrel]$). 
To conclude the proof, it is thus enough to prove the opposite inequality, namely 
$(\relone \vee \idrel)[\idrel] \leq \relone[\idrel] \vee \idrel$. 
The latter
is equivalent to $\relone \vee \idrel \leq \idrel \gg (\relone[\idrel] \vee \idrel)$, 
which follows from 
$\relone \leq \idrel \gg (\relone[\idrel] \vee \idrel)$ 
(i.e. $\relone[\idrel] \leq \relone[\idrel] \vee \idrel$, which trivially holds) 
and 
$\idrel \leq \idrel \gg (\relone[\idrel] \vee \idrel)$ 
(i.e. $\idrel[\idrel] \leq\relone[\idrel] \vee \idrel$, which 
trivially follows from $\idrel[\idrel] = \idrel$).
\end{proof}

\begin{proposition}
\label{prop:howe-equals-full-reduction}
For an \expr-system $(\signature, \vars, \relone)$, we have:
\begin{enumerate}
    \item $\relone\sff = \relone\shh = \mu x.\ideta \vee \compreff{x};\relone\eqq[\idrel]$.
    \item $\relone\sffdd = \relone\shhdd \mu x.\ideta \vee \relone\eqq[\idrel]\dd;\compreff{x}$.
\end{enumerate}
\end{proposition}
\begin{proof}
We prove item 1, as item 2 is a direct consequence of it.
By \autoref{lemma:howe-equals-full-reduction}, we have 
$$\relone\sff = \relone\shh = \relone[\idrel]\hh = \mu x. \compref{x}; \relone[\idrel]\eqq.$$ 
    We then calculate:
\begin{align*}
\mu x. \compref{x}; \relone[\idrel]\eqq
    &= \mu x. \compref{x}; \relone\eqq[\idrel]
    & & \text{(\autoref{lemma:reflexive-closure-substitutive})}
    \\
    &= \mu x. (\ideta \vee \compreff{x}); \relone\eqq[\idrel]
    \\
    &= \mu x. \ideta;\relone\eqq[\idrel] \vee \compreff{x}; \relone\eqq[\idrel]
    \\
    &= \mu x. \ideta;\relone[\idrel]\eqq \vee \compreff{x}; \relone\eqq[\idrel]
    & & \text{(\autoref{lemma:reflexive-closure-substitutive})}
     \\
    &= \mu x. \ideta;(\relone[\idrel] \vee \idrel) \vee \compreff{x}; \relone\eqq[\idrel]
     \\
    &= \mu x. \ideta;\relone[\idrel] \vee \ideta \vee \compreff{x}; \relone\eqq[\idrel]
    \\
    &= \mu x. \bot \vee \ideta \vee \compreff{x}; \relone\eqq[\idrel]
    & & \text{(\autoref{remark:variables-not-left-rule})}
    \\
    &= \mu x.\ideta \vee \compreff{x}; \relone\eqq[\idrel].
\end{align*}
\end{proof}

We are now ready to prove substitutivity (and compatibility) of 
full reduction.

\begin{proposition}
\label{corollary:sbstitutivity-howe}
$\relone\sff$ is compatible and substitutive. 
\end{proposition}

\begin{proof}
We first show that $\relone\sff$ is substitutive, i.e. 
$\relone\sff[\relone\sff] \leq \relone\sff$. 
We prove the equivalent inequality 
$\relone\sff \leq \relone\sff \gg \relone\sff$
by $\omega$-continuous fixed point induction 
(recall that $-[=]$ is $\omega$-continuous in the first argument).
using \autoref{prop:howe-equals-full-reduction}. 
We thus assume $x \leq \relone\sff$ 
and $x \leq \relone\sff \gg \relone\sff$ --- 
i.e. $x[\relone\sff] \leq \relone\sff$ --- and show 
$\ideta \vee \compreff{x};\relone\eqq[\idrel] \leq \relone\sff \gg \relone\sff$.
Proving the latter amounts to prove 
$\ideta \leq \relone\sff \gg \relone\sff$ and 
$\compreff{x};\relone\eqq[\idrel] \leq \relone\sff \gg \relone\sff$. 
The former is equivalent to $\ideta[\relone\sff] \leq \relone\sff$, which follows from 
\autoref{proposition:lassen-algebra-2}. 
For the latter, it sufficient to prove 
$(\compreff{x};\relone\eqq[\idrel])[\relone\sff] \leq \relone\sff$. 
We calculate:\footnote{
By associativity of relation substitution, i.e. $x[y][z] = x[y[z]]$, 
we have $\relone^=[\idrel][\idrel] 
= \relone^=[\idrel[\idrel]] \leq \relone^=[\idrel]$.}
\begin{align*}
    (\compreff{x};\relone\eqq[\idrel])[\relone\sff]
    &= (\compreff{x};\relone\eqq[\idrel])[\relone\sff;\idrel]
    \\
   % &= \compreff{x}[\relone\sff]; \relone\eqq[\idrel][\idrel]
    %\\
    &\leq \compreff{x}[\relone\sff]; \relone\eqq[\idrel]
    \\
    &\leq \compreff{x[\relone\sff]}; \relone\eqq[\idrel]
    \\
    &\leq \compreff{\relone\sff}; \relone\eqq[\idrel]
    && (\text{since } x[\relone\sff] \leq \relone\sff)
    \\
    &\leq \ideta \vee \compreff{\relone\sff}; \relone\eqq[\idrel]
     \\
    &= \relone\sff.
\end{align*}

% it is sufficient to prove 
% substitutivity and compatibility of $\relone\shh$.
% Let $\relone^=$ be $\relone \vee \idrel$.
% For substitutivity, we have to show
% $\relone\shh[\relone\shh] \leq \relone\shh$, i.e. 
% $\relone\shh \leq \relone\shh \gg \relone\shh$. We proceed by 
% fixed point induction showing 
% $(\compref{\relone\shh \gg \relone\shh}); \relone^=[\idrel] 
% \leq \relone\shh \gg \relone\shh$. 
% Recall that $\compref{x \gg y} \leq x \gg \compref{x \vee y}$, so that 
% $\compref{\relone\shh \gg \relone\shh} 
% \leq \relone\shh \gg \compref{\relone\shh \vee \relone\shh} 
% = \relone\shh \gg \compref{\relone\shh}$. 
% Since $\relone^=[\idrel]$ is trivially substitutive\footnote{
% By associativity of relation substitution, i.e. $x[y][z] = x[y[z]]$, 
% we have $\relone^=[\idrel][\idrel] 
% = \relone^=[\idrel[\idrel]] \leq \relone^=[\idrel]$.} 
% and thus $\relone^=[\idrel] \leq \idrel \gg \relone^=[\idrel]$, we calculate: 
% \begin{align*}
%     (\compref{\relone\shh \gg \relone\shh}); \relone^=[\idrel] 
%     &\leq (\relone\shh \gg \compref{\relone\shh}); \relone^{=}[\idrel]
%     \\
%     &\leq (\relone\shh \gg \compref{\relone\shh}); (\idrel \gg \relone^{=}[\idrel])
%     \\
%     &\leq (\relone\shh; \idrel) \gg (\compref{\relone\shh}; \relone^{=}[\idrel])
%     \\
%     &\leq \relone\shh \gg \relone\shh.
% \end{align*}
For compatibility, i.e. 
$\compref{\relone\shh} \leq \relone\shh$, we notice that since $\relone^=[\idrel]$ is reflexive, 
we have $\compref{\relone\shh} = \compref{\relone\shh}; \idrel \leq 
\compref{\relone\shh}; \relone^=[\idrel] \leq \relone\shh$.
\end{proof}

\begin{remark}
Notice that substitutivity of $\relone\shh$ holds independently of
reflexivity of $\relone\eqq[\idrel]$ (whereas compatibility 
actively uses it). This is a consequence of \autoref{remark:variables-not-left-rule} 
--- which gives \autoref{prop:howe-equals-full-reduction} --- whereby we 
do \emph{not} have to account for the case 
$\ideta;\relone\eqq[\ideta] \leq \relone\sff \gg \relone\sff$. 
To handle such a case, we indeed need to rely on reflexivity of 
$\relone\eqq[\ideta]$.
\end{remark}

Putting together \autoref{corollary:sbstitutivity-howe} and 
\autoref{lemma:substitutivity-implies-nesting} we obtain 
the nesting property for $\relone\sff$. 

\begin{corollary}
\label{corollary:full-reduction-has-the-nesting-property}
Full reduction $\relone\sff$ has the nesting property.
\end{corollary}

We now have all he ingredients to prove that full reduction has the diamond property. 
Before that, however, we mention another candidate definition of a `full reduction' that 
makes actively use of substitutivity. Such a definition is usually called multi-step 
reduction in term rewriting \cite{terese} and it is sometimes used in the 
context of the so-called Tait-Martin-L\"of technique~\cite{Barber96,aczel-general-church-rosser,Takahashi1995} --- notice, however, 
that in concrete calculi, such as the $\lambda$-calculus, such a technique uses 
(concrete instances of) full reduction, rather than multi-step reduction. 

As for parallel and full reduction, also multi-step reduction can be (re)discovered 
in the literature on program equivalence, where it goes under the name of 
 \emph{substitutive context closure} of a relation \cite{Lassen/PhDThesis,Lassen/RelationalReasoning}. 
 
 \begin{definition}
Let $(\signature, \vars, \relone)$ be an \expr-system.
The \emph{substitutive context closure} of $\relone$ is defined 
as $\relone\scccc \defeq \lfp{x}{\relone[x] \vee \compref{x}}$.
 \end{definition}
 
 Since $\relone\shh$ is substitutive and compatible, $\relone\scccc$ is 
 contained in it ($\relone\scccc \leq \relone\shh$), and 
 the two relations give the same reduction sequences. 
 We do not investigate multi-step reduction any further, although 
 we observe (without giving a formal proof) that 
 $\relone\scccc$ enjoys the so-called \emph{triangle property} 
 (i.e. $x \leq x; x\dd)$ from which confluence follows.

\subsection{The Tait-Martin-L\"of Technique} 
We are finally ready to prove the main result of this section, 
namely that in orthogonal systems full reduction has the diamond 
property, and thus it is confluent. 
Our result provides an abstract and relational version of the so-called 
Tait-Martin-L\"of technique~\cite{Barber96,aczel-general-church-rosser,Takahashi1995}, 
whereby confluence of a system (originally the $\lambda$-calulus) 
is proved showing the diamond property of 
full reduction. 

% Before that, we exploit 

% \begin{lemma}
% \label{lemma:tml-helper}
% \begin{enumerate}
%     \item $\relone\sffdd = \mu x. \ideta \vee \compreff{x} \vee \relone\eqq[\idrel]\dd;\compreff{x}$.
%     \item If $\relone[\idrel]\dd; \relone[\idrel] \leq \idrel$, then 
%         $\relone\eqq[\idrel]\dd; \relone\eqq[\idrel] \leq \idrel$
%     \item If $\relone[\idrel]\dd; \compreff{\relone\sff} \leq \relone\dd[\relone\sff]$, then 
%         $\relone\eqq[\idrel]\dd; \compreff{\relone\sff} \leq \relone\eqqdd[\relone\sff]$
% \end{enumerate}
% \end{lemma}

% \begin{proof}
%  \begin{enumerate}
%      \item Straightforward. 
%      \item Since $\relone\eqq[\idrel] = (\relone \vee \idrel)[\idrel] = \relone[\idrel] \vee \idrel$, 
%         (and $\relone[\idrel]\dd = \relone\dd[\idrel]$), it is sufficient to prove 
%         $(\relone\dd[\idrel] \vee \idrel); (\relone[\idrel] \vee \idrel) \leq \idrel$. 
%         This amounts to show:
%         \begin{enumerate}
%             \item  $(\relone\dd[\idrel]; (\relone[\idrel] \leq \idrel$, 
%             \item  $\relone\dd[\idrel]; \idrel \leq \idrel$, 
%             \item  $\idrel; \relone[\idrel]  \leq \idrel$, 
%             \item $\idrel;\idrel \leq \idrel$.
%         \end{enumerate}
%  \end{enumerate}
% \end{proof}

\begin{theorem}
\label{theorem:tait-martin-lof}
Say that an \expr-system $(\signature, \vars, \relone)$ is 
orthogonal if
$\relone[\idrel]\dd; \relone[\idrel] \leq \idrel$ 
and
$\relone[\idrel]\dd; \compreff{\relone\sff} \leq \relone\dd[\relone\sff]$. 
Then, in an orthogonal system full reduction $\relone\sff$ has the diamond 
property: $\relone\sffdd; \relone\sff \leq \relone\sff; \relone\sffdd$.
\end{theorem}

\begin{proof}
First of all, we notice that using orthogonality and 
the nesting property (\autoref{corollary:full-reduction-has-the-nesting-property}), 
we obtain:
\begin{align}
    \relone[\idrel]\dd; \compreff{\relone\sff} 
    \leq \relone\dd[\relone\sff] 
    \leq \relone\sff; \relone\dd[\idrel]
    \label{eq:orthogonality-helper}
    \tag{ortho-nesting}
    \\
    \compreff{\relone\sff}\dd; \relone[\idrel]
    \leq \relone[\relone\sffdd] 
    \leq \relone[\idrel]; \relone\sffdd
    \label{eq:orthogonality-helper-dual}
    \tag{ortho-nesting$\dd$}
\end{align}

 Let $\relx \defeq  \relone\sff; \relone\sffdd$. We prove 
 $\relone\sffdd; \relone\sff \leq \relx$ by $\omega$-continuous fixed point induction 
 on $\relone\sffdd$ (\autoref{prop:howe-equals-full-reduction}).
We thus assume 
$x \leq \relone\sffdd$ and $x;\relone\sff \leq \relx$ and show
(we tacitly exploit distributivity of composition over join 
 and the universal property of the latter):
 \begin{align}
     \ideta; \relone\sff &\leq \relx 
     \label{tml-eq-1}
     \\
     \relone\eqq[\idrel]\dd;\compreff{x};\relone\sff &\leq \relx 
    \label{tml-eq-2}
 \end{align}
 For \eqref{tml-eq-1}, we first observe that $\ideta; \compreff{z} = \bot$, for any 
 $z$. Consequently, we have:
 \begin{align*}
          \ideta; \relone\sff &= \ideta;(\ideta \vee \compreff{\relone\sff};\relone\eqq[\idrel])
          \\
          &= \ideta;\ideta \vee \ideta;\compreff{\relone\sff};\relone\eqq[\idrel]
          \\
          &= \ideta \vee \bot
          \\
          &\leq \relx.
 \end{align*}
 Let us now move to \eqref{tml-eq-2}.
Proceeding as for \eqref{tml-eq-1}, we have:
\begin{align*}
\relone\eqq[\idrel]\dd;\compreff{x};\relone\sff 
&= \relone\eqq[\idrel]\dd;\compreff{x};(\ideta \vee \compreff{\relone\sff};\relone\eqq[\idrel]) 
\\
&= \relone\eqq[\idrel]\dd;\compreff{x};\ideta \vee  \relone\eqq[\idrel]\dd;\compreff{x};\compreff{\relone\sff};\relone\eqq[\idrel]
\\
&= \relone\eqq[\idrel]\dd;\compreff{x};\compreff{\relone\sff};\relone\eqq[\idrel].
\end{align*}
We now exploit the induction hypothesis and obtain:
\begin{align*}
 \relone\eqq[\idrel]\dd;\compreff{x};\compreff{\relone\sff};\relone\eqq[\idrel]
 &= \relone\eqq[\idrel]\dd;\compreff{x;\relone\sff};\relone\eqq[\idrel]
 \\
  &\leq \relone\eqq[\idrel]\dd;\compreff{\relone\sff; \relone\sffdd};\relone\eqq[\idrel]
  \\
  &\leq \relone\eqq[\idrel]\dd;\compreff{\relone\sff}; \compreff{\relone\sff}\dd;\relone\eqq[\idrel].
\end{align*}
Consequently, to conclude the thesis it is sufficient to prove 
$\relone\eqq[\idrel]\dd;\compreff{\relone\sff}; \compreff{\relone\sff}\dd;\relone\eqq[\idrel] \leq \relx$. 
We use \autoref{lemma:reflexive-closure-substitutive} 
(which also gives $\relone\eqq[\idrel]\dd = \relone\dd[\idrel]\eqq = \relone[\idrel]\ddeqq$) 
and reduce the proof of the above inequality to the proofs of the following ones:
\begin{align*}
    \compreff{\relone\sff}; \compreff{\relone\sff}\dd
    &\leq \relx
    \\
    \relone[\idrel]\dd;\compreff{\relone\sff}; \compreff{\relone\sff}\dd
    &\leq \relx
    \\
    \compreff{\relone\sff}; \compreff{\relone\sff}\dd;\relone[\idrel] 
    &\leq \relx
    \\
    \relone[\idrel]\dd;\compreff{\relone\sff}; \compreff{\relone\sff}\dd;\relone[\idrel] 
    &\leq \relx.
\end{align*}
The first is tautological. For the second, we calculate: 
\begin{align*}
\relone[\idrel]\dd;\compreff{\relone\sff}; \compreff{\relone\sff}\dd;\relone[\idrel]
& \leq \relone\sff;\relone\dd[\idrel];\compreff{\relone\sff}\dd;\relone[\idrel]
&& \eqref{eq:orthogonality-helper}
\\
&= \relone\sff;\relone[\idrel]\dd;\compreff{\relone\sff}\dd;\relone[\idrel]
\\
&= \relone\sff;(\compreff{\relone\sff}; \relone[\idrel])\dd
\\
&\leq \relone\sff;(\compreff{\relone\sff}; \relone\eqq[\idrel])\dd
\\
&\leq \relone\sff;\relone\sffdd \text{ }(= \relx).
\end{align*}
For the third inequality we proceed as for the second one, but in a dual fashion 
(hence relying on \eqref{eq:orthogonality-helper-dual}).
Finally, for the fourth inequality we have:
\begin{align*}
\relone[\idrel]\dd;\compreff{\relone\sff}; \compreff{\relone\sff}\dd;\relone[\idrel]
& \leq \relone\sff;\relone\dd[\idrel];\compreff{\relone\sff}\dd;\relone[\idrel]
&& \eqref{eq:orthogonality-helper}
\\
& \leq \relone\sff;\relone\dd[\idrel];\relone[\idrel]; \relone\sffdd
&& \eqref{eq:orthogonality-helper-dual}
\\
& \leq \relone\sff; \relone\sffdd \text{ }(= \relx).
&& \text{(Orthogonality)}
\end{align*}

\end{proof}

\section{Sequential Reduction: A Few Words Only}
\label{sect:sequential-reduction}
The theory developed so far shows that parallel and full reduction are remarkably natural, 
at least from a 
structural and algebraic perspective. When it comes to think about reduction 
computationally, however, \emph{sequential} (or \emph{linear}) reduction is usually considered more 
fundamental. In fact, almost all textbooks in rewriting theory first define sequential reduction, 
and then introduce parallel (and full)
reduction on top of that. 

Even if the relational analysis of sequential reduction is still work in progress, we 
mention that sequentiual reduction can be recovered in the allegorical framework, both 
structurally and algebraically. In the former case, 
one relies on the \emph{derivative}~\cite{derivative-joyal,derivative-1,derivative-2,derivative-3} of the signature functor 
(as well as of the corresponding monad). 
In fact, the derivative of a functor $F: \topos \to \topos$, if it exists, is 
the functor $\partial F: \topos \to \topos$ coming with a plug-in \emph{weakly cartesian} natural 
transformation 
$dF_X: \partial F X \times X \to FX$ satisfying the following universal mapping property:
for any functor $G$ with a weakly cartesian\footnote{
A natural transformation is weakly cartesian if its naturality squares are weak pullbacks~\cite{Clementino2014TheMO}. Intuitively, we can think about 
such natural transformations as \emph{linear} maps between functors~\cite{Faggian-2019}.
} natural transformation 
$\vartheta: GX \times X \to FX$, there exists a unique weakly cartesian natural transofmation 
$\vartheta': GX \to \partial FX$ satisfying the following diagram.
\[
\xymatrix{
\partial F X \times X \ar[r]^{\theta} & FX 
\\
GX \times X \ar@{.>}[u]^{\vartheta' \times \idrel} \ar[ru]_{\vartheta} &
}
\]

Derivatives of simple polynomial functors and of their free monads, for instance, always 
exist~\cite{derivative-1,derivative-2,derivative-3}. In those cases, one clearly sees 
that such derivatives provides contexts with one \emph{linear} hole, which is exactly 
what is required to define sequential reduction. 
Consequently, one possible definition of the sequential reduction (on an \expr-system 
$(\signature, \vars, \relone)$ is the relation 
$\relone\lin \defeq d\freemonad\dd; (\idrel \times \relone);d\freemonad$. 

In a similar fashion, it is possible to define a \emph{linear} compatible refinement operator 
$\widetriangle{\relone} \defeq d\functor\dd; (\idrel \times \relone);d\signature$ 
and use the latter to give an inductive characterisation of $\relone\lin$ 
as $\mu x.\relone \vee \widetriangle{x}$. 
At this point, it is possible to proceed following the methodology of the augmented 
calculus of relations isolating the algebraic laws defining $\widetriangle{-}$. 
Notice, however, that such laws largely differ from those of $\compref{-}$. 
For instance, we have $\widetriangle{\relone;\reltwo} \leq \widetriangle{\relone};\widetriangle{\reltwo}$ 
but not the vice versa.

Another option to capture forms of sequentiality (albeit not sequential reduction itself) 
is to think about parallel and full reduction as primitives, and 
to regard sequential-like reductions as their restrictions. 
Following this direction, we may introduce the notion of a 
\emph{sequentialisation} of a relator $\signature$, namely a family of 
finitary maps 
$\srelator: \allegory(A,A) \to \allegory(\signature A, \signature A)$ 
such that: 
\begin{align*}
\idrel &= \seqq \idrel
\\
\srelator(\relone\dd) &= (\srelator \relone)\dd
\\
\srelator(\relone \vee \reltwo) &= \srelator \relone \vee \srelator \reltwo
\\
\srelator \relone &\leqq \signature \relone 
\\
\signature \relone &\leqq (\srelator \relone)^*.
\end{align*}
For instance, if $\signature$ is the functor induced by a first-order 
signature, then a sequantialisation of (the Barr extension of) $\signature$ 
is the following inductively defined map:
\[
\infer{\val{x} \mathrel{\srelator \relone} \val{y}}
{x \mathrel{\relone} y}
\quad
\infer{\op(\termthree_1, \hhh, \termone, \hhh, \termthree_n) \mathrel{\srelator \relone} 
\op(\termthree_1, \hhh, \termtwo, \hhh, \termthree_n)}
{\termone \mathrel{\srelator \relone} \termtwo 
& n \geq 0
}
\]
Notice that (the reduction induced by) $\srelator$ does \emph{not} coincide with the usual 
sequential reduction. In fact, $\srelator$ allows to always reduce  
$0$-ary operations (i.e. constants) to themselves (otherwise, it behaves 
 as sequential reduction).
% This is not the case for traditional rewriting which, in fact, is defined by requiring 
% $n \geq 1$ in the aforementioned rule. Theoretically, such a form of 
% sequential rewriting is obtained by taking the lax version of \eqref{seq-relator-id}, hence 
% requiring $\srelator \idrel \leqq \idrel$. 

At this point, we can proceed as in the previous sections, simply working with 
(fixed) a sequentialisation $\srelator$ of $\signature$ in place of the latter. 
We do not go any further but simply remark that a weak Kleisli-like lemma 
along the lines of \autoref{lemma:parallel-moves-weak} can be easily proved 
for sequentialised reductions.

\section{Conclusion}
\label{section:conclusion}
In this work, we have outlined a general relational theory of 
symbolic manipulations in rewriting style. The theory is 
given in the framework of allegory theory and goes 
in tandem with the so-called mathematical theory of syntax: remarkably, 
these two theories build upon the same collection of concepts, the theory of syntax 
implementing them in a categorical way, the theory of symbolic manipulation 
implementing them in an allegorical, relational way. 
We have then pushed the relational approach even further by noticing how 
the aforementioned relational counterparts of syntactic notions define
new operators on relations subject to specific algebraic laws. Such operators and 
their laws turned out to be all that matters to study symbolic manipulation, and thus 
give raise to a syntax-independent augmented calculus of relations within which we have 
defined classic reduction relations (viz. parallel and full reduction) and proved nontrivial 
properties about them.

\subsection{Relational Rewriting and Operational Semantics}
The results presented in this paper give (first) evidences that 
the relational approach to rewriting goes considerably beyond abstract reduction systems. 
The author hopes that such results
will contribute to a renewed interest in the relational approach to 
rewriting (an outline of a research program for that is given in the next 
section). 

(Relational) Rewriting, however, is just one 
 piece in the (operational) jigsaw. 
In fact, the augmented calculus of relations and, most importantly, its 
underlying methodology, suggest that (part of) operational reasoning 
can be developed in an axiomatic and syntax-independent fashion. 
Indeed, one way to read the results of this work, together with 
previous results on program equivalence, is that
the augmented calculus of relations is expressive enough to account for two main forms 
of operational reasoning: 
\emph{rewriting} and \emph{program equivalence} (and refinement).\footnote{
Calculi subsumed by the augmented calculus of relations have been employed to 
give relational accounts of
logical relations~\cite{DBLP:journals/pacmpl/LagoG22a}, contextual and CIU 
equivalence~\cite{Lassen/PhDThesis,DBLP:phd/basesearch/Gavazzo19}, and
applicative and normal bisimilarity~\cite{Gordon-1995,Lassen/PhDThesis,DalLagoGavazzoLevy/LICS/2017,Lassen/BismulationUntypedLambdaCalculusBohmTrees,Lassen/EagerNormalFormBisimulation/2005,DBLP:conf/esop/LagoG19}.} 
Furthermore, the author conjectures that much more operational reasoning, such as theories 
of program dynamics, can be 
developed within such a calculus (or variations thereof). 

This perspective, which we may refer to as \emph{relational} or \emph{allegorical 
operational semantics}, aims to achieve a systematic development of operational techniques 
within a (truly) relational paradigm, whereby program relations,\footnote{I.e. 
suitable notions of relation on programs texts, rather than
on their abstract denotations.} their operations, and 
algebraic properties are first-class citizens.
The successful application of relational calculi to the field of
program equivalence and rewriting  
hints that the relational approach to operational semantics has the potential to 
achieve a general axiomatic and largely syntax-independent basis for operational reasoning.\footnote{
Notice also that such a basis seems to be also well suited for machine formalisation.}

\subsection{Future Work} 
Following the discussion made so far, future work can be divided into 
two research directions. The first one is devoted to the development of 
relational rewriting, hence complementing the confluence results 
proved in previous sections; the second, instead, focuses on 
extending the relational framework to cover more operational behaviours, hence 
going towards the aforementioned allegorical operational semantics. 

Beginning with the former and omitting the already discussed issue of 
sequential reduction (\autoref{sect:sequential-reduction}), here is a 
possible research agenda.

\begin{varenumerate}
    \item \emph{Termination.} 
       The work by Hasegawa~\cite{hasegawa} shows how syntax-based \emph{termination techniques} 
        (such as multiset and recursive path ordering~\cite{terese}) on \trs{s} can be 
        abstractly recovered in terms of lifting of (analytic) functors~\cite{derivative-joyal}, 
        and thus suggests that 
        syntax-based termination can be indeed analysed in a relational framework.\footnote{
        Relational analysis of termination for abstract systems have already been given, especially 
        concerning modularity results~\cite{backshouse-calculational-approach-to-mathematical-induction,doornbos-1,doornbos-2}.
        }
        It thus seems natural to incorporate and 
        extend Hasegawa’s results in the allegorical framework.
    \item \emph{Strategies and Factorisation.}
        Together with confluence and termination, another crucial property of rewriting systems is 
        \emph{factorisation}~\cite{Curry/Combinatory-logic/1958}.
        The Kleisli-like lemmas proved in the paper can be already generalised to 
        factorisation techniques (simply replace $\relone\dd$ and alike with arbitrary relations 
        $\reltwo$). These, however, provide only a superficial account of factorisation 
        and it is thus interesting to ask whether deeper analyses of factorisation can be given 
        relationally, perhaps along the line of the recent work by 
        Accattoli et al.\cite{faggian-factorization}.
    \item \emph{Analytic Functors and Rewriting Modulo.} In this work, our examples were intended to model 
        notions of syntax-based symbolic systems. Consequently, we focused on syntax-like 
        finitary functors (and free monads), polynomial functors being a prime examples of those. 
        Another interesting class of examples that we have not studied is the one of 
        \emph{analytic functors}~\cite{derivative-joyal}. In a first approximation, 
        analytic functors can be seen as polynomial functors modulo an equivalence obtained via a group of symmetries. From a rewriting perspective, working with analytic functors we recover notions of 
        syntax modulo permutations, in a very general sense. 
        Looking at analytic functors, consequently, we may apply relational rewriting 
        to rather liberal notions of syntax carrying a nontrivial semantic import.
    \item \emph{Infinitary and Coinductive Rewriting.} The theory developed in this paper 
        applies to  
        finitary syntax and rewriting. This naturally leads to asking whether the 
        allegorical account scales to infinitary syntax and coinductive rewriting. 
        An educated guess in this direction is to replace finitary syntax 
        with infinitary one, modelling the latter through iterative algebras~\cite{iterative-algebras}, 
        or structures alike. 
        Notice that doing so, reduction relations will be still defined via relators,
        although they would be recasted in the extended calculus of relations not as 
        inductive relations (viz. least fixed points), but as \emph{coinductive} 
        (viz. greatest fixed points) or 
        \emph{mixed inductive-conductive} (viz. nested least and greatest fixed points) relations~\cite{endrullis-infinitary-rewriting}.
    \item \emph{Quantitative Rewriting.} Last but not least, an interesting limitation 
        of the allegorical framework is that it cannot cope with quantitative forms of 
        rewriting~\cite{fuzzy-rewriting-1,fuzzy-rewriting-2,fuzzy-rewriting-3,10.1145/3571256}. 
        The problem is foundational, in the sense that categories of quantitative relations 
        do \emph{not} form an allegory, as they fail to satisfy the modular law. However, 
        they form Frobenius quantaloids~\cite{Schubert2006} and it is natural to explore whether 
        an allegorical-like theory of rewriting can be given on top of such structures 
        (very likely enriched with structures such as power objects).  
        Interestingly, from the axiomatic perspective of the augmented calculus of 
        relations, one observes that relational calculi for modal and quantitative program equivalence 
        (viz. program metrics) have already been defined~\cite{Gavazzo/LICS/2018,DBLP:journals/pacmpl/LagoG22a}. 
        Such calculi are impressively close
        to the augmented calculus of relations except for the addition of a 
        graded comonadic modality acting as scaling~\cite{Orchard-icfp-2019,dagnino-1}. 
        Consequently, one promising direction to approach quantitative rewriting is 
        to proceed axiomatically by extending the augmented calculus of relations 
        with suitable modalities.
\end{varenumerate}

Let us now move to allegorical operational semantics. In this case, outlining 
a research agenda is more difficult, as the subject is considerably vast. 
Nonetheless, we can fix a couple of general research-goals aiming to
explore the potential of the relational approach (and, more specifically, 
of the augmented calculus of relations) 
as a foundational formalism for operational reasoning. 
\begin{varenumerate}
    \item \emph{Reduction-Based Semantics}. 
        A first, natural question to answer in order to test 
        the robustness of relational calculi is: 
        \emph{can theories of program 
        dynamics be given in such calculi?}
       A possible path towards an answer is showing 
       that reduction-based operational semantics can be given 
       inside suitable extensions of the augmented calculus of relations. 
       Such extensions should be obtained in a rather uniform way
       by defining a relational counterpart of Felleisen-style evaluation contexts~\cite{felleisen-evaluation-context} methodology. 
        Accordingly, the specification of an operational dynamics, such as a call-by-name one, 
        is given not syntactically by means of suitable evaluation contexts, 
        but relationally throughout context operators 
        defining the action of evaluation contexts on relations. 
        Notice that this approach closely relates to reduction strategies,
        and we can see context operators as refining the compatible refinement operator 
        used in this work
        (morally, the latter operator would be recovered as the context operator 
        regarding any context
        as an evaluation context). 
        Following this idea, it becomes interesting to focus \emph{not} on explicit definitions
        of operational dynamics, but on suitable axiomatics on 
        context operators ensuring desirable semantic properties: 
        for instance, rather than giving an explicit definition of a call-by-name semantics 
        (which is language specific), we may prove that any lax functorial context operator 
        behaves in such and such way, showing only in a second moment that, on suitable families 
        of languages, a call-by-name dynamics induces such an operator.
    \item \emph{Computational Effects.} Operational behaviours being oftentimes 
        effectful, it is desirable to have extensions of the relational framework accounting 
        for the production of computational effects. One way to 
        introduce them in operational semantics is 
        by means of monadic evaluation 
        semantics~\cite{PlotkinPower/FOSSACS/01,DalLagoGavazzoLevy/LICS/2017}; another, 
        approach, that seems better suited for our purposes, is the one of \emph{monadic rewriting}~\cite{Gavazzo-Faggian-2021}. 
        The latter develops a general relational theory of \emph{abstract} reduction 
        systems with (monadic) computational effects relying on monadic relations which, 
        roughly, can be seen as arrows in the Kleisli allegory of the monad modelling computational 
        effects. Unfortunately, monadic rewriting has been developed for 
        abstract systems only, and no extension of the theory to syntax-based systems 
        is currently available. The theory developed in this paper suggests 
        that the 
        key to account for both syntax-based and effectful rewriting relies  
        on the combination of monads for syntax and for computational effects, at an allegorical level. 
\end{varenumerate}

\section*{Acknowledgment}
The author would like to thank 
the anonymous reviewers for their 
helpful observations. Special thanks go to
Filippo Bonchi, Francesco Dagnino, Ugo Dal Lago, 
and Simone Martini.

\bibliographystyle{IEEEtran}
\bibliography{main}

\end{document}

%% file: macros.tex
\usepackage{amsthm}
\usepackage{sansmath}
\usepackage{mathtools}
\usepackage{mathrsfs}
\usepackage{stackengine}
\usepackage[bbgreekl]{mathbbol}
\usepackage{multicol}
\usepackage{proof}
\usepackage[all]{xy}
\usepackage{xcolor}
\usepackage{epigraph}
\usepackage{qtree}
\usepackage{framed}

\usepackage{accents}

\usepackage{bm}
\usepackage{url}
\usepackage{stmaryrd}
\usepackage{thm-restate}
\usepackage{float}
\usepackage{caption}
\usepackage{graphicx}
\usepackage{tikz}
\usepackage{tcolorbox}

\makeatletter
\newcommand{\circlearrow}{}% just in case
\DeclareRobustCommand{\circlearrow}{%
  \mathrel{\vphantom{\rightarrow}\mathpalette\circle@arrow\relax}%
}
\newcommand{\circle@arrow}[2]{%
  \m@th
  \ooalign{%
    \hidewidth$#1\circ\mkern1mu$\hidewidth\cr
    $#1\longrightarrow$\cr}%
}
\makeatother

\makeatletter
\DeclareRobustCommand{\circleleftarrow}{%
  \mathrel{\vphantom{\leftarrow}\mathpalette\circleleft@arrow\relax}%
}
\newcommand{\circleleft@arrow}[2]{%
  \m@th
  \ooalign{%
    \hidewidth$#1\circ\mkern1mu$\hidewidth\cr
    $#1\longleftarrow$\cr}%
}
\makeatother

\newcommand{\code}[1]{{\fontfamily{cmtt}\selectfont \text{#1}}}

\newtheorem{theorem}{Theorem}
\newtheorem{proposition}{Proposition}
\newtheorem{lemma}{Lemma}
\newtheorem{corollary}{Corollary}

\theoremstyle{definition}
\newtheorem{definition}{Definition}
\newtheorem{axiom}{Assumption}

\newtheorem{example}{Example}

\newtheorem{remark}{Remark}

\usepackage{hyperref}

\makeatletter

 \def\Autoref#1{%
   \begingroup
   \edef\reserved@a{\cpttrimspaces{#1}}%
   \ifcsndefTF{r@#1}{%
     \xaftercsname{\expandafter\testreftype\@fourthoffive}
       {r@\reserved@a}.\\{#1}%
   }{%
     \ref{#1}%
   }%
   \endgroup
 }
 \def\testreftype#1.#2\\#3{%
   \ifcsndefTF{#1autorefname}{%
     \def\reserved@a##1##2\@nil{%
       \uppercase{\def\ref@name{##1}}%
       \csn@edef{#1autorefname}{\ref@name##2}%
       \autoref{#3}%
     }%
     \reserved@a#1\@nil
   }{%
     \autoref{#3}%
   }%
 }
\makeatother

\definecolor[named]{ACMBlue}{cmyk}{1,0.1,0,0.1}
\definecolor[named]{ACMYellow}{cmyk}{0,0.16,1,0}
\definecolor[named]{ACMOrange}{cmyk}{0,0.42,1,0.01}
\definecolor[named]{ACMRed}{RGB}{235,83,60}
\definecolor[named]{ACMLightBlue}{cmyk}{0.49,0.01,0,0}
\definecolor[named]{ACMGreen}{cmyk}{0.20,0,1,0.19}
\definecolor[named]{ACMPurple}{RGB}{134,72,143}
\definecolor[named]{ACMDarkBlue}{RGB}{35,85,147}
\newenvironment{varitemize}
{
\begin{list}{\labelitemi}
{\setlength{\itemsep}{0pt}
 \setlength{\topsep}{0pt}
 \setlength{\parsep}{0pt}
 \setlength{\partopsep}{0pt}
 \setlength{\leftmargin}{10pt}
 \setlength{\rightmargin}{0pt}
 \setlength{\itemindent}{0pt}
 \setlength{\labelsep}{5pt}
 \setlength{\labelwidth}{10pt}
}}
{
 \end{list}
}
\newcounter{numberone}
\newenvironment{varenumerate}
{
\begin{list}{\arabic{numberone}.}
{
  \usecounter{numberone}
  \setlength{\itemsep}{0pt}
  \setlength{\topsep}{0pt}
  \setlength{\parsep}{0pt}
  \setlength{\partopsep}{0pt}
  \setlength{\leftmargin}{15pt}
  \setlength{\rightmargin}{0pt}
  \setlength{\itemindent}{0pt}
  \setlength{\labelsep}{5pt}
  \setlength{\labelwidth}{15pt}
}}
{
\end{list}
} 

\newcommand{\valone}{\mathtt{v}}
\newcommand{\valtwo}{\mathtt{w}}
\newcommand{\varone}{\mathtt{x}}
\newcommand{\vartwo}{\mathtt{y}}

\newcommand{\termone}{\mathtt{t}}
\newcommand{\termtwo}{\mathtt{s}}
\newcommand{\termthree}{\mathtt{u}}

\newcommand{\vars}{V}

\newcommand{\vect}[1]{\bar{#1}}

\newcommand{\signatureop}{\mathtt{\Sigma}}
\newcommand{\signature}{\Sigma}

\newcommand{\op}{o}

\newcommand{\val}[1]{\mathtt{#1}}

\newcommand{\cc}{\cdots}
\newcommand{\hhh}{\hdots}

\newcommand{\allegory}{\mathcal{A}}
\newcommand{\mapp}[1]{\bm{\mathit{Map}}(#1)}

\newcommand{\category}{\mathcal{C}}
\newcommand{\rell}[1]{\bm{\mathit{Rel}}(#1)}

\newcommand{\monad}{\mathcal{R}}
\newcommand{\unit}{\eta}
\newcommand{\multiplication}{\rho}

\newcommand{\relone}{a}
\newcommand{\reltwo}{b}
\newcommand{\relthree}{c}

\newcommand{\barr}[1]{\overline{#1}}

\newcommand{\join}{\bigvee}

\def\tobar{\mathrel{\mkern3mu  \vcenter{\hbox{$\scriptscriptstyle+$}}%
            \mkern-12mu{\to}}}
\newcommand{\torel}{\tobar}

\newcommand{\idrel}{\Delta}

\newcommand{\strength}{\varpi}

\newcommand{\laxcommuterel}{\ar @{} [dr] |\subseteq}

\newcommand{\dual}[1]{#1^{\scriptstyle \circ}}

\newcommand{\relator}{\Gamma}

\newcommand{\substmap}{\varsigma}

\newcommand{\numeral}[1]{\bm{#1}}

\newcommand{\defeq}{\triangleq}

\newcommand{\terms}[2]{\mathcal{T}(#2)}
\newcommand{\variables}{X}

\newcommand{\stepto}{\mapsto}

\definecolor{cmykorange}{cmyk}{0,0.5,1,0}
\definecolor{cmykred}{cmyk}{0,1,1,0}
\definecolor{cmykyellow}{cmyk}{0,0,1,0}
\definecolor{cmykblue}{cmyk}{1,1,0,0}
\definecolor{cmykgreen}{cmyk}{1,0,1,0}
\definecolor{cmykviolet}{cmyk}{0.6,0.9,0,0.2}
\definecolor{cmykblack}{cmyk}{0,0,0,1}

\newcommand{\G}{\texttt{G}}

\newcommand{\relbot}{\bot}

\newcommand{\trs}{TRS}
\newcommand{\ars}{ARS}

\newcommand{\bext}[1]{\barr{#1}}
\newcommand{\bextt}[1]{\overline{#1}}

\newcommand{\srelator}{\Gamma}

\newcommand{\functor}{\Sigma}

\newcommand{\catone}{\mathcal{C}}
\newcommand{\nom}{\mathit{Nom}}
\newcommand{\set}{\mathit{Set}}
\newcommand{\rel}{\mathit{Rel}}
\newcommand{\nats}{\mathcal{N}}
\newcommand{\alg}{\mathit{Alg}}
\newcommand{\ctxcat}{\mathcal{F}}

\newcommand{\cata}[1]{\llparenthesis #1 \rrparenthesis}
\newcommand{\freemonad}{\mathcal{S}}
\newcommand{\lfp}[2]{\mu #1.#2}
\newcommand{\lfpp}[1]{\mu#1}

\newcommand{\seqq}{\Gamma}

\makeatletter
\let\save@mathaccent\mathaccent
\newcommand*\if@single[3]{%
  \setbox0\hbox{${\mathaccent"0362{#1}}^H$}%
  \setbox2\hbox{${\mathaccent"0362{\kern0pt#1}}^H$}%
  \ifdim\ht0=\ht2 #3\else #2\fi
  }
\newcommand*\rel@kern[1]{\kern#1\dimexpr\macc@kerna}
\newcommand*\widebar[1]{\@ifnextchar^{{\wide@bar{#1}{0}}}{\wide@bar{#1}{1}}}
\newcommand*\wide@bar[2]{\if@single{#1}{\wide@bar@{#1}{#2}{1}}{\wide@bar@{#1}{#2}{2}}}
\newcommand*\wide@bar@[3]{%
  \begingroup
  \def\mathaccent##1##2{%
    \let\mathaccent\save@mathaccent
    \if#32 \let\macc@nucleus\first@char \fi
    \setbox\z@\hbox{$\macc@style{\macc@nucleus}_{}$}%
    \setbox\tw@\hbox{$\macc@style{\macc@nucleus}{}_{}$}%
    \dimen@\wd\tw@
    \advance\dimen@-\wd\z@
    \divide\dimen@ 3
    \@tempdima\wd\tw@
    \advance\@tempdima-\scriptspace
    \divide\@tempdima 10
    \advance\dimen@-\@tempdima
    \ifdim\dimen@>\z@ \dimen@0pt\fi
    \rel@kern{0.6}\kern-\dimen@
    \if#31
      \overline{\rel@kern{-0.6}\kern\dimen@\macc@nucleus\rel@kern{0.4}\kern\dimen@}%
      \advance\dimen@0.4\dimexpr\macc@kerna
      \let\final@kern#2%
      \ifdim\dimen@<\z@ \let\final@kern1\fi
      \if\final@kern1 \kern-\dimen@\fi
    \else
      \overline{\rel@kern{-0.6}\kern\dimen@#1}%
    \fi
  }%
  \macc@depth\@ne
  \let\math@bgroup\@empty \let\math@egroup\macc@set@skewchar
  \mathsurround\z@ \frozen@everymath{\mathgroup\macc@group\relax}%
  \macc@set@skewchar\relax
  \let\mathaccentV\macc@nested@a
  \if#31
    \macc@nested@a\relax111{#1}%
  \else
    \def\gobble@till@marker##1\endmarker{}%
    \futurelet\first@char\gobble@till@marker#1\endmarker
    \ifcat\noexpand\first@char A\else
      \def\first@char{}%
    \fi
    \macc@nested@a\relax111{\first@char}%
  \fi
  \endgroup
}
\makeatother

\newcommand\symNet{\lower3pt\hbox{$\includegraphics[width=20pt]{images/symmetryalt.pdf}$}}
\newcommand\Idnet{\lower3pt\hbox{$\includegraphics[width=20pt]{images/id.pdf}$}}
\newcommand{\ZeronetT}{\lower4pt\hbox{$\includegraphics[width=14pt]{images/idzerocircuit.pdf}$}}

\newcommand{\topos}{\mathcal{E}}
\newcommand{\relext}[1]{\barr{#1}}
\newcommand{\lcp}{LCP}

\newcommand{\fp}[1]{\mu x.#1}
\newcommand{\fpp}[1]{\mu#1}

\newlength{\dhatheight}

\newcommand{\fpinitiall}[1]{\mu#1}

\newcommand{\new}{\nu}
\newcommand{\nameabs}[2]{#2^#1}

\newcommand{\compref}[1]{\widehat{#1}}

\newcommand{\parr}[1]{#1^{\scriptscriptstyle \mathbf{P}}}

\newcommand{\pp}{^{\scriptscriptstyle \textbf{P}}}
\newcommand{\ppdd}{^{{\scriptscriptstyle \textbf{P}}\circ}}
\newcommand{\ddpp}{^{\circ {\scriptscriptstyle \textbf{P}}}}
\newcommand{\ppstar}{^{{\scriptscriptstyle \textbf{P}}*}}

\newcommand{\ppstardd}{^{{\scriptscriptstyle \textbf{P}}* \circ}}

\newcommand{\spp}{^{\scriptscriptstyle \textbf{SP}}}
\newcommand{\sppdd}{^{{\scriptscriptstyle \textbf{SP}}\circ}}

\newcommand{\sppstar}{^{{\scriptscriptstyle \textbf{SP}}*}}
\newcommand{\sppstardd}{^{{\scriptscriptstyle \textbf{SP}}* \circ}}

\newcommand{\ff}{^{\scriptscriptstyle \textbf{F}}}

\newcommand{\ffstar}{^{{\scriptscriptstyle \textbf{F}}*}}

\newcommand{\sff}{^{\scriptscriptstyle \textbf{SF}}}
\newcommand{\sffdd}{^{{\scriptscriptstyle \textbf{SF}}\circ}}

\newcommand{\hh}{^{\scriptscriptstyle \textbf{H}}}

\newcommand{\shh}{^{\scriptscriptstyle \textbf{SH}}}
\newcommand{\shhdd}{^{{\scriptscriptstyle \textbf{SH}}\circ}}

\newcommand{\scccc}{^{\scriptscriptstyle \textbf{SCC}}}

\newcommand{\lin}{^{\scriptscriptstyle \textbf{L}}}

\newcommand{\ccc}{^{\scriptscriptstyle \textbf{C}}}
\newcommand{\scc}{^{\scriptscriptstyle \textbf{SC}}}

\renewcommand{\gg}{\mathbin{\text{\guillemetright}}}

\newcommand{\expr}{\textit{E}}

\newcommand{\relx}{s}
\newcommand{\compreff}[1]{\widetilde{#1}}
\newcommand{\ideta}{\mathtt{I}_\eta}

\DeclareSymbolFont{yhlargesymbols}{OMX}{yhex}{m}{n}

\DeclareMathAccent{\widetriangle}{\mathord}{yhlargesymbols}{"E6}

\newcommand{\eqq}{^{\scriptscriptstyle =}}
\newcommand{\ddeqq}{^{\scriptscriptstyle {\circ}{=}}}